%% file: causal_optimization.tex
\documentclass[twoside,11pt]{article}

 \usepackage{jmlr2e}
\usepackage[usenames,dvipsnames,table,xcdraw]{xcolor}
\usepackage{microtype}
\usepackage{graphicx}
\usepackage{booktabs} %
\usepackage[all]{hypcap}
\hypersetup{colorlinks = true}
\hypersetup{citecolor=MidnightBlue}
\hypersetup{linkcolor=black}
\hypersetup{urlcolor=Violet}
\usepackage{mathtools}
\usepackage{hyperref}
\usepackage{amssymb}
\usepackage{amsmath}
\usepackage{dsfont}

\usepackage{placeins}
\usepackage{booktabs}

\input{preamble/preamble}
\input{preamble/preamble_acronyms}

\usepackage{cleveref}
\usepackage{enumitem}
\usepackage{setspace}
\graphicspath{ {images/} }
\usepackage{algorithm}
\usepackage{algorithmic}
\usepackage[algo2e, ruled]{algorithm2e}

\usepackage[caption=false]{subfig}
\usepackage[normalem]{ulem}
\usepackage{pifont}
\usepackage{makecell}
\usepackage{nccmath}
\newcolumntype{C}[1]{>{\centering\arraybackslash}p{#1}}

\newtheorem{assumption}{Assumption}

\jmlrheading{1}{2022}{1-48}{4/00}{10/00}{meila00a}{Mingzhang Yin, Yixin Wang and David M. Blei}

\firstpageno{1}

\DeclareRobustCommand{\parhead}[1]{\textbf{#1}~}

\PassOptionsToPackage{table,xcdraw}{xcolor}

\usepackage{lastpage}
\jmlrheading{25}{2024}{1-\pageref{LastPage}}{8/21; Revised
12/22}{4/24}{21-1028}{Mingzhang Yin, Yixin Wang, David M. Blei}
\ShortHeadings{Optimization-based Causal Estimation}{Yin, Wang, Blei}

\begin{document}

\title{Optimization-based Causal Estimation \\from Heterogeneous Environments}

\author{\name Mingzhang Yin  \email mingzhang.yin@warrington.ufl.edu \\
       \addr Warrington College of Business\\
       University of Florida\\
       Gainesville, FL, 32611, USA
       \AND
       \name Yixin Wang \email yixinw@umich.edu \\
       \addr Department of Statistics \\
       University of Michigan\\
       Ann Arbor, MI, 48109, USA
       \AND
       \name David M. Blei  \email david.blei@columbia.edu \\
       \addr Department of Computer Science and Department of Statistics\\
       Columbia University\\
       New York, NY, 10027, USA
     }

\editor{Pradeep Ravikumar}
\maketitle

\begin{abstract}%

  This paper presents a new optimization approach to causal
  estimation. Given data that contains covariates and an outcome,
  which covariates are causes of the outcome, and what is the strength
  of the causality? In classical machine learning (ML), the goal of
  optimization is to maximize predictive accuracy. However, some
  covariates might exhibit a non-causal association with the outcome. Such
  spurious associations provide predictive power for classical ML, but
  they prevent us from causally interpreting the result. This paper
  proposes \gls{CoCO}, an optimization algorithm that bridges the gap
  between pure prediction and causal inference. \gls{CoCO} leverages
  the recently-proposed idea of environments~\citep{peters2016causal,arjovsky2019invariant}, datasets of
  covariates/response where the causal relationships remain invariant
  but where the distribution of the covariates changes from
  environment to environment. Given datasets from multiple
  environments---and ones that exhibit sufficient
  heterogeneity---\gls{CoCO} maximizes an objective for which the only
  solution is the causal solution. We describe the theoretical
  foundations of this approach and demonstrate its effectiveness on
  simulated and real datasets. Compared to classical ML and existing
  methods, \gls{CoCO} provides more accurate estimates of the causal
  model and more accurate predictions under interventions.

\end{abstract}

\section{Introduction}
\label{sec:intro}
\input{intro}

\section{Causal optimization from heterogeneous environments}
\label{sec:method}

\input{method}

\section{Connections to invariant risk minimization}
\label{sec:connection}
\input{connection}

\section{Identification with heterogeneous environments}
\label{sec:identify}
\input{identification}

\section{Extensions to the nonlinear model}
\label{sec:nonlinear}
\input{extension}

\section{Empirical studies}
\label{sec:experiment}
\input{empirical}

\section{Conclusion} This paper formulates causal estimation as a constrained
optimization problem. Applying directional derivative methods to this optimization problem, we propose the
\gls{CoCO} objective, a computationally tractable optimization method for
estimating causal coefficients with datasets from multiple
environments. Theoretically, we discussed the necessary and sufficient conditions by which the causal coefficients are identified by optimizing the CoCo objective. We discuss the mathematical connection between
\gls{CoCO} and \gls{IRM}. In empirical studies, we find that
\gls{CoCO} produces accurate causal estimation and distributionally
robust predictions. \gls{CoCO} is applicable to high-dimensional data
and to linear and nonlinear models.

Looking ahead, we think several problems are worthy of further exploration. One direction is to consider the situations when there is an unobserved confounder as a direct cause. In this case, a potential approach is to connect environments and instrumental variables. Another direction is to further understand the interplay between the type of interventions, the number of environments, and the identification of causal coefficients, especially for nonlinear models. Such understanding can enable causal estimation with a minimal number of environments by methods like active learning. 

\section*{Acknowledgments}
The authors thank the Editor, Action Editors, and anonymous
reviewers for helpful and constructive comments. Mingzhang
Yin thanks the support from the Data Science Institute and the Irving Institute for Cancer Dynamics, Columbia University. The authors acknowledge the fundings from Warrington College of Business, Office of Naval Research under grant number N00014-23-1-2590, the National Science Foundation under Grant No. 2231174 and No. 2310831, NSF IIS-2127869, NSF DMS-2311108, ONR N000142412243, and Simons Foundation.

\appendix

\input{appendix}

\FloatBarrier

\bibliography{references}

\end{document}

%% file: preamble/preamble.tex
\usepackage[nameinlink]{cleveref}

\usepackage[acronyms,nowarn]{glossaries-extra}
\glsdisablehyper{}

\crefname{figure}{fig.}{figs.}
\Crefname{figure}{Fig.}{Figs.}
\crefname{equation}{eq.}{eqs.}
\Crefname{equation}{Eq.}{Eqs.}
\crefname{algorithm}{alg.}{algs.}
\Crefname{algorithm}{Alg.}{Algs.}
\crefname{section}{\S}{\S\S}
\Crefname{section}{\S}{\S\S}
\crefname{lemma}{lemma}{lemmas}
\Crefname{lemma}{Lemma}{Lemmas}
\crefname{theorem}{theorem}{theorems}
\Crefname{theorem}{Theorem}{Theorems}
\crefname{proposition}{proposition}{propositions}
\Crefname{proposition}{Proposition}{Propositions}

\DeclareRobustCommand{\parhead}[1]{\textbf{#1}~}

\usepackage{lineno}

\usepackage{ragged2e}

\newcounter{parcount}

\DeclareRobustCommand{\parhead}[1]{\textbf{#1}~}

\DeclareMathOperator*{\argmin}{arg\,min}

\newcommand{\norm}[1]{\left\lVert#1\right\rVert}
\newcommand{\ip}[2]{\langle #1,#2 \rangle}
\newcommand{\idx}[1]{\{1,2,\cdots,#1\}}

\newcommand{\pde}[1]{\frac{\partial}{\partial #1}}
\usepackage{stackengine}

\makeatletter
\newcommand{\distas}[1]{\mathbin{\overset{#1}{\kern\z@\sim}}}%

\newcommand{\bas}[1]{\begin{align*}#1\end{align*}}
\newcommand{\bac}[1]{\begin{align}\begin{split}#1\end{split}\end{align}}
\newcommand{\ba}[1]{\begin{align}#1\end{align}}

\newcommand{\bE}{\mathbb{E}}
\newcommand{\bR}{\mathbb{R}}

\newcommand{\cL}{\mathcal{L}}

\newcommand{\cN}{\mathcal{N}}
\newcommand{\cP}{\mathcal{P}}

\newcommand{\cD}{\mathcal{D}}
\newcommand{\cE}{\mathcal{E}}
\newcommand{\cI}{\mathcal{I}}

\newcommand{\cC}{\mathcal{C}}

\newcommand{\bzero}{\mathbf{0}}

\newcommand{\beqs}{\vspace{0mm}\begin{eqnarray}}
\newcommand{\eeqs}{\vspace{0mm}\end{eqnarray}}
\newcommand{\barr}{\begin{array}}
\newcommand{\earr}{\end{array}}
\newcommand{\Amat}[0]{{{\bf A}}}
\newcommand{\Bmat}[0]{{{\bf B}}}

\newcommand{\Dmat}{{\bf D}}

\newcommand{\Imat}{{\bf I}}

\newcommand{\Wmat}[0]{{{\bf W}}}
\newcommand{\Xmat}[0]{{{\bf X}}}
\newcommand{\Ymat}[0]{{{\bf Y}}}

\newcommand{\gv}[0]{{\boldsymbol{g}} }

\newcommand{\uv}{\boldsymbol{u}}
\newcommand{\vv}{\boldsymbol{v}}

\newcommand{\xv}{\boldsymbol{x}}

\newcommand{\zv}{\boldsymbol{z}}

\newcommand{\alphav}{\boldsymbol{\alpha}}
\newcommand{\betav}[0]{{\boldsymbol{\beta}}}
\newcommand{\gammav}[0]{{\boldsymbol{\gamma}} }
\newcommand{\deltav}[0]{{\boldsymbol{\delta}} }
\newcommand{\epsilonv}{\boldsymbol{\epsilon}}

\newcommand{\thetav}{\boldsymbol{\theta}}

\newcommand{\muv}[0]{{\boldsymbol{\mu}}}

\newcommand{\E}{\mathbb{E}}

\newcommand{\half}{\frac{1}{2}}

\newcommand{\one}{\mathbf{1}}
\newcommand{\zero}{\mathbf{0}}
\newcommand{\indep}{\!\perp\!\!\!\perp}

\newcommand{\rmp}{\mathrm{p}}
\newcommand{\g}{\, | \,}
\newcommand{\coco}{\mathrm{coco}}

\makeatletter
\providecommand{\leftsquigarrow}{%
 \mathrel{\mathpalette\reflect@squig\relax}%
}
\newcommand{\reflect@squig}[2]{%
 \reflectbox{$\m@th#1\rightsquigarrow$}%
}
\makeatother

%% file: preamble/preamble_acronyms.tex
\newacronym{DGP}{DGP}{data generating process}
\newacronym{SEM}{SEM}{structural equation model}
\newacronym{OLS}{OLS}{ordinary least square}
\newacronym{IRM}{IRM}{invariant risk minimization}
\newacronym{ERM}{ERM}{empirical risk minimization}
\newacronym{CoCO}{CoCo}{constrained causal optimization}

%% file: intro.tex
We consider the problem posed in \cite{peters2016causal}. An outcome $y$ is
generated according to a linear structural equation model (\gls{SEM})
based on a set of covariates $\xv^*$~\citep{pearl2009causality},
\begin{align}
  y \leftarrow (\betav^*)^\top \xv^* + \varepsilon,  ~~\xv^* \indep \varepsilon.
  \label{eq:lSEM}
\end{align}
Here, $\xv^*$ are the causal parents of the outcome $y$, which are independent of the unobserved noise $\varepsilon$,  and the \textit{causal coefficient} $\betav^*$ represents the direct causal effects of $\xv^*$.  In practice, we might measure a potentially large set of covariates $\xv = \{x_1, \cdots, x_p\}$, such that the causes $\xv^*$ are a subset of $\xv$. More precisely, there is a subset $S \subset \{1,2, \cdots, p\}$ with $\xv^* = \xv_S$, but we do not know this set of causes $S$.  Our goal is to infer the causal relationship between each covariate $x_j$ and the outcome in the data generating process (DGP), and estimate their direct causal effects.

The challenge to solving this problem is \textit{spurious association}. Though the causes $\xv_S$ are assumed to be exogenous, the other observed covariates might depend on the noise term with $\xv_{\backslash S} \not\!\perp\!\!\!\perp \varepsilon$.  The dependency may arise because of confounding, where there is an omitted common cause of $y$ and $\xv_{\backslash S}$, i.e., $y \leftarrow \epsilon \to \xv_{\backslash S}$;  the dependency may also arise from a collider, reverse causality, or selection bias, e.g., $y  \to \xv_{\backslash S}$ or $y \to x_1 \leftarrow x_2,~x_1,x_2 \in \xv_{\backslash S}$.  It leads to spurious association because the covariates $\xv_{\backslash S}$ will not be the genuine causal parents of the outcome in the DGP, even though they correlate with the outcome. In the face of spurious association, typical inference methods, such as regression based on empirical risk minimization (ERM) or maximum likelihood estimation (MLE), will lead to a biased estimate of the causal effect and an incorrect interpretation of the causal relationships \citep{angrist1996identification, peters2016causal, efron2020prediction}. From the predictive perspective, a model that captures spurious association will not generalize to non-i.i.d. data under perturbations of the DGP \citep{arjovsky2019invariant,rothenhausler2018anchor}.  ~\looseness=-1

We will develop \textit{constrained causal optimization} (CoCo), an
optimization-based method to solve the spurious association problem. The key idea behind
CoCo is to leverage datasets from \textit{multiple environments} \citep{peters2016causal}. The environments are a set of heterogeneous DGPs. In each, the causal mechanism of the outcome generation %
remains invariant but the distribution of the
covariates changes from environment to environment. 
While classical ML methods cannot distinguish the direct causes from spurious
association, simultaneously analyzing data from multiple environments
allows us to triangulate on the correct causal coefficients. 
This work builds on recent research about multi-environment
estimation, beginning with the seminal work of \citet{peters2016causal}
and continuing with the risk-based algorithms 
of~\citet{arjovsky2019invariant}. The method developed in this paper
builds and improves on the risk-based algorithms.

We will describe the method here and derive it in the subsequent
sections. To begin, consider a single data-generating distribution
$\rmp(\xv, y) = \rmp(\xv) \rmp(y \g x)$. For instance, the conditional
distribution of the outcome could come from \Cref{eq:lSEM}. Consider a
 predictor $\hat{y}(\xv; \alphav)$ with coefficients $\alphav$ and define a
non-negative loss function to measure the fidelity of a prediction,
$\ell: \mathcal{Y} \times \mathcal{Y} \to \bR$ (e.g., squared loss).
Finally, define the risk of the predictor to be the expectation of the
loss relative to the data-generating distribution, \begin{align}
  R(\alphav) = \E_{\xv,y \sim \rmp(\xv,y)}[\ell(\hat{y}(\xv; \alphav),
  y)].
\label{eq:riskfun}
\end{align}
Classical machine learning (ML) provides methods that analyze data
from $\rmp(\xv, y)$ to  minimize the risk. The pure prediction methods leverage all information in data, causal and non-causal, to reduce the predictive error quantified by the risk function.  The spurious associations
improve predictions, but they bias the resulting estimates of the
coefficients away from the true causal coefficients.

We consider
distributions of data from a set of environments $\cE$. The
data from environment $e$ form a joint distribution
$\rmp^e(\xv, y) = \rmp^e(\xv) \rmp(y \g \xv)$. In this joint, the
distribution of covariates $p^e(\xv)$ changes from environment to environment,
but the conditional expectation of the outcome given its causal parents $\E(y \g \text{Pa}(y))$ is the same across environments, governed by the same SEM.  %
Notice each
environment is associated with its own risk $R^e(\alphav)$, since the
risk is an expectation with respect to the per-environment
distribution.

Given a set of datasets from multiple environments, the CoCo algorithm
solves the following optimization,
\begin{align}
  \alphav_{\coco} =
  \argmin_{\alphav}
    \frac{1}{|\cE|}
    \sum_{e \in \cE} \big(\norm{\nabla R^e(\alphav) \circ \alphav}_2 \big),
  \label{eq:coco-intro}
\end{align}
where $\circ$ is the Hadamard product and each environment's risk $R^e(\alphav)$ can be approximated by its
empirical estimate.

As we describe below, this objective stems from
the idea of the directional
derivative~\citep{rudin1964principles,marban1969directional}, and its
role in the first-order conditions for the optimizer of each
environment's risk function. Under invariance assumptions, the solution to \Cref{eq:coco-intro} is
the intersection across environments of all points that minimize
each term $\norm{\nabla R^e(\alphav) \circ \alphav}_2$. When the environments are sufficiently heterogenous---that is,
when there is enough variety in different $\rmp^e(\xv)$ and outcome noise---this
optimization is solved by the causal coefficients.

\Cref{eq:coco-intro} is the basic optimization problem behind CoCo.  The
rest of this paper presents its theoretical foundations, the
assumptions under which its solution is the true causal coefficients,
algorithms building on and solving \Cref{eq:coco-intro} from multi-environment data, and
studies about the performance of CoCo on several simulated and real
datasets, both with linear predictors and with nonlinear predictors.
When compared to classical ML and Invariant Risk Minimization (IRM) \citep{arjovsky2019invariant}-related methods, CoCo improves the estimation accuracy
of causal coefficients and the predictive accuracy in new
environments.

Broadly, CoCo represents progress towards multi-environment
optimization for causal estimation, and it helps explain the empirical
the success of IRM when data is linear-Gaussian or
linear-Bernoulli. Compared to IRM, CoCo requires fewer environments to
identify causal coefficients and enjoys a more stable training
procedure.  Practically, CoCo is compatible with general graph
structures, can be used with flexible ML tools such as deep neural
networks, scales well to high-dimensional problems, and is easy to
implement.

The paper is organized as follows.  \Cref{sec:related} situates this
paper in the broader landscape of the research literature on
multi-environment analysis.  \Cref{sec:method} describes the
methodology and theoretical basis for CoCo. \Cref{sec:connection}
presents the connections between CoCo and IRM in mathematical
detail. \Cref{sec:identify} studies the identification properties in
detail, particularly in the setting of a linear SEM, and
\Cref{sec:nonlinear} extends CoCo to nonlinear models.
\Cref{sec:experiment} presents a study of CoCo with synthetic,
semi-synthetic and real-world data.

\section{Related work}
\label{sec:related}

Formulating the goals of data analysis, and their relationship to
optimization, has long been discussed in the research literature in
statistics and machine learning.  As early as
\citet{wright1921correlation}, researchers recognized that the goals
of prediction and estimation are not always aligned in optimization.
\citet{shmueli2010explain} and \citet{efron2020prediction} provide
insightful discussions about the differences between prediction
problems, association problems, and causal estimation problems.

This paper builds on the body of research around causal estimation
with multiple environments. Methodologies developed in this area rest
on the invariance property of causality, also known as autonomy
\citep{haavelmo1944probability}, modularity
\citep{scholkopf2012causal} and stability
\citep{dawid2010identifying}.  In a pioneering work,
\citet{peters2016causal} uses statistical hypothesis testing to
estimate causal structures by exploiting invariance across multiple
environments. Subsequent research extends this idea, for example, to
nonlinear models \citep{heinze2018invariant} and sequential data
\citep{pfister2019invariant}. See \citet{buhlmann2020invariance} for a
comprehensive review.

To improve its flexibility and scalability, researchers have begun to
use optimization over multi-environment data for causal estimation
\citep{rothenhausler2019causal,rothenhausler2018anchor}. One
influential paper along these lines is \citet{arjovsky2019invariant},
which introduces invariant risk minimization (IRM) as a method that
adapts modern predictive models to this task. The objective function
of IRM includes an additional penalty term to empirical risk function,
which encourages a predictor to be invariant across environments.  The
work presented here provides a contribution to optimization-based
causal estimation.

Since its introduction, IRM has been extended in several ways. It has been formulated as game
theory problem \citep{ahuja2020invariant}, combined with meta-learning
methods \citep{bae2021meta}, applied to reinforcement learning
\citep{zhang2020invariant}, and applied to causal inference
\citep{shi2020invariant,lu2021nonlinear}.  Its conditions and
limitations have been studied
\citep{rosenfeld2020risks,kamath2021does,guo2021out}.

Besides IRM, other objectives also aim to improve
predictive accuracy by encouraging invariance of empirical risks
across environments.  These objectives are often built on equal
noise variance assumption, which is a strong version of invariance \citep{peters2014identifiability}. 
Some objectives regularize the variance of the
empirical risks
\citep{xie2020risk,krueger2020out,heinze2021conditional} and control
the worst case risk across training environments
\citep{sagawa2019distributionally}. When the condition of equal
noise variance is met, such regularizations can be combined with  CoCo.

The idea of invariance and environments has also been adapted to
causal discovery
\citep{tian2001causal,yu2019learning,yu2019multi,brouillard2020differentiable,mooij2020joint,
  muller2020learning}. Invariance enables the discovery of causal
structures within Markov equivalence classes
\citep{ghassami2020causal}, which cannot be reconstructed from
traditional single environment data \citep{spirtes2000causation}.
Nevertheless, existing methods might encounter problems of limited model
flexibility and high computational cost. For example, some methods rely on
linear data generating process
\citep{ghassami2018multi,huang2019causal, huang2020causal}, require
regression over multiple subsets of covariates
\citep{ghassami2017learning}, and/or involve multiple independence
testings
\citep{ghassami2017learning,huang2020causal,brouillard2020differentiable}.  CoCo provides an optimization-based method to assist causal discovery by identifying direct causes for observed variables.  

Lastly, CoCo is loosely related to the variable selection literature \citep{hastie2009elements}. It can be viewed as selecting causal variables or learning a causal representation from the observed covariates, but the motivation and the targeted problems are  different. Variable selection methods, such as LASSO \citep{tibshirani1996regression} and best subset selection \citep{bertsimas2016best,yin2020probabilistic}, focus on the situations where the number of data points is small compared to the number of the observed covariates. The model selection under the traditional variable selection setting can generally recover the true sparsity pattern asymptotically \citep{zhao2006model}. In contrast, the estimation bias problem due to the spurious association considered in this paper cannot be solved by increasing the number of data points even to infinity. %
Unlike classic regularized regression with the regularizers often imposing shrinkage on the parameter space \citep{golub1999tikhonov}, the proposed method imposes regularizations in the derivative space with the restricted derivative directions.

%% file: method.tex
We discuss \gls{CoCO}, a method that estimates causal effects via optimization. In \Cref{sec:assumption}, we set up the problem and assumptions. In \Cref{sec:optimality}, we introduce an idealized optimization objective, which produces causal coefficients as the solution, but is intractable. In \Cref{sec:estimation}, we derive a relaxed objective function; it is tractable with observable data but contains extra solutions besides the causal coefficients. In \Cref{sec:invariance}, we aggregate the relaxed objective over multiple environments to whittle down the set of optima to the causal coefficients.

\subsection{Setup and assumptions}
\label{sec:assumption}

Consider an observed multi-environment dataset. Denote $\cE$ as a discrete set of environments.  Each environment $e \in \cE$ specifies a DGP  similar to \Cref{eq:lSEM},
\ba{
y^e \leftarrow \betav^\top\xv^e + \epsilon^e, \quad \xv^e \sim p^e(x_1^e,\cdots,x_p^e). 
\label{eq:lSEM-e}
} 
We absorb the intercept term into $\xv^e$ and $\betav$ and do not write it explicitly. Comparing to  \Cref{eq:lSEM}, here $\betav_S = \betav^*$, $\beta_j \neq 0$ for $j \in S$, and $\betav_{\backslash S} = 0$ where $S \subset \{1,2,\cdots, p\}$. It means only $\xv^e_S = \xv^*$ are the true causes for the set $S$.  The SEM in \Cref{eq:lSEM-e} includes all the observed covariates to reflect the fact that we do not know the set of causes $S$ a priori.  We overload the notation to denote $\betav$ as the \emph{causal coefficient} and call the set $S$ as the \emph{causal structure}.  There could be potential endogeneity with $\xv \not\!\perp\!\!\!\perp\epsilon$.  For each environment, the observed data $\cD^e = (\Xmat^e, \Ymat^e)$ consists of $n^e$ i.i.d. data points, where $\Ymat^e \in \bR^{n^e}$ are the outcomes, and $\Xmat^e = [X^e_1, \cdots, X^e_p] \in \bR^{n^e \times p}$ are the
covariates.  Each column $X_j^e \in \bR^{n^e}, ~j \in \idx{p}$, contains the
observations of the $j$-th covariate for $n^e$ units. ~\looseness=-1

\paragraph{Assumptions for each environment.} First, we specify the assumptions for the data in each environment. For notational simplicity, we suppress the superscript $e$ for now and state the assumptions for any environment $e \in \cE$.  The support set and magnitude of the causal coefficients $\betav$  are unknown. For the covariates $\xv$, we do not specify the DGP and allow an arbitrary functional form of the joint distribution $p(\xv)$.  We will specify the additional assumptions on the joint distribution for identification in \Cref{sec:identify}.

We assume the noise $\epsilon$ is zero-mean, and the covariates and noise have finite variance. The noise term is assumed to be independent of the observed direct causes $\xv_S$. To summarize, the assumptions for each environment are
\begin{assumption}   (i) Linear DGP as \Cref{eq:lSEM-e}; (ii) Moment conditions: $\E[\epsilon] = 0$,  $\text{Var}[\epsilon], \text{Var}[x_j] \linebreak<
\infty$ for all $j \in \idx{p}$; (iii)  Independence: the observed direct causes $\xv_{S} \indep \epsilon$.
\label{assp:sem}
\end{assumption}

\noindent \paragraph{Remark.} \Cref{assp:sem} (i) assumes linearity; we will study the  nonlinear causal models
in \Cref{sec:nonlinear}. \Cref{assp:sem} (ii) is a standard regularity assumption.  
\Cref{assp:sem} (iii)  assumes $\xv_{S} \indep \epsilon$, where the noise $\epsilon$   incorporates the unobserved causes of $y$ as long as they are independent of the observed direct causes $\xv_S$. The noise and the observed covariates $\xv_{\backslash S}$, however,  can be dependent, i.e., $\xv_{\backslash S} \not\!\perp\!\!\!\perp\epsilon$.  Hence, \Cref{assp:sem} (iii) does not imply that the covariates $\xv$ as a whole satisfy the back-door criterion \citep{pearl2009causality}. %
Accordingly, standard inference strategies, such as regression adjustment for all covariates, may produce biased causal estimates \citep{elwert2014endogenous}.  Our goal is to develop a method that can automatically identify the set $S$ as the causal structure, estimate the causal coefficient $\betav$, and make prediction based on the causes of the outcome. 

\Cref{assp:sem} (iii) distinguishes the setting of this paper from that of instrumental variables (IV) which allows the dependency between $\xv_S$ and $\epsilon$ \citep{angrist1996identification}. However, we allow the index of endogenous covariates in $\xv$ to be unknown; such information is usually required by the IV methods such as the two-stage least square.  
\Cref{assp:sem} (iii)  is a common assumption in the invariance-based causal inference  \citep[Theorem 9]{arjovsky2019invariant} \citep{rojas2018invariant,pfister2019invariant,krueger2020out}, and causal discovery literature \citep[Assumption 1]{peters2016causal}\citep{ghassami2017learning,yu2019learning,brouillard2020differentiable}.

\paragraph{Assumptions across environments.} Now, we consider the assumption across environments.  The key property of causality we exploit from multi-environment data is the invariance \citep{peters2016causal, arjovsky2019invariant}. Invariance means that conditional on the same value of the direct causes,  the expectation of the outcome is the same across environments. 
\begin{assumption} [Invariance]
The index set of direct causes of $y^e$ is the same across environments. And given a possible value $\mathbf{c}$ of the direct causes, %
\begin{equation}
  \begin{gathered}
      \E[y^e | \text{Pa}(y^e)=\mathbf{c}] = \E[y^{e'} | \text{Pa}(y^{e'})=\mathbf{c}], 
  \end{gathered}
  \label{eq:invariance}
\end{equation}
for all $e, e' \in \cE$.  
\label{assp:invariance}
\end{assumption} 
The invariance condition in \Cref{eq:invariance} is weaker than that in \citet{peters2016causal}, which requires strong invariance over the conditional outcome distribution
\ba{
p(y^e | \text{Pa}(y^e)=\mathbf{c}) = p(y^{e'} | \text{Pa}(y^{e'})=\mathbf{c}).
 \label{eq:strong-invariance}
}
When  the distributions of covariates $\xv^e$ and noise $\epsilon^e$ change across environments, we call environments $\cE$  \emph{heterogeneous}.  The heterogeneity will be important to the method we derive.

\subsection{Idealized causal optimization}
\label{sec:optimality}

We start by considering the data set of a single environment, such as one that is generated by \Cref{eq:lSEM}. Suppose for the moment that we do not know the causal
coefficients $\betav$, but we do know which covariates are direct
causes of the outcome, i.e., the set $S$. A key observation is that
among the models that share the true causal structure (i.e. the support set of coefficients as $S$), the causal
model  (i.e. the model with causal coefficients) is the best predictive model. This observation, though conceptually straightforward,  makes the goals of causal estimation and optimization consistent. We can then obtain the causal coefficients by solving a constrained optimization problem. 

\begin{lemma}[Causal Optimality]  \label[lemma]{lem:optimal}
Under Assumptions~\ref{assp:sem}, for the squared risk function $R(\alphav) =
  \bE[(1/2)(\hat{y}(\xv; \alphav) - y)^2]$, and linear predictor $\hat{y}(\xv; \alphav)=\alphav^\top\xv$, the following constrained
optimization problem 
\bac{
\min_{\alphav} & ~R(\alphav)  \\
\mathrm{s.t.} & ~\alpha_j = 0, \quad j \notin S
\label{eq:constrained}
}
has the causal coefficients $\alphav = \betav$ as the unique solution.
\end{lemma}
The proof is in \Cref{sec:proof}.

\Cref{lem:optimal}  is the ideal situation where the set of causes $S$ in \Cref{eq:constrained} is known. Of course, in practice, we do not know which covariates are causal and which are not.  
This paper aims to build on this idealized
optimization problem in \Cref{eq:constrained}  to construct a tractable objective for causal
estimation. We first introduce a tractable objective with observed data. Then we aggregate this objective over multiple environments to isolate the causal coefficients. 

\subsection{Derivation of a tractable objective}
\label{sec:estimation}

In this section, we derive an optimization objective for causal
estimation. It only involves the observable data  and
is a relaxation of the idealized optimization in
\Cref{eq:constrained}. In this relaxation, the causal coefficient
$\betav$ is one of the optima, though it is not the only one.

We will first review directional derivatives and feasible directions, which are used to characterize the extreme points of the idealized optimization in
\Cref{eq:constrained}. We then relax the optimization problem to a practical one and characterize its extreme points. 

\paragraph{Directional derivatives and feasible directions.} The directional derivative $\mathbf{D}_{\vv}$ in the direction of a unit-length vector $\vv$ is defined as the change rate of a function in that direction \citep{rudin1964principles}.  It can be computed as the inner product of the gradient and the direction vector $ \mathbf{D}_{\vv}R(\alphav)
  \vcentcolon= \lim_{t \to 0}  (R(\alphav + t \vv ) - R(\alphav)) /t =  \ip{\nabla\, R(\alphav)}{\vv}$, where $\ip{\cdot}{\cdot}$   denotes  an inner product.  ~\looseness=-1

Directional derivatives provide the first-order condition for the optima of a constrained optimization. Denote the constraints in \Cref{eq:constrained} as $g_j(\alphav) =
\alpha_j = 0$ for $j \notin S$. (Recall $S$ is the support set of
$\betav$, the indices of the non-zero causal coefficients.) The points that satisfy these constraints form a surface in $\bR^p$.  To find a stationary point while obeying to the constraints, an algorithm should search the parameters in the directions tangent to this surface rather than an arbitrary direction. Hence, the tangent directions are called the  \emph{feasible direction}. The first-order condition for optimality is that the directional derivative in all feasible directions vanishes \citep{marban1969directional}. Intuitively, satisfying the condition means the algorithm cannot improve the objective value, at least locally,  without violating the constraints. ~\looseness=-1

\paragraph{Optimality condition for causal optimization.} We derive the feasible directions for the optimization problem in \Cref{eq:constrained}. We then derive an important property of the causal coefficients from the first-order condition of the directional derivative.

Geometrically, the feasible directions are perpendicular to the normal vectors of the surface defined by the constraints. The normal vectors point in the directions that  violate the constraints at
the maximum rate, which can be computed by taking the gradient of the constraints
\ba{
  d g_j(\alphav) / d \alphav = \mathbf{e}_j,\quad j \notin S. 
  \label{eq:perp}
  } 
The feasible directions are perpendicular to the vector space spanned by the basis vectors in \Cref{eq:perp}. Therefore, all  the feasible directions  form a linear space $\mathcal{U} = \text{span} \{\mathbf{e}_j: j \in S\}$.

The first-order condition for the causal optimization in \Cref{eq:constrained} requires the directional derivative vanishing in all the feasible directions, that is, $\mathbf{D}_{\vv } =0$ for each $\vv \in \mathcal{U}$. This condition can be equivalently stated using the basis vectors of $\mathcal{U}$ as 
\ba{
  \mathbf{D}_{\mathbf{e}_j }R(\alphav)
  = \ip{\nabla\, R(\alphav)}{\mathbf{e}_j} = 0, ~\text{for}~ j \in S.
  \label{eq:first_cond}
} 
This is because any $\vv  \in \mathcal{U}$ is a linear combination of the basis
$\{\mathbf{e}_j\}_{j\in S}$. More compactly, since $\beta_j \neq 0$ for $j \in S$, \Cref{eq:first_cond} can be
written equivalently as
\ba{
  \norm{\nabla  R(\alphav) \circ \betav}_2 = 0
  \label{eq:v2_1}
} where $ \circ $ is the Hadamard product. 

Traditional empirical risk minimization (ERM) searches for $\alphav$ with $\nabla  R(\alphav) = \zero$, or equivalently  $\norm{\nabla  R(\alphav) \circ \one}_2 = 0$. A problem of the gradient-based ERM is its greedy nature: the optimization follows a derivative pointing at the steepest descent direction. Such greedy optimization may leverage the spurious associations and fails to converge to the causal coefficient. 
In comparison, \Cref{eq:v2_1} 
reveals that the key to establish the optimality for the causal model is to regularize the derivative's direction by restricting it to the subspace $\mathcal{U}$.  The regularization of the derivative direction distinguishes CoCo from the classic regression regularizations that often control the parameter magnitudes.

The first order condition in \Cref{eq:v2_1} is an explicit condition that the causal coefficients $\betav$ satisfy. That's to say, solving $\min_{\alphav} \norm{\nabla  R(\alphav) \circ \betav}_2$ would return the causal coefficients for convex risk functions. However, there is no free lunch. Similar to the idealized causal optimization in \Cref{eq:constrained} with an unknown set $S$, the condition in \Cref{eq:v2_1} includes $\betav$, the unknown causal coefficients.  Hence \Cref{eq:v2_1} is not a tractable objective yet.

\paragraph{Relaxing the first-order condition.} Lemma~\ref{lem:optimal} states
that $\alphav = \betav$ is an optimum of the problem in
\Cref{eq:constrained}. Therefore, the vector $\betav $ must satisfy the first-order
condition of \Cref{eq:v2_1}. Plugging $\alphav = \betav$ into \Cref{eq:v2_1}, we have $\norm{\nabla R( \betav) \circ
  \betav}_2 = 0$. This fact reveals that the causal
coefficient $\betav$  is an optimum of the following optimization
problem,
\ba{
  \min_{\alphav} \norm{\nabla R(\alphav) \circ \alphav}_2.
  \label{eq:v2}
} Notice that \Cref{eq:v2} is an optimization problem based entirely
on the observational data. %
It states that  the  causal coefficient $\betav$ satisfies 
\ba{
\betav \in \argmin_{\alphav} \norm{\nabla R(\alphav) \circ \alphav}_2.
} 
\Cref{eq:v2} is an important step towards an optimization-based causal estimation. The derivation based on the directional derivative theory also reveals an intrinsic connection and distinction between the CoCo objective and the invariance risk minimization (IRM) objective \citep{arjovsky2019invariant}, which we will discuss in \Cref{sec:connection}. 

We call the set of points that minimizes \Cref{eq:v2} the
\emph{plausible set} $\mathcal{F}$. We have shown that causal
coefficients belong to it. What other
points are in $\mathcal{F}$?

For each $j \in \{1, 2, \ldots, p\}$ the objective of \Cref{eq:v2} is
minimized when either $\alpha_j = 0$ or $\nabla \, R(\alphav)_j = 0$.
Accordingly, two additional points are the all-zero vector $\mathbf{0}$ and the ERM solution for which $\nabla \, R(\alphav)=\zero$. %
Finally, the plausible set contains the points
``in-between'' these solutions, those that set the parameters to zero at a subset of indices, and set the elements of the gradient vector to zero at the remaining indices.  For convex risk
functions, there are at most $2^p$ such solutions, one for each
subset; note that the causal coefficient $\betav$ is one of them.

In summary, we began with a constrained optimization in
\Cref{eq:constrained}. Its single optimum is the causal coefficient, but it requires the knowledge of which covariates are causal. We
relaxed that optimization to \Cref{eq:v2}, which
only relies on the observable data. %
The causal coefficient remains to be an
optimum of \Cref{eq:v2}, but there are other solutions too, those between (and including) the
$\zero$-vector and the ERM solution. Hence, solving \Cref{eq:v2} alone does not identify the causal coefficient yet.

In the next section, we will restore identifiability by appealing to
the invariance property of causality under interventions
\citep{peters2016causal, arjovsky2019invariant,scholkopf2021toward}. With data from
multiple environments---each one coming from a different
intervention on the DGP---we can define an optimization problem that whittles
down the plausible set to one that only contains the causal
coefficient.

\begin{figure}[htbp]
\centering{
\subfloat[IRM optima (Env.\#1)]
   {{\includegraphics[width=0.35\textwidth]{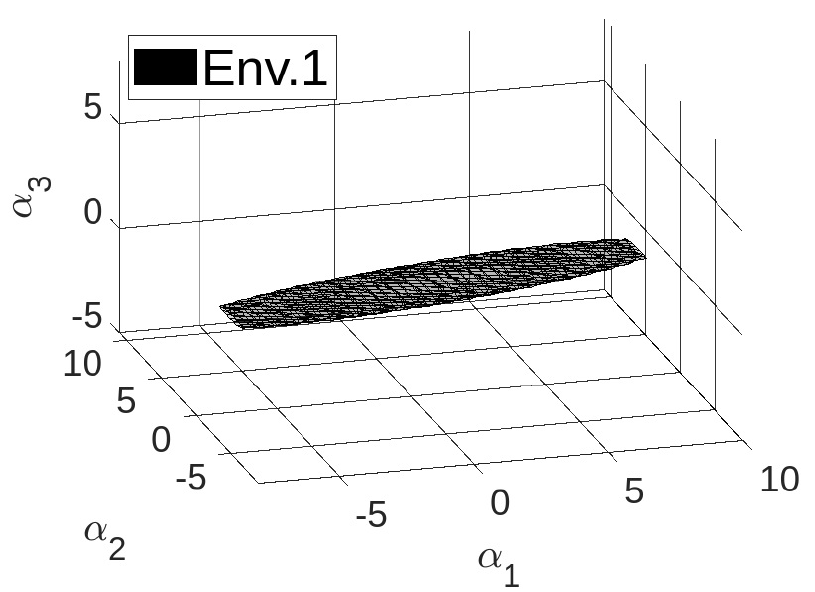} }}  
    \subfloat[IRM optima (Env.\#2)]
    {{\includegraphics[width=0.32\textwidth]{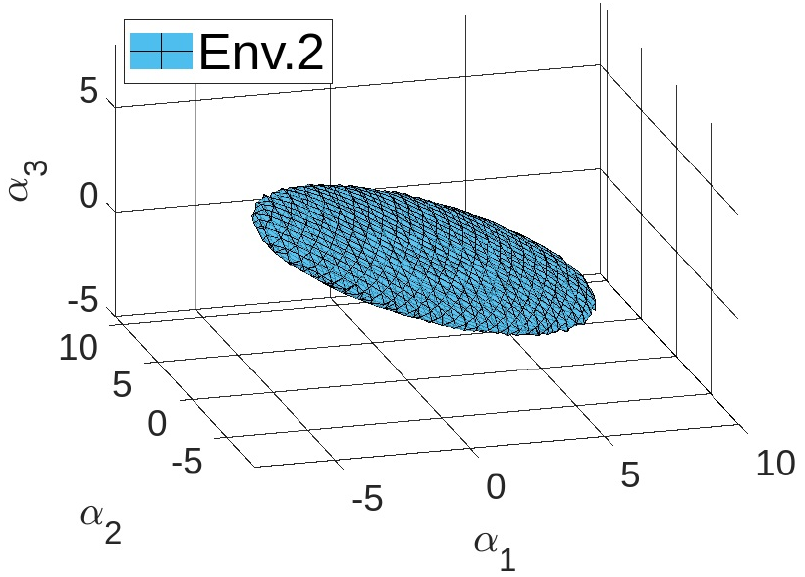} }} 
    \subfloat[IRM optima (Two envs.)]
   {{\includegraphics[width=0.32\textwidth]{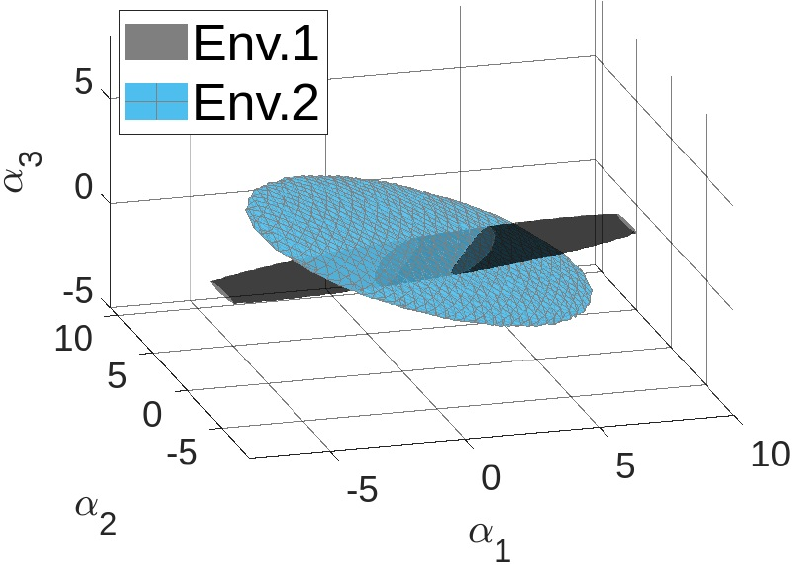} }} \\
    \subfloat[CoCo optima (Two envs.)]
    {{\includegraphics[width=0.33\textwidth]{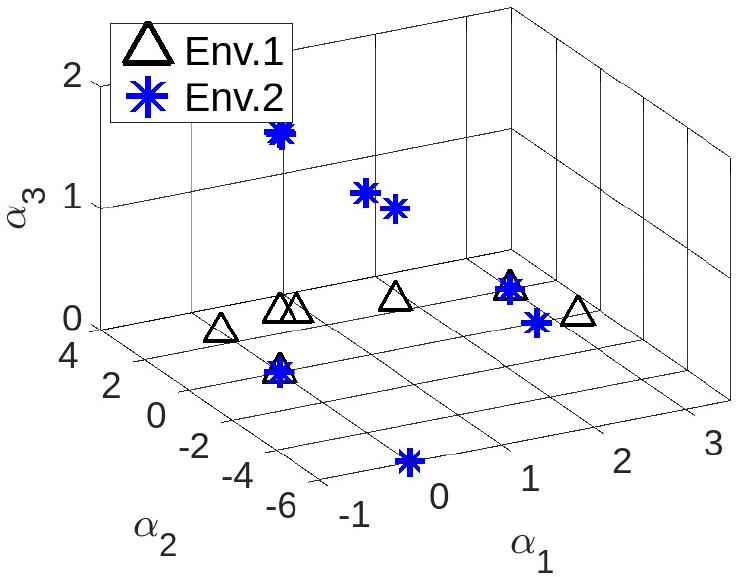} }} ~~~~~~~~
    \subfloat[IRM \& CoCo optima (Two envs.)]
   {{\includegraphics[width=0.33\textwidth]{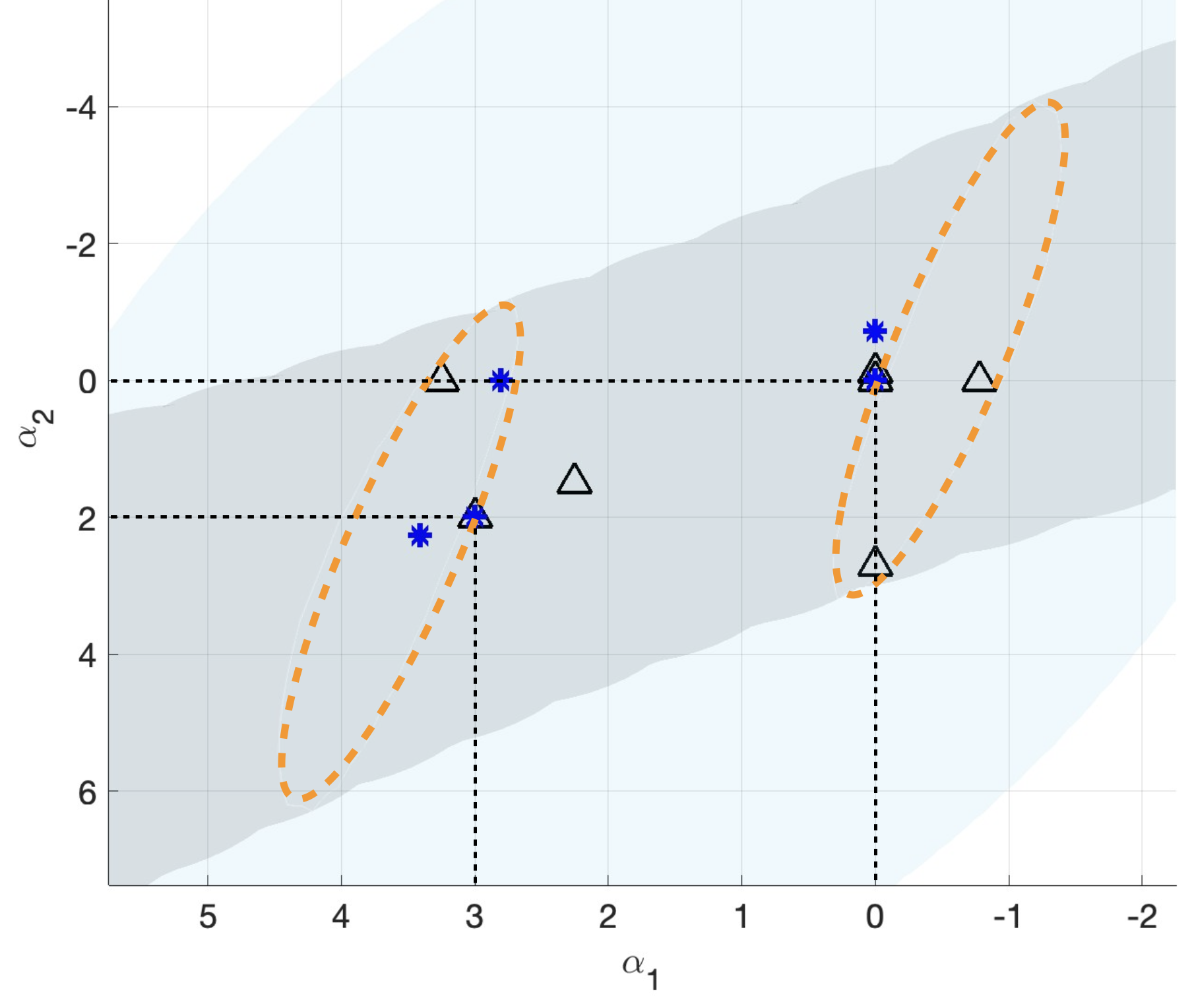} }} 
} 
    \caption{Geometry of the analytic optima sets for the IRM regularization and CoCo objectives in the 3D space. The causal coefficient is $\betav = (3,2,0)$. (a),(b): the optima of the IRM regularization for each environment form a 3D quadric  surface; (c): the optima of the IRM regularization with two environments  is the intersection of the two surfaces, which forms two elliptic curves; (d): The optima of CoCo objective is a discrete finite set for each environment. The CoCo optima over two environments is the intersection consisting of the zero point and the causal coefficient (the overlap of the black triangle and blue star points); (e): The top view of the IRM regularization and CoCo optima for the two environments. The optima set by CoCo ($\alphav \in \{(0,0,0),(3,2,0)\}$) is a strict subset of that by the IRM regularization (the dashed orange elliptic curves). Better viewed in color. }
    \label{fig:exp_irm}%
\end{figure}

\subsection{Optimization with multiple environments}
\label{sec:invariance}

In this section, we leverage multi-environment data to restore 
the uniqueness of the causal coefficient as the solution of an optimization problem.

\paragraph{Narrowing down the optima set by environments.} As discussed in \Cref{sec:estimation}, the causal coefficient is nonidentifiable by optimizing \Cref{eq:v2} with i.i.d. data.  We turn to the setting where we observe data from multiple environments to achieve identifiability.  %

We assume the invariance property as in \Cref{eq:invariance} across environments and that the environments are heterogeneous. Heterogeneous  environments can be constructed by (hard) interventions which actively fix a variable at a specific value during the data generation. They can  
also be constructed by (soft) interventions where the changes in the DGP are passively observed rather than actively executed \citep{eberhardt2007interventions}. 
For example, the heterogeneity might come from varied physical factors such as space and time. When studying the effect of health measurements on the chance of cancer,
the environments can be different hospitals from which the data are collected  \citep{winkler2019association}. %

Consider the relaxed causal optimization problem in
\Cref{eq:v2}. Due to the invariance of the conditional $\E(Y \g \text{Pa}(Y))$, for each environment, its feasible set include the same causal coefficient $\betav$.  But the other optima that utilize the spurious associations will be different across environments  because
 the joint distribution of the covariates $p^e(\xv)$ is
not invariant.

The invariance property  motivates us to aggregate the optimization problems across environments,
\ba{
\min_{\alphav}f_{\cE}(\alphav) \vcentcolon=  \frac{1}{|\cE|}\sum_{e\in \cE} \big(\norm{\nabla R^e(\alphav) \circ \alphav}_2 \big).
\label{eq:coco}
} 
\Cref{eq:coco} is the CoCo objective. We call this optimization framework as \emph{double-gradient  causal optimization} as it computes gradient twice: it first computes the gradient of the risk function $R^e(\alphav)$, using it to obtain the CoCo objective, and then computes the gradient of CoCo objective to update the model parameters. %

Denote the optima of each single environment objective as $\mathcal{F}^e
\vcentcolon= \argmin_{\alphav} \allowbreak \norm{\nabla R^e(\alphav) \circ
\alphav}_2$. The optima of the \gls{CoCO}  objective~\Cref{eq:coco} consist of
the intersection of all $\mathcal{F}^e$s, 
\bas{
\mathcal{F}^{\cE}  \vcentcolon= \argmin_{\alphav} f_{\cE}(\alphav) = \bigcap_{e \in \mathcal{E}} \mathcal{F}^e,
} as long as the intersection is not empty.  \Cref{assp:sem},\ref{assp:invariance} guarantee the nonemptiness because the causal coefficient $\betav \in
\mathcal{F}^{e}$ for all $e$.

Because $\mathcal{F}^{\cE} $ is expressed as an intersection, its size  shrinks with an increasing number of
environments, i.e., $|\mathcal{F}^{\cE_1}|
\leq |\mathcal{F}^{\cE_2}|$ if  $\cE_2 \subset \cE_1$. 
The multiple environments and heterogeneity therein induce differences between the
optima  sets of each environment and, as a result, narrow down the optima set $\mathcal{F}^{\cE} $ of the
\gls{CoCO} objective. For instance, the sets $\mathcal{F}^{e} $ and $\mathcal{F}^{\cE} $ are visualized with an example DGP in \Cref{fig:exp_irm}.   \Cref{fig:exp_irm} (d) illustrates how two environments and the intersection of single-environment optima sets help with the identification of the CoCo objective. 
\begin{algorithm}[t]
\small{
\SetKwData{Left}{left}\SetKwData{This}{this}\SetKwData{Up}{up}
\SetKwFunction{Union}{Union}\SetKwFunction{FindCompress}{FindCompress}
\SetKwInOut{Input}{input}\SetKwInOut{Output}{output}
\Input{
Data $\Dmat^e = \{\Ymat^e, \Xmat^e\}$,  $ \Xmat^e \in \bR^{n^e\times p}$; the risk function $R^e$ for each
environment  $e \in \cE$; the set of known non-descendant variables
$\cC$; the predictor $f(\cdot)$. }
\Output{ Coefficient estimation $\alphav$ with causal interpretation.}
Initialize $\alphav$ randomly

\While{not converged}{

 \For{$e$ in $\cE$}
 {
 Compute the gradient of the empirical risk: $$\gv^e(\alphav) = \frac{1}{n_e} \pde{\alphav} \sum _{i=1}^{n_e}  \allowbreak R^e(\alphav ; y_i^e, \hat{y}_i^e), ~\hat{y}_i^e = f(\xv_i^e;\alphav)$$

 Set $\tilde{\alphav} = \alphav \circ (\mathbf{1} - \mathbf{1}_{\cC}) + \mathbf{1}_{\cC}$

 Compute the optimization objective:  $$\cL^e( \alphav) =\norm{  \gv^e(\alphav)  \circ \tilde{\alphav}}_2$$
 }

 Update $\alphav \leftarrow \alphav - \eta \pde{\alphav} \sum_{e \in \cE} \cL^e(\alphav)$ with step size $\eta$
}
\caption{CoCo with known exogenous variables}
\label{alg:coco}}
\end{algorithm}

\paragraph{Removing the non-informative solution from the optima set.} While the environments help remove the points except for the causal coefficient from the optima of CoCo objective, the zero vector remains a solution. %
We propose two modifications of the objective to avoid the zero vector being an optimum when $\betav \neq \zero$. 

\begin{algorithm}[t]
\small{
\SetKwData{Left}{left}\SetKwData{This}{this}\SetKwData{Up}{up}
\SetKwFunction{Union}{Union}\SetKwFunction{FindCompress}{FindCompress}
\SetKwInOut{Input}{input}\SetKwInOut{Output}{output}
\Input{
Data $D^e = \{\Ymat^e, \Xmat^e\}$, $ \Xmat^e \in \bR^{n^e\times p}$, the risk function $R^e$ for each environment  $e \in \cE$; predictor $f_{\alphav}(\cdot)$; regularizer coefficients $\lambda_r$, $\lambda_w$; annealing scheme ANNEAL($\cdot$)
}
\Output{ Predictor $f_{\alphav}(\cdot)$ that is robust to interventions}
Initialize $\alphav$ randomly

\While{not converged}{

 \For{$e$ in $\cE$}
 { Compute the gradient of the empirical risk: $$\gv^e(\alphav) =
\frac{1}{n_e}  \pde{\alphav} \sum _{i=1}^{n_e}  \allowbreak R^e(\alphav; \hat{y}_i^e), ~~\hat{y}_i^e = f(\xv_i^e;\alphav)$$

 Compute:  $\cL^e( \alphav) =\norm{  \gv^e(\alphav)  \circ \alphav}_2$

 (Optional step:) add weak condition to the objective:  $$\cL^e( \alphav) \mathrel{+}=
 \lambda_w (\ip{ \gv^e(\alphav)}{\alphav})^2$$

Add risk function as a regularization term: $$\cL^e( \alphav) \mathrel{+}=
 \lambda_r \frac{1}{n_e} ( \sum _{i=1}^{n_e}  \allowbreak R^e(\alphav ; \hat{y}_i^e))$$ }

 Update $\alphav \leftarrow \alphav - \eta \pde{\alphav} \sum_{e \in
\cE} \cL^e(\alphav)$ with step size $\eta$ 

 $\lambda_r \leftarrow \text{ANNEAL}(\lambda_r)$
 }
\caption{CoCo with regularization}
\label{alg:coco2}}
\end{algorithm}

We first consider the scenario when certain covariate $X_{j^*}$ is known as an exogenous variable independent of the random noise of the outcome, i.e., $X_{j^*} \indep \epsilon$. Under \Cref{assp:sem}, any ancestor variable of the outcome is a proper exogenous variable. 
With such information, we can modify the \gls{CoCO}  objective \Cref{eq:coco} to be
\ba{
\min_{\alphav} \frac{1}{|\cE|} \sum_{e \in \mathcal{E}} \norm{\nabla R^e(\alphav) \circ \tilde{\alphav}]}_2, \label{eq:thm1}
} 
where $\tilde{\alpha}_{j^*}= 1$ and $\tilde{\alpha}_j = \alpha_j $ for $j \neq j^*$. In other words, the elements in $\tilde{\alphav}$ corresponding to the known exogenous variables are fixed as one. The following lemma demonstrates the properties of \Cref{eq:thm1}'s  optima set. 

\begin{lemma}
The optima set of  \Cref{eq:thm1} contains the causal coefficient and is a subset of \Cref{eq:coco}'s optima set.  If $\beta_{j^*} \neq 0$, The vector $\zero$ is not an optimum of \Cref{eq:thm1} almost surely. 
\label{lem: optima1}
\end{lemma}
The proof of Lemma~\ref{lem: optima1} is in \Cref{sec:proof}. \vspace{1mm}

When there is more than one known exogenous variable, we generalize $\tilde{\alphav}$ as
$\tilde{\alphav} = \alphav \circ (\mathbf{1} - \mathbf{1}_{\cC}) +
\mathbf{1}_{\cC}$ where $\cC$ is the set of known exogenous variables. It further reduces the number of non-causal optima for  \Cref{eq:thm1}. %
The algorithm is summarized in \Cref{alg:coco}. We will discuss what conditions guarantee its output to be the causal coefficients in \Cref{sec:identify} when the environments are sufficiently heterogeneous.  We find in theory (\Cref{sec:identify}) and simulation  (\Cref{sec:experiment}) that the optima set of \Cref{eq:thm1} can shrink to the causal coefficient.  %

In the scenarios when no variables are known as exogenous, an alternative method is to use the risk function as an additive regularization of \Cref{eq:coco},
\ba{ 
\min_{\alphav}~ \frac{1}{|\cE|}\sum_{e\in \cE} \big\{  \norm{\nabla R^e(\alphav) \circ \alphav}_2  + \lambda_r R^e(\alphav)\big\},
\label{eq:coco-erm}
} 
where $  \lambda_r  \geq 0$ controls the regularization strength.  Note that when there is only a single environment $|\cE|=1$, the ERM solution minimizes both terms in \Cref{eq:coco-erm}, so the minimizers of \Cref{eq:coco-erm} are identical to the minimizers of the ERM. %

The mechanism behind the regularizer in \Cref{eq:coco-erm} is that, though the causal coefficient and the $\zero$ vector both minimize \Cref{eq:coco}, the risk of the $\zero$ vector is higher than that of the causal coefficient. So the risk function introduces inductive bias to disfavor the $\zero$ vector. To see this, for the DGP in \Cref{eq:lSEM-e} and the squared risk, under \Cref{assp:sem}, the risk $R^e(\zero) =  \betav^\top \E[\xv\xv^\top]\betav +  \E[\epsilon^2] + 2\E[\xv^\top\betav\epsilon]$,  and $\E[\xv^\top\betav\epsilon] = \E[\xv_S^\top \betav_S \epsilon]=0$ since $\betav_{\backslash S} = \zero$ and $\xv_S \indep \epsilon$; and we have $R^e(\betav) =  \E[\epsilon^2]$. Thus, $R^e(\zero) \geq  R^e(\betav)$, which means the risk regularizer in \Cref{eq:coco-erm} would encourage the convergence to the causal coefficient $\betav$ rather than the $\zero$ vector. 
Optimizing \Cref{eq:coco-erm} with the risk regularizer yields a predictive model that is robust to perturbations of spurious associations.  However, the risk regularizer may introduce potential estimation bias when the focus is causal estimation. In practice, this bias can be reduced by annealing the strength parameter $\lambda_r$ during the optimization. We summarize the complete algorithm in \Cref{alg:coco2}. %

%% file: connection.tex
\label{sec:irm}
\citet{arjovsky2019invariant} introduces \gls{IRM} that can learn

Invariant Risk Minimization (IRM) is a popular app
robust representation in the presence of spurious associations between the covariates and outcome. In particular, \gls{IRM} considers a
predictor $f(\xv;\alphav): \bR^p \mapsto \bR$ with parameter
$\alphav$. In a setting similar to \gls{CoCO} , it considers a set of heterogeneous
environments $
\mathcal{E}$  and a risk function
$R^e(\alphav; \hat{y})$ for each $e \in \mathcal{E}$ (for notational clarity, we explicitly write the prediction $\hat{y}$ in the notation of the risk function).  Based on
the intuition that invariant predictor induces invariant features,
\gls{IRM} proposes the following objective to find an invariant model
\ba{
&\min_{\alphav, w}~\sum_{e \in \mathcal{E}}~R^e(\alphav; w (f(\xv_i^e;\alphav))) \\
&s.t.~  w \in \argmin_{\bar{w}}~R^e(\alphav; \bar{w} (f(\xv_i^e;\alphav) )),~\text{for all }e \in \cE, \notag
}
where $w(\cdot)$ is a mapping from the range of $f(\cdot)$ to
$\hat{y}$. For tractable computation, \citet{arjovsky2019invariant} further
introduces the
\gls{IRM}v1 objective:
\begingroup
\allowdisplaybreaks
\bac{
\min_{\alphav}~\sum_{e \in \mathcal{E}} \Big[ \underbrace{R^e(\alphav;  f(\xv_i^e;\alphav))}_\text{Empirical risk} +  \lambda\underbrace{ \big(\nabla_{w|w=1.0} R^e(\alphav;  w\cdot f(\xv_i^e;\alphav))\big)^2}_\text{IRM regularization}\Big],
\label{eq:irm}
} 
\endgroup
where $\lambda > 0$ and $w$ is simplified as a dummy scalar
variable. The \gls{IRM}v1 objective in \Cref{eq:irm} consists of an
empirical risk term and an IRM regularization term that encourages invariance. 
Since the empirical risk does not produce the correct causal estimate, we focus on analyzing the IRM regularization term. ~\looseness=-1

\vspace{1mm} \noindent \parhead{Analytic and geometric connections.} The IRM regularization and  CoCo are connected by the constrained optimization problem in \Cref{eq:constrained}. Specifically, the IRM regularization can be derived from optimization in \Cref{eq:constrained} for several common models. In \Cref{sec:estimation}, we use the directional derivative to obtain the set of feasible
directions $\mathcal{U} = \text{span}
\{\mathbf{e}_j: j \in S\}$ of the idealized causal optimization and the first-order optimality condition in
\Cref{eq:v2_1}. An interesting observation is that the causal parameter itself is a feasible direction with $\betav \in \mathcal{U}$. Since $\betav$ is a feasible direction and is also an optimum of the optimization in \Cref{eq:constrained}, Lemma~\ref{lem:optimal} implies that the directional derivative with direction $\betav$ vanishes at the causal coefficient $\betav$, which leads to the following lemma.

\begin{lemma}
\label{lem:opt1}
For any partition
$\mathcal{P}$ of the set $\idx{p}$, the optimality of the causal coefficient $\betav$ is an optimum of 
\ba{
\min_{\alphav}  ~ \sum_{A \in \mathcal{P}}(\ip{\nabla
R(\alphav)_A}{\alphav_A})^2.
\label{eq:general}
}
Specifically,  $\betav$ is an optimum of
\ba{
\min_{\alphav}  ~(\ip{\nabla  R(\alphav)}{\alphav})^2.
\label{eq:v1} 
} 
\end{lemma}
Lemma~\ref{lem:opt1} gives the general form of optimization that admits causal optimality.  The proof of Lemma~\ref{lem:opt1} is in \Cref{sec:proof}.

When the outcome model is Linear-Gaussian or
Linear-Bernoulli, minimizing the IRM regularization is equivalent to  solving \Cref{eq:v1}.  To see this, suppose a linear  \gls{DGP} as in \Cref{eq:lSEM-e}, linear predictor $\hat{y}^e = \alphav^\top \xv^e$, and a squared risk
function, then
\ba{
\big(\nabla_{w|w=1.0} R^e(\alphav;  w \alphav^\top \xv^e)\big)^2 &=  (\E[(y^e - \hat{y}^e)\alphav^\top\xv^e])^2 \notag \\
&=(\ip{\nabla R^e(\alphav;\hat{y})}{\alphav})^2. 
\label{eq:linear-gaussian} 
} 
The left side of \Cref{eq:linear-gaussian} is the invariant regularization term and the right side is \Cref{eq:v1}.

Similarly, suppose the outcome is generated by $y^e \leftarrow
\text{Bernoulli}(\sigma(\betav^\top \xv^e))$, the predictor is
$\hat{y}^e = \sigma(\alphav^\top \xv^e)$ where $\sigma(x) =
1/(1+\exp(-x))$ is the sigmoid function,  and the risk function is the
cross entropy loss $R^e(\alphav; \allowbreak \hat{y}^e ) = -
\bE[y^e
\log(\hat{y}^e)+ (1-y^e)\log(1-\hat{y}^e)]$. Then\footnote{The form $R^e(\alphav;  \sigma(w \alphav^\top \xv^e ))$ follows the IRM implementation of binary classification \emph{``loss = mean\_nll(logits * scale, y)''}  at \url{https://github.com/facebookresearch/InvariantRiskMinimization/blob/main/code/colored_mnist/main.py}. }
\ba{
\big(\nabla_{w|w=1.0} R^e(\alphav;  \sigma(w \alphav^\top \xv^e ))\big)^2 &= (\E[(\hat{y}^e - y^e)\alphav^\top\xv^e])^2 \notag \\
&= (\ip{\nabla R^e(\alphav;\hat{y})}{\alphav})^2.
\label{eq:linear-bernoulli}
} 
The comparison is summarized in \Cref{tab:compare}. 

\begin{table}[t] 
    \centering
    \begin{tabular}{C{4cm}C{4cm}C{4cm}} 
        \toprule
        ERM  & IRM regularization & CoCo \\ \midrule
        $\min_{\alphav} \norm{\nabla_{\alphav} R^e(\alphav) \circ \one}_2 $ & $\min_{\alphav} (\nabla_{\alphav} R^e(\alphav) \cdot \alphav)^2$ & $\min_{\alphav} \norm{\nabla_{\alphav} R^e(\alphav) \circ \alphav}_2$ \\ %
        \bottomrule
    \end{tabular}
    \caption{Comparison of the ERM (by gradient methods), IRM regularization (for the linear Gaussian and Bernoulli models), and CoCo objectives for one environment. }
    \label{tab:compare}
\end{table}

The connections in 
\Cref{eq:v1,eq:linear-gaussian,eq:linear-bernoulli} reveal that IRM regularization implicitly imposes a first order condition based on the directional derivative with the direction $\alphav$, which offers an explanation to 
 the mechanism behind \gls{IRM}v1. Lemma~\ref{lem:opt1} and \Cref{eq:v1,eq:linear-gaussian,eq:linear-bernoulli} can prove that, for linear Gaussian or Bernoulli models, the causal coefficient $\betav$ is a minimizer of the IRM regularization for one environment.  The data from multiple environments narrow down the optima set to the causal coefficient under the invariance assumption. The additional empirical risk term in the IRMv1 objective plays the role of further regularization by encouraging the solutions with high predictive performance.

 The connection also illustrates the sub-optimality of the IRM regularization. Geometrically, the IRM regularization, rewritten as the inner product
between the gradient and parameter vectors, only considers a single
feasible direction $\betav$ for the constrained optimization problem
\Cref{eq:constrained}. In contrast,  CoCo finds the optimum across all the feasible directions that form a $(p-|S|)$ dimensional space $\mathcal{U}$. The spectrum in \Cref{eq:general} shows that the objective 
of CoCo corresponds to the finest partition, imposing the strongest constraint, and has the smallest optima set. %
Algebraically, we have $p \norm{\nabla R(\alphav) \circ \alphav}^2_2 \geq  (\ip{\nabla  R(\alphav)}{\alphav})^2 $ by the Cauchy–Schwarz inequality. The inequality indicates that IRMv1 potentially minimizes a loose lower bound. The following proposition shows that the CoCo objective is guaranteed to have a better identification result than the IRM regularization. 

\begin{proposition}
\label{prop:irm}
For linear-Gaussian or linear-Bernoulli outcome generating distribution, if minimizing the IRM regularization in \Cref{eq:v1} identifies the causal coefficient (excluding point $\zero$) over environments $\cE$, then minimizing the CoCo objective \Cref{eq:v2} will identify the causal coefficient (excluding point $\zero$). The inverse statement does not hold. 
\end{proposition}

Because of an excessive number of optima in the IRM regularization for a single environment, it has a high requirement on the number of environments and the type of heterogeneity to sieve out the causal coefficient, or relies on the causal model to have the lowest empirical risk among all the optima. These requirements are especially challenging for high-dimensional problems. In a nutshell, connecting IRM regularization with causal optimization in \Cref{eq:constrained} indicates the potential looseness of IRMv1 objective in  the \emph{linear-Gaussian} and \emph{linear-Bernoulli} DGPs. Another potential issue with the IRM regularization is that it might not optimize in a feasible direction for other types of  DGPs, for example, with a categorical outcome.

\vspace{1mm} \noindent \parhead{A case study.}  We present a brief case study to illustrate the connections. We consider a linear-Gaussian DGP as the Case 1 of \Cref{tab:sems}. The two environments correspond to parameters  $(m_1^{(1)}, m_2^{(1)}, \gamma^{(1)}) = (2,0.5,2)$, $(m_1^{(2)}, m_2^{(2)}, \gamma^{(2)}) = (3,-1,0.5)$ in the DGP. The invariant causal coefficient is $\betav = (3,2,0)$.   \Cref{fig:exp_irm} visualizes the optima set of CoCo and the IRM regularization where all the optima are computed analytically. 

\Cref{fig:exp_irm} (a) (b) show the set of optima for IRM regularization in each of the two environments; each set forms a quadric surface.  When considering the two environments together, the IRMv1 optima are the intersection of two 3D surfaces in  \Cref{fig:exp_irm} (c). Geometrically, as shown  in \Cref{fig:exp_irm} (e), the intersection forms two continuous elliptic curves consisting of an infinite number of points. The IRMv1 objective cannot identify the causal coefficient unless $\betav$ has the lowest empirical risk among all the optima of the IRM regularization, which, as we will see, is not the case here.  
In contrast, \Cref{fig:exp_irm} (d) shows the optima set of CoCo that contains only a finite number of points per environment.  With two environments, the optima of CoCo becomes the zero point and the causal coefficient $\betav$. The causal coefficient can be identified by the modified objectives in \Cref{eq:thm1,eq:coco-erm} that remove the zero point. 

We examine this case study 
empirically by implementing CoCo in \Cref{eq:thm1} and IRMv1 in \Cref{eq:irm} with a large sample size $N=10^5$ of each environment.  CoCo objective with $j^*=1$ converges to $\hat{\alphav}^{\text{CoCo}} = (3.001, 2.004, 0.000)$ with random initialization.  
IRMv1 with $\lambda=1$ converges to $\hat{\alphav}^{\text{IRM}}_1 = (-0.842, -0.390,  0.621)$ with initialization by a random seed and converges to $\hat{\alphav}^{\text{IRM}}_2 = (2.932, 1.847, 0.017)$ with an idealized initialization manually set at the true causal coefficient $\betav = (3,2,0)$. Furthermore, $\hat{\alphav}^{\text{IRM}}_2$ has the empirical risk as $0.97634$ and the IRM regularization as $0.00028$, while those of $\betav$ are $0.99852$ and $0.00036$, respectively. 
Thus, among the optima of the IRM regularization for the two environments, the causal coefficient $\betav$ does not have the lowest risk, and the minimizer of the IRMv1 objective is a biased causal estimate. We provide detailed computation and additional results with different tuning parameters of IRMv1 for this case study in the \Cref{sec:example}.

%% file: identification.tex
Now we establish the causal identification for CoCo with the linear SEM in \Cref{eq:lSEM-e}. Identification
requires the causal quantity of interest to be expressed as a functional of
the observed data distribution. This functional is also known as 
the causal identification strategy. In the context of CoCo, we consider
the functional that maps the joint distributions $p(\xv^e, y^e)$ over
a set of environments to the risk function, then to the optima of  CoCo objective. To establish identification, we prove that the optimum of the CoCo objective is unique and equals the causal coefficient of interest.  In \Cref{sec:identification,sec:eff-intervene}, we present two assumptions on the environment heterogeneity that guarantee the identification. In \Cref{sec:suff}, we provide two sufficient conditions  implying the identification assumption.

\subsection{Identification of the causal coefficients}
\label{sec:identification}

The causal identification of an optimization-based approach is to demonstrate the existence and uniqueness of its optimum, and it equals the causal coefficient. By fully characterizing the optima set of the modified CoCo objective in  \Cref{eq:thm1}, we explore the assumption on the environments that bestows identification.

The optimum of the CoCo objective has a unique characteristic. It is a point $\alphav$ for which $\nabla R^e(\alphav)_H = \zero$ and $\alphav_{\backslash H} = \zero$ with certain $H \subset \idx{p}$.  Denote the risk function with a subset of covariates $\xv^e_H$ as the predictor by $R^e_H(\alphav_H) = \half \E[(y - \alphav_H^\top \xv^e_H)^2]$. We have the following proposition.
\begin{lemma}
For the linear DGP as \Cref{eq:lSEM-e}, a linear predictor and squared risk, if $\nabla R^e(\alphav)_H = \zero$ and $\alphav_{\backslash H} = \zero$, then $\nabla R^e_H(\alphav_H) = \zero$; if $\nabla R^e_H(\alphav_H) = \zero$, then $\alphav = (\alphav_H, \alphav_{\backslash H} = \zero)^\top$ satisfies  $\nabla R^e(\alphav)_H = \zero$. 
\label{prop:subset}
\end{lemma}
Lemma~\ref{prop:subset} demonstrates that a CoCo optimum is an ERM minimizer over a subset of covariates. When restricted to this subset of covariates, the ERM minimizer is shared by all the environments.  

Lemma~\ref{prop:subset} offers an interpretation of CoCo optimum. To keep the notation consistent, denote $\cC$ as the set of exogenous covariates and $S$ as the unknown set of direct causes for the outcome. For any set $H$ with $\cC \subset H \subset \idx{p}$, we fit a  regression model on $X_{H}^e$ in each environment, and collect the regression coefficients as $\{\hat{\alphav}^e_{H}\}_{e \in \cE}$. We call the set $H$ an \emph{invariant set}, if the estimates
\ba{
\hat{\alphav}^e_{H} = \hat{\alphav}^{e'}_{H} \vcentcolon=\hat{\alphav}_{H}, \quad \forall e, e' \in \cE.
\label{eq:invariant-set}
} 
If $H$ is an invariant set, we define a length $p$ vector as an \emph{invariant vector} by equating it to $\hat{\alphav}_{H} $ when restricting to the set $H$ and padding it with zeros at other elements. By Lemma~\ref{prop:subset}, the optima of the objective \Cref{eq:thm1} consist of all the invariant vectors. 

Based on the interpretation of CoCo optimum, we introduce the following assumption for sufficiently heterogeneous environments.  ~\looseness=-1
\begin{assumption} [Effectiveness]
There is only one invariant vector (defined in \Cref{eq:invariant-set}) for all the sets $H$ with $\cC \subset H \subset \idx{p}$. $\cC$ is the set of known exogenous variables.
\label{assp:eff}
\end{assumption}
Then we have the following identification result.

\begin{theorem}
Under Assumptions \ref{assp:sem}, \ref{assp:invariance} and \ref{assp:eff},
assuming the Gram matrix $\Wmat^e = \E[\xv^e(\xv^e)^\top] \succ 0$ for all $e \in \mathcal{E}$,  the causal coefficient $\betav$ is identifiable, and is given by %
\ba{
\betav = \argmin_{\alphav} \frac{1}{|\cE|} \sum_{e \in \mathcal{E}} \norm{\nabla R^e(\alphav) \circ \tilde{\alphav}]}_2,
}
which is the optimum of the objective in \Cref{eq:thm1}.
\label{thm:lm}
\end{theorem}

The invariance \Cref{assp:invariance} ensures the existence of a point that reaches the minimum value of the objective \Cref{eq:thm1}, and this point is the causal coefficient, while the effectiveness \Cref{assp:eff} induces uniqueness of the solution.  The following corollary provides practical guidance in collecting multi-environment data. %

\begin{corollary}
Assume the environment sets  $\cE_1 \subset \cE_2$, then in \Cref{thm:lm},   (i) if the invariance \Cref{assp:invariance} holds for $\cE_2$,  it holds for $\cE_1$; (ii) if the effectiveness \Cref{assp:eff} holds for $\cE_1$, it holds for $\cE_2$.
\label[corollary]{cor:expand}
\end{corollary}

\subsection{Identifying the effects of the exogenous variables}
\label{sec:eff-intervene}

The effectiveness assumption can be checked with the observed data, but it needs regression over the power sets of covariates across the environments, which can be computationally expensive. In practice, we might be interested in estimating the causal effect of specific variables instead of the whole causal coefficient. This motivates us to explore identification results under a relaxed \Cref{assp:eff}. 

Suppose the goal is to estimate the causal effect of a known exogenous variable $x_{j^*}$. For example, such a variable could be the standard treatment variable under the unconfoundedness assumption \citep{imbens2015causal}. Though $x_{j^*}$ is assumed exogenous, i.e.,$x_{j^*} \indep \epsilon$, regressing over $x_{j^*}$ alone may produce biased estimation because $x_{j^*}$ may correlate with other causes $\xv_{S\backslash j*}$ in the DGP. %

We adopt the following notations. Recall that $S$ is the set of direct causes, $\cC $ is the known exogenous variables, $j^* \in \cC$,  and $H$ is a set $ \cC \subset H \subset \idx{p}$. Denote $W_{AB}$ to be a sub-matrix of the Gram matrix $\Wmat  = \E[\xv\xv^\top]$ with rows in the index set $A$ and columns in the index set $B$.  For the environments $ \cE = \{e_1,\cdots, e_m\}$, denote $\Wmat_H^{\cE} \in  \bR^{(m\cdot|H|)\times |H|}$ as a stacking matrix that  stacks $W^e_{HH}$ by row, i.e. 
$\Wmat_H^{\cE} \coloneqq [W^{e_1}_{HH}| W^{e_2}_{HH} | \cdots | W^{e_m}_{HH}]^\top$.

\Cref{assp:eff} excludes a non-causal invariant vector when regressing over a set of covariates $\xv_H^e$.   Here, we analyze the property of an invariant vector. By Proposition~\ref{prop:subset}, the invariant vector can be computed by zeroing out the gradient 
\ba{
 \nabla_{\alphav_H} R_H(\alphav_H) = W^e_{HH}(\alphav_H - \betav_H) - W^e_{HH^c} \betav_{H^c} - \mathbf{s}_H^e, 
\label{eq:part-gradient}
}
where $H^c$ is the complement of $H$ in $\idx{p}$ and $s_j^e  \vcentcolon= \E[x_j^e\epsilon^e]=\text{cov}(x_j^e\epsilon^e)$. Denote a stacking vector $\thetav_H^{\cE} \in \bR^{m\cdot|H|}$ as  $\thetav_H^{\cE} \coloneqq [W^{e_1}_{HH^c} \betav_{H^c}  + \mathbf{s}_H^{e_1}, \cdots, W^{e_m}_{HH^c} \betav_{H^c}  + \mathbf{s}_H^{e_m} ]^\top$. By setting the gradient in \Cref{eq:part-gradient} to zero and by Proposition~\ref{prop:subset}, if  
\ba{
\Wmat_H^{\cE} \deltav = \thetav_H^{\cE} 
 \label{eq:inconsistent}
}
holds for $\deltav \neq \mathbf{0}$,  $H$ is an invariant set with invariant vector $\alphav$,  $\hat{\alphav}_H = \betav_H + \deltav$, $\hat{\alphav}_{\backslash H} = \zero$.  This violates \Cref{assp:eff} because $\hat{\alphav}$ and $\betav$ are two different invariant vectors that both minimize the CoCo objective. We formulate the negating statement as the following assumption.

\begin{assumption}[Weak effectiveness]
$\forall~H \neq S$, $\cC \subset H \subset \idx{p}$, if the stacking vector $\thetav_H^{\cE} \neq \mathbf{0}$, it is not in the column space of the stacking matrix $\Wmat_H^{\cE}$, i.e., $\thetav_H^{\cE} \notin \mathds{C}(\Wmat_H^{\cE})$.
\label{assp:eff-w}
\end{assumption}

In sum, we have shown ($\neg$ \Cref{assp:eff-w}) $\Rightarrow$ ($\neg$ \Cref{assp:eff}), thus the contrapositive implies   (\Cref{assp:eff}) $\Rightarrow$  (\Cref{assp:eff-w}).  On the other hand,  cases exist when \Cref{assp:eff-w} hold but \Cref{assp:eff} doesn't, so \Cref{assp:eff-w} is a weaker assumption. For example, if two subsets of causes $S_i \subset S, i=1,2$, $\mathcal{C} \subset S_i$, and $\xv^e_{S_i} \indep \xv^e_{\backslash S_i}$, 
then both $S_1$ and $S_2$ are invariant sets when $\E[\xv^e_{\backslash S_i}] = \zero$. Hence it violates \Cref{assp:eff} due to the existence of multiple invariant sets. But in such a case, \Cref{assp:eff-w} can still hold because setting $H=S_i$  in \Cref{eq:inconsistent} makes $\thetav_H^{\cE}=0$ and $\deltav = \zero$. The CoCo optimum might be $(\betav_{S_1}, \zero)$ and $(\betav_{S_2}, \zero)$, which share the common unbiased part $\betav_{\cC}$. 

With the weak effectiveness assumption, we have the following identification result.
\begin{theorem}
Under Assumptions \ref{assp:sem}, \ref{assp:invariance} and \ref{assp:eff-w}, the causal effect of $\xv_{\cC}$ on $y$ is identifiable and is given by the optimum of objective \Cref{eq:thm1}. That is, for
\ba{
\alphav^* \in \argmin_{\alphav} \frac{1}{|\cE|} \sum_{e \in \mathcal{E}} \norm{\nabla R^e(\alphav) \circ \tilde{\alphav}]}_2,
}
$\alphav^*_{\cC} = \betav_{\cC}$, where $\betav$ is the causal coefficient.  
\label{thm:weak}
\end{theorem}

Theorem~\ref{thm:lm} guarantees the identification of the whole causal coefficient vector $\betav$. In comparison,  \Cref{thm:weak} identifies the effects of the known exogenous variables $\betav_{\cC}$ based on a weaker effectiveness assumption. In the next section, we leverage the explicit analytic form of \Cref{eq:inconsistent} and discuss several scenarios where weak effectiveness can be guaranteed.

\subsection{Sufficient conditions for identification}
\label{sec:suff}

We show two approaches that ensure sufficient heterogeneity of the environments. One is by actively taking do-interventions, and the other is by passively checking the rank conditions.

\paragraph{Effectiveness by do-interventions.} %
Consider the \emph{do}-intervention \citep{pearl2009causality} as 
\bas{
X_j^e \leftarrow a_j^e, \quad a_j^e \sim p(a), \quad  j \in \cI^e,
}
where $e$ is environment index, $e \in \cE$, $\cI^e$ is the index set of intervened variables, and $p(a)$ is a continuous distribution defined on $\bR$.  $do(X_j^e=a_j^e)$ means all the samples of $X_j$ in the environment $e$ are fixed at the constant $a_j^e$ during the  data generation. Suppose the intervention on one variable is independent of other interventions and variables.  The following corollary gives a sufficient condition that guarantees \Cref{assp:eff-w}. 
\begin{corollary}
\label[corollary]{thm:suff}
For the linear SEM  in \Cref{eq:lSEM-e} and predictor in \Cref{lem:optimal}, and for do-interventions, suppose $\forall j \in \idx{p}$, $\exists ~e, e' \in \cE$ s.t. $\cI^e = \cI^{e'} = \{j\}$, and $\exists k \in \cC$, $\E[X_k]\neq 0$, then the optimum $\alphav^*$ of the optimization in \Cref{eq:thm1}  satisfies $\alphav^*_{\cC} = \betav_{\cC}$, where $\betav$ are the causal coefficients.
\end{corollary}

\Cref{thm:suff} is closely related to the identification results  of ICP in \citet[Theorem 2]{peters2016causal}. Comparing to \Cref{thm:suff}, ICP asks for one less do-intervention on each covariate,  but it crucially relies on the invariance assumption in \Cref{eq:strong-invariance} that is stronger than \Cref{assp:invariance} of CoCo.

\paragraph{Effectiveness by rank checking.}   
In many cases, we collect data from environments where the heterogeneity is introduced not by the manual interventions but by the natural factors such as spatial and temporal differences. We propose a checking method to ensure a sufficient but necessary condition for the \Cref{assp:eff-w}, which is easy to compute. 

As  \Cref{sec:eff-intervene},  denote $\cC \subset \cP = \idx{p}$ as the known exogenous variables.  For each environment $e \in \cE$, the Gram matrix using the observed data is $\Wmat^e = \E[(X^e)^TX^e ] \in \bR^{p\times p}$ and  the submatrix  $\Wmat^e_{\cC \cP}$ is the rows of $\Wmat^e$ with index in $\mathcal{C}$.  Let $\Wmat_{\mathcal{C} \cP}^{\cE} \in \bR^{(|\cE|\cdot |\mathcal{C}|) \times p}$ be a matrix that stacks $\Wmat^e_{\mathcal{C} \cP}$ by the rows for all $e \in \cE$. 

\Cref{assp:eff-w} assumes  the linear system $\Wmat_H^{\cE} \deltav \neq \thetav_H^{\cE}$ for any $\deltav \neq \zero$.  A key observation is that for $j \in \cC$,   $\E[X_j \epsilon] = 0$ due to exogeneity, so $\mathbf{s}^e_{\cC} = \bzero$ for the gradient computed in \Cref{eq:part-gradient}.  
Therefore, \Cref{assp:eff-w}  is guaranteed if $\forall \deltav \neq \mathbf{0}, \Wmat_{\cC H}^{\cE} \deltav \neq \Wmat_{\cC H^c}^{\cE} \betav_{H^c}$, which can be further guaranteed if 
the homogeneous linear system $\Wmat^{\cE}_{\mathcal{C} \cP} \vv= \mathbf{0}$ only has trivial solution for the variable $\vv$. This linear system only depends on observed data. 
Then we have the following corollary.  ~\looseness=-1
\begin{corollary}
\Cref{assp:eff-w}  is guaranteed if the linear system $\Wmat^{\cE}_{\mathcal{C} \cP} \vv= \mathbf{0}$ only has the trivial solution $\vv = \zero$. 
\label[corollary]{prop:checking}
\end{corollary}
In practice, we first collect data from environments $\cE$ where the heterogeneity may come from do-interventions or soft interventions that change the distribution of the covariates  \citep{eberhardt2007interventions}. Then we check Corollary~\ref{prop:checking} by equivalently checking if the matrix $\Wmat^{\cE}_{\mathcal{C} \cP}$ has full column rank. If the rank is full, we proceed to Algorithm \ref{alg:coco} with environment set $\cE$; otherwise, we collect data from a new environment. The algorithm is summarized in \Cref{alg:ico}.

\begin{algorithm}[t] 
\small{
\SetKwData{Left}{left}\SetKwData{This}{this}\SetKwData{Up}{up}
\SetKwFunction{Union}{Union}\SetKwFunction{FindCompress}{FindCompress}
\SetKwInOut{Input}{input}\SetKwInOut{Output}{output}
\Input{
The  environmental set $\cE$, the set of known exogenous variable $\cC$
}
\Output{ Coefficients $\alphav$ with causal interpretation}
Initial $\cE = \varnothing$

\Repeat{rank$(\Wmat_{\cC\cP}^{\cE}) =  p$}
{ Randomly choose a valid intervention

Let $\cE \leftarrow \cE \cup \{e\}$ with $e$ as the index of new environment

Collect data in environment $e$, $\mathcal{D}^e = \{y_i^e, \xv_i^e\}_{i=1}^{n_e}$ 

Update $\Wmat_{\cC\cP}^{\cE}$
}

Run Algorithm~\ref{alg:coco} or \ref{alg:coco2} with $\{\cE, \{\mathcal{D}^e\}_{e \in \cE},\{R^e(\cdot)\}_{e \in \cE}\}$ where $R^e(\cdot)$ is the risk function.
\caption{Heterogeneity checking}
\label{alg:ico}
}
\end{algorithm}

%% file: extension.tex
So far, we focus on the linear \gls{SEM}s. Here we generalize these results to a nonlinear SEM and a predictor that maps linear combinations of covariates to the outcome, i.e. $y = f(\Amat \xv)$. 
For example, the fully connected neural network is a special case in such a functional form. 
The generalization of the algorithm to the nonlinear model closely follows the derivation in \Cref{sec:method}.  The first step is to build a constrained optimization problem similar to causal optimality in 
\Cref{lem:optimal}  and show that it admits the causal coefficient as a solution. 
The analysis presented in \Cref{sec:estimation} based on the directional derivative can then be applied to nonlinear models. 

Suppose we have a collection of environments $\cE$, and for each
$e\in\cE$, we observe i.i.d. data for variables $(\xv^e,y^e)$, $\xv^e
\in \bR^p$, $y^e \in \bR$. Suppose the underlying DGP is
\ba{
y^e \leftarrow f(\tilde{\Bmat}^*\xv_S^e; \gammav^*) + \epsilon^e
\label{eq:nonlinear}
}
where $S \subset \idx{p}$, $\epsilon^e \indep \xv_S^e$ and
$\E[\epsilon^e] = 0$. $f: \bR^K \to \bR$ is an arbitrary function
mapping with causal parameters $ \betav^* = (\tilde{\Bmat}^*,\gammav^*)$ where $\tilde{\Bmat}^* \in
\bR^{K\times |S|}$ and $ \gammav^* \in \bR^M$. %
When $K=1$ and $f(\cdot)$ is an identity mapping,
\Cref{eq:nonlinear} reduces to the linear \gls{SEM}.
\Cref{eq:nonlinear}  can represent a flexible DGP that generates
 the outcome by a fully-connected neural network, where $K$ and $\tilde{\Bmat}^*$
are the width and weights of the first hidden layer, respectively.

Assume the nonlinear predictor is
\ba{
\hat{y}^e = f(\Bmat \xv^e; \gammav),
\label{eq:nl-pred}
} where $\Bmat \in \bR^{K\times p}$, $ \gammav \in \bR^M$ and $\alphav
= (\Bmat,  \gammav )$ are the parameters to optimize. We can re-write
$\tilde{\Bmat}^* \xv_S^e= \tilde{\Bmat}^* \Lambda \xv^e$ where $\Lambda \in \bR^{|S|
\times p}$ has the i-th row as $\mathbf{e}_i^\top$ if $i \in S$ and as
$\mathbf{0}^\top_p$ if $i \notin S$. Let $\Bmat^* =  \tilde{\Bmat}^* \Lambda$ where the
$j$-th column of $\Bmat^*$ is $\mathbf{0}_K$ if $j \notin S$.  Then for
square error $R^e(\alphav)$ we have the following proposition.

\begin{proposition}[Causal Optimality, Nonlinear]
With the DGP in \Cref{eq:nonlinear} and with Assumptions~\ref{assp:sem} (ii) (iii), the causal coefficient $\alphav = (\Bmat, \gammav) = ( \Bmat^*, \gammav^*)$  is an
optima of the following problem
\bac{
&  \min_{\Bmat,\gammav}~R(\Bmat,\gammav),~\hat{y} = f(\Bmat \xv; \gammav) \\
& s.t. ~B_{kj} = 0 ~\text{if}~ B^*_{kj} = 0,   ~~1\leq k\leq K,~ 1 \leq j \leq p \\
& \qquad \gamma_m = 0  ~\text{if}~ \gamma^*_m = 0,  ~~1\leq m \leq M.
}
\label[proposition]{prop:nonlinear}
\end{proposition}

\Cref{prop:nonlinear} greenlights the analysis in
\Cref{sec:method}. The \gls{CoCO}  objective
in \Cref{eq:coco-erm} can then be used for
nonlinear models in \Cref{eq:nonlinear,eq:nl-pred}, 
\ba{ 
\min_{\Bmat, \gammav}~ \frac{1}{|\cE|}\sum_{e\in \cE} \big\{  \norm{\nabla R^e(\alphav) \circ \alphav}_2  + \lambda_r R^e(\alphav)\big\},
} 
where $\alphav = (\Bmat, \gammav)$. When $\Bmat = \Bmat^*$, the multiplication $\Bmat
\xv$ zeros out the non-causal covariates $\xv_{
\backslash S}$ so that the prediction $\hat{y}$ is independent of the spurious covariates. 

In situations when the causes and non-causes are not disentangled in the covariate space,  there could be a representation $\zv(\xv)$ of the covariates where $\zv_{S}(\xv)$ are the causes of the outcome and $\zv_{\backslash S}(\xv)$ are the non-causes. Based on the representation $\zv$, the DGP becomes $y^e \leftarrow f(\tilde{\Bmat}^*\zv_S(\xv^e); \gammav^*) + \epsilon^e$, and the predictor is $\hat{y}^e = f(\Bmat \zv(\xv^e;\gammav_1); \gammav_2)$ with parameters $\alphav = (\Bmat, \gammav_1, \gammav_2)$. As will be shown in \Cref{sec:mnist,sec:animal}, empirically we find CoCo in such situations can produce predictions not depending on the spurious associations. For example, \Cref{sec:mnist} studies the Color-MNIST data set where the input is a  digit image $\xv \in \bR^{14\times14\times2}$ with $14\times14$ pixels and 2 color channels.  Across the data points, the causal variables (the  digit pixels) change their positions in the image and color channels. The position change makes the selection of causes infeasible in the covariate space but possible in the representation space.  We find that for such cases, CoCo can still learn a robust predictor  relying on the causal information such as the digit pixels. We leave the theoretical analysis of causal representation learning by CoCo as an interesting future direction. 

In the nonlinear regime, due to the high flexibility of the predictor, identification in the high-dimensional parameter space can be difficult.  Different parameterizations can represent a similar
mapping from the input to the output. Thus the same data generation might correspond to an equivalent class of points in the parameter space \citep{heinze2018invariant,christiansen2020causal}. Consequently, the optima of CoCo may not be unique. However, the optima points enjoy robust predictive properties.

\begin{proposition}[Local optimality]
Suppose $\alphav'$ minimizes CoCo objective \Cref{eq:coco} with $f_{\cE}(\alphav') = 0$. Suppose a new environment $\iota$ satisfies $p^\iota(x,y) = \sum_{e \in \mathcal{E}} w_e p^e(x,y)$, $\sum_{e \in \mathcal{E}} w_e = 1$, then $\frac{\partial}{\partial \alpha_{\pi}} R^{\iota}(\alphav)|_{\alphav = \alphav'} = 0$, $\pi = \text{supp}(\alphav')$.
\label[proposition]{prop:pred}
\end{proposition}

\Cref{prop:nonlinear} and \Cref{assp:invariance}  guarantee that the causal coefficient $\betav$ satisfies $f_{\cE}(\betav) = 0$ in \Cref{eq:coco}. Therefore, if there exists another global optima $\alphav'$ under these assumptions, it satisfies the condition $f_{\cE}(\alphav') = 0$ in \Cref{prop:pred}. \Cref{prop:pred} shows that if the data distribution of a new environment is a mixture of those for the training environments, the optimum by CoCo already minimizes the predictive risk of this new environment locally for its nonzero elements. This means CoCo optima can transfer the predictive accuracy, quantified by the risk  value, from the training environments to their mixtures. As we will show in empirical studies, a predictor optimized by 
\gls{CoCO}  generalizes the predictive accuracy to non-i.i.d. data in the new environments where the spurious associations change.

%% file: empirical.tex
In the empirical study, we aim to answer the following questions: (1) Can \gls{CoCO} accurately estimate causal effects when some covariates are spuriously associated with the outcome? (2) Can \gls{CoCO} make an accurate prediction in new environments by relying on causal information? (3) How sensitive is \gls{CoCO} to its assumptions and tuning parameter? (4) Are the empirical results aligned with theoretical analysis? To answer these questions, we study \gls{CoCO} and its comparable methods on simulated data, semi-synthetic data, and real data. Code implementations for the empirical studies are available at \url{https://github.com/mingzhang-yin/CoCo}.

\subsection{Causal inference on the synthetic data}
\label{sec:linear-sync}

In this section, we apply CoCo for causal inference with linear synthetic data.  Consider the scenario where we know one exogenous variable, but the other variables are of unknown status; some might be spurious variables, some might be direct causes, some might be neither.  As discussed in \Cref{sec:method}, running ERM with such data will end up with biased estimates of the causal coefficients.  %
But if we have data from multiple environments, we can use \gls{CoCO} to estimate the causal coefficients.  %

\begin{figure}[t]
    \centering
    \includegraphics[width=0.75\textwidth]{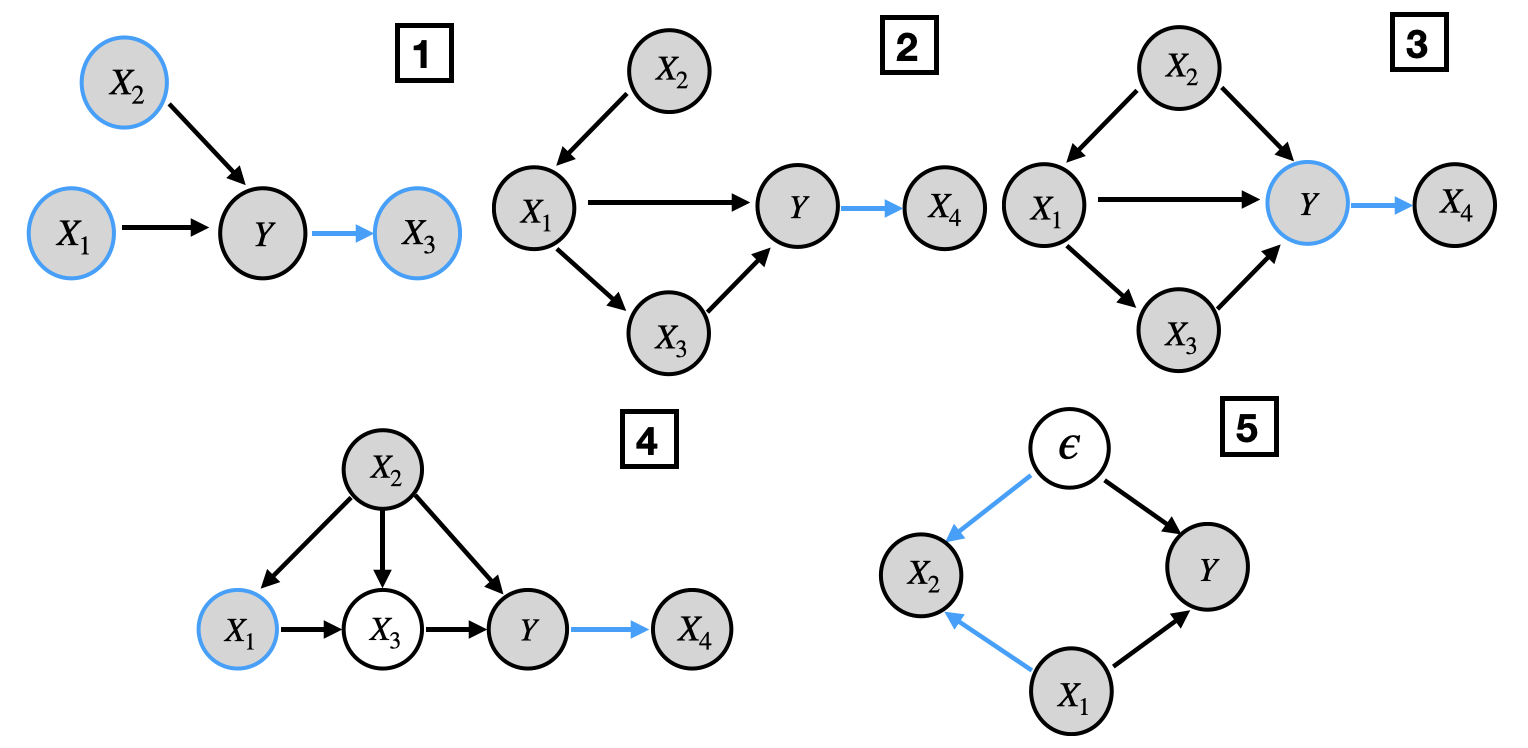}
    \caption{The graphs for the simulation studies in
    \Cref{sec:linear-sync}. The case ID of each graph is in the rectangle box. The blue arrow represents a path whose parameter varies across environments, the blue circle of a covariate means its distribution given parents varies across environments, and the blue circle of outcome means the variance of its additive noise varies across environments.  Invariance \Cref{eq:invariance} holds in all cases. The shaded nodes are the variables that are observed.}
    \label{fig:dags}%
\end{figure}

We generate data from five different graphs in \Cref{fig:dags} with \gls{SEM}s in \Cref{tab:sems}. In cases 1 to 5, variables $x_3$, $x_4$, $x_4$, $x_4$, $x_2$ are  spuriously associated with the outcome, respectively.  The mapping from the causes to the outcome is linear with additive noise. We specify $x_1$ as a known exogenous variable (for the use of the method in \Cref{alg:coco}) and run \gls{CoCO}, IRM, and ERM to estimate the causal coefficients.  To generate data from different environments, we set the parameter $\gamma^e$ in \gls{SEM}s of  \Cref{tab:sems} with environment-specific parameters $m_1^e, m_2^e \sim  \text{Unif}(0,1)$, $m^e \sim  \text{Unif}(1,2)$. We generate two environments with $\gamma^e \in \{0.5, 2.0\}$, each environment with 10,000 data points.  As required, the DGPs leave the causal coefficient invariant.   %

The five graphs test different scenarios: (1) independent causes; (2)
observed mediator; (3) observed confounder and mediator; (4)
observed confounder and unobserved mediator; (5) unobserved confounding with an omitted common cause.  The data generation covers the following circumstances:  variables except the outcome might be generated from nonlinear models (Case 2, 3); the distribution of the causes of the outcome might shift across environments (Case 1, 4);  the variance of the outcome distribution conditional on its causes might vary across environments (Case 3).  Whether a method can produce accurate estimation in all of these situations reflects its generalizability.  We evaluate with the mean absolute error (MAE) between the estimation
$\alphav$ and the true coefficients $\betav$. We compare \gls{CoCO} with ERM, IRM \citep{arjovsky2019invariant}, V-REx \citep{krueger2020out}, RVP \citep{xie2020risk} and Causal Dantzig \citep{rothenhausler2019causal}. The properties and assumptions of these methods are summarized in \Cref{tab:assumptions} in \Cref{sec:summary} for comparison. For the algorithms with tuning parameter $\lambda$, we report the best result for IRM with $\lambda \in \{2,20,200\}$, for V-REx and RVP with $\lambda \in \{10, 10^2,10^3,10^4\}$. We choose stepsize from $\{0.01, 0.1\}$ that produces the lowest objective for each method. For all methods, the algorithm is considered to converge if the mean absolute difference between the parameters in consecutive iterations is less than $10^{-3}$ and the total iterations are over $10^4$. ~\looseness=-1

\setlength{\extrarowheight}{3pt}
\begin{table}[]
\centering
\caption{The mean absolute error of the estimations for
causal parameters $\betav$ (lower the better).   The estimate by \gls{CoCO} (this paper)  is close to the true causal coefficients across
the DGPs.  \gls{CoCO} has a more accurate estimation comparing to RVP \citep{xie2020risk}, V-REx \citep{krueger2020out}, Causal Dantzig \citep{rothenhausler2019causal}, IRM \citep{arjovsky2019invariant} and ERM. The reported mean and the standard deviation in the parentheses are computed across ten independent trials. }
\setlength{\tabcolsep}{0pt}
\begin{tabular*}{\textwidth}{@{\extracolsep{\fill}\quad}lccccc@{}}
\toprule
Case &   1&    2&    3&    4&   5  \\ \midrule
ERM & 0.31 (0.06) &  0.16 (0.00) &  0.32 (0.00) &  0.19 (0.03) &  0.38 (0.01) \\
V-REx &  0.16 (0.06) &  0.11 (0.01)&  0.44 (0.01)&  0.13 (0.04)&  0.06  (0.10)\\
RVP &  0.10 (0.04)&  0.10 (0.01)&  0.43 (0.01)&  0.11 (0.04)&  0.05 (0.04)\\
Dantzig  &  0.54 (0.62)&  3.23 (2.64)&  4.95 (3.06)&  0.43 (0.05)&  0.20 (0.01) \\ 
IRMv1 & 2.12 (0.70) &   \cellcolor[HTML]{d1d0d0}  0.01 (0.00) &  0.02 (0.01)&  2.17 (0.65)&  0.72 (0.35)  \\
CoCo & \cellcolor[HTML]{d1d0d0}  0.01 (0.00) &  0.02 (0.01) &  \cellcolor[HTML]{d1d0d0}   0.01 (0.01) &   \cellcolor[HTML]{d1d0d0}  0.01 (0.01)&   \cellcolor[HTML]{d1d0d0}  0.01 (0.00)\\ \bottomrule
\end{tabular*}%
\label{tab:synthetic}
\end{table}

The results are presented in \Cref{tab:synthetic}. The mean MAE shows that the estimates of pure prediction \gls{ERM}  are biased when the covariates have spurious associations with the outcome. IRMv1 with proper hyper-parameter performs well in cases 2, 3, while it has large MAEs in other cases.  It suggests the IRMv1 performance is affected by a limited number of environments and the type of intervention. The MAE by IRM has a high standard deviation across random trials, indicating IRM converges to different global optima. This echoes the discussion in \Cref{sec:connection} that the IRMv1 objective is over-relaxed and has excessive non-causal optima.  V-REx and RVP perform better than ERM  except in Case 3 when the variance of the exogenous noise of outcome is not invariant. This means their performance largely relies on whether a strong assumption of invariance in \Cref{eq:strong-invariance} is satisfied.  Causal Dantzig has large MAE in these cases which might be because the DGPs do not satisfy the inner-product invariance it requires \citep{rothenhausler2019causal}.   In comparison to these methods, \gls{CoCO} estimates have the lowest or equally lowest error uniformly over all the cases.  The MAEs with four environments are presented in \Cref{fig:4env} in Appendix.

\begin{table}[ht]
 \caption{  SEMs for the simulation study in \Cref{sec:linear-sync}. The environments are indexed by $e$.  $\gamma^e, m_1^e, m_2^e, m^e$ are fixed scalars in an environment.\label{tab:sems} }
 \centering
\begin{tabular}{ccc}
\toprule
Case 1 & Case 2 & Case 3 \\
\midrule
\centering \parbox{0.26\textwidth}{
\begin{equation*}
\begin{split}
    x_2^e \leftarrow &\cN(m_2^e, (\gamma^e)^2) \\
    x_1^e \leftarrow & \cN(m_1^e, (\gamma^e)^2) \\
    y^e \leftarrow &3x_1^e + 2x_2^e  + \cN(0,1) \\
    x_3^e  \leftarrow &\gamma^ey^e + \cN(0, (\gamma^e)^2)    
\end{split}
\end{equation*}
}
& 
\parbox{0.28\textwidth}{
\begin{equation*}
\begin{split}
    x_2^e \leftarrow &\cN(1, (\frac 12)^2) \\
    x_1^e \leftarrow &x_2^e + \text{Unif}(-1,1) \\
    x_3^e \leftarrow &\sin(x_1^e) + \cN(0, (\frac 12)^2) \\
    y^e \leftarrow &2x_1^e +  1.5 x_3^e + \cN(0,1) \\
    x_4^e  \leftarrow &\gamma^ey^e + \cN(0,1)
\end{split}
\end{equation*}
}
&
\parbox{0.38\textwidth}{
\begin{equation*}
\begin{split}
    x_2^e \leftarrow &\cN(1, (\frac 12)^2) \\
    x_1^e \leftarrow &x_2^e + \text{Unif}(-1,1) \\
    x_3^e \leftarrow &\sin(x_1^e) + \cN(0, (\frac 12)^2) \\
    y^e \leftarrow &2x_1^e + x_2^e +  1.5 x_3^e + \cN(0,(\gamma^e)^2) \\
    x_4^e  \leftarrow &\gamma^ey^e + \cN(0,1)
\end{split}
\end{equation*}
}  \\
\midrule
Case 4 & Case 5 & \\
\midrule
\parbox{0.26\textwidth}{
\begin{equation*}
\begin{split}
    x_2^e \leftarrow &\cN(1, (\frac 12)^2) \\
    x_1^e \leftarrow &x_2^e + \text{Unif}(0,m^e) \\
    x_3^e \leftarrow &x_1^e + x_2^e + \cN(0, (\frac 12)^2) \\
    y^e \leftarrow &x_2^e + 2x_3^e + \cN(0,1) \\
    x_4^e  \leftarrow &\gamma^ey^e + \cN(0,1)
\end{split}
\end{equation*}
}
&
\parbox{0.28\textwidth}{
\begin{equation*}
\begin{split}
	\epsilon^e \leftarrow & \cN(0,1) \\
    x_1^e \leftarrow &\cN(1, \frac12) \\
    y^e \leftarrow & 2x_1^e +  \epsilon^e \\
    x_2^e  \leftarrow &0.5\gamma^e\epsilon^e + (0.5 + \gamma^e)x_1^e  + \cN(0,1)
\end{split}
\end{equation*}
}
& \\
\bottomrule
\end{tabular}
\end{table}

As an ablation study, we replace the elementwise produce in \gls{CoCO} objective by the inner product and minimize $\sum_{e \in
\mathcal{E}} ( \ip{\nabla R^e(\alphav)}{\tilde{\alphav}})^2$. We call this  method 
Naive-\gls{CoCO}, which is an intermediate between the CoCo objective \Cref{eq:thm1} and the IRMv1 objective in \Cref{eq:linear-gaussian}.  Naive-\gls{CoCO}  has high MAEs $1.17, 0.53,  0.51, 1.31, 0.03$ across the five cases, respectively. It means the key element for the CoCo objective is the Hadamard product and the norm derived from optimizing in all the feasible directions (\Cref{sec:estimation}), instead of knowing a variable is exogenous.   ~\looseness=-1

To test the model checking method proposed in \Cref{sec:suff}, we generate heterogeneous data with $\gamma^e \sim \text{Unif}(0,5)$ for each environment. When the set of known non-descendant of outcome is $\cC=\{1\}$, Cases 1, 4, 5 pass the checking condition with the number of environments 3, 3, 2 respectively, while cases 2, 3 do not pass. We further check the nonidentifiable case in \Cref{sec:nonidentify}. It cannot pass the checking step with any number of environments.  In practice, CoCo can accurately estimate the causal coefficients with two environments, as shown in \Cref{tab:synthetic}. It validates our analysis in \Cref{sec:suff} that 
the checking step is a sufficient condition for identification but is not a necessary condition.  

\paragraph{Model Mismatch.} Using data from Case 5, we further study the performance of \gls{ERM} and \gls{CoCO} when the predictive model does not exactly match the data-generating model. The data in Case 5 is generated from a linear model. We compare two predictors, one is a linear model that matches the DGP, and the other is a nonlinear neural network. Both models are trained with \gls{ERM} and \gls{CoCO}. In Case 5,  $X_1$ is the cause, and $X_2$ is a predictive but non-causal covariate. 

\begin{figure}[htbp]
\centering{
    \subfloat[ERM, linear]
   {{\includegraphics[width=0.25\textwidth]{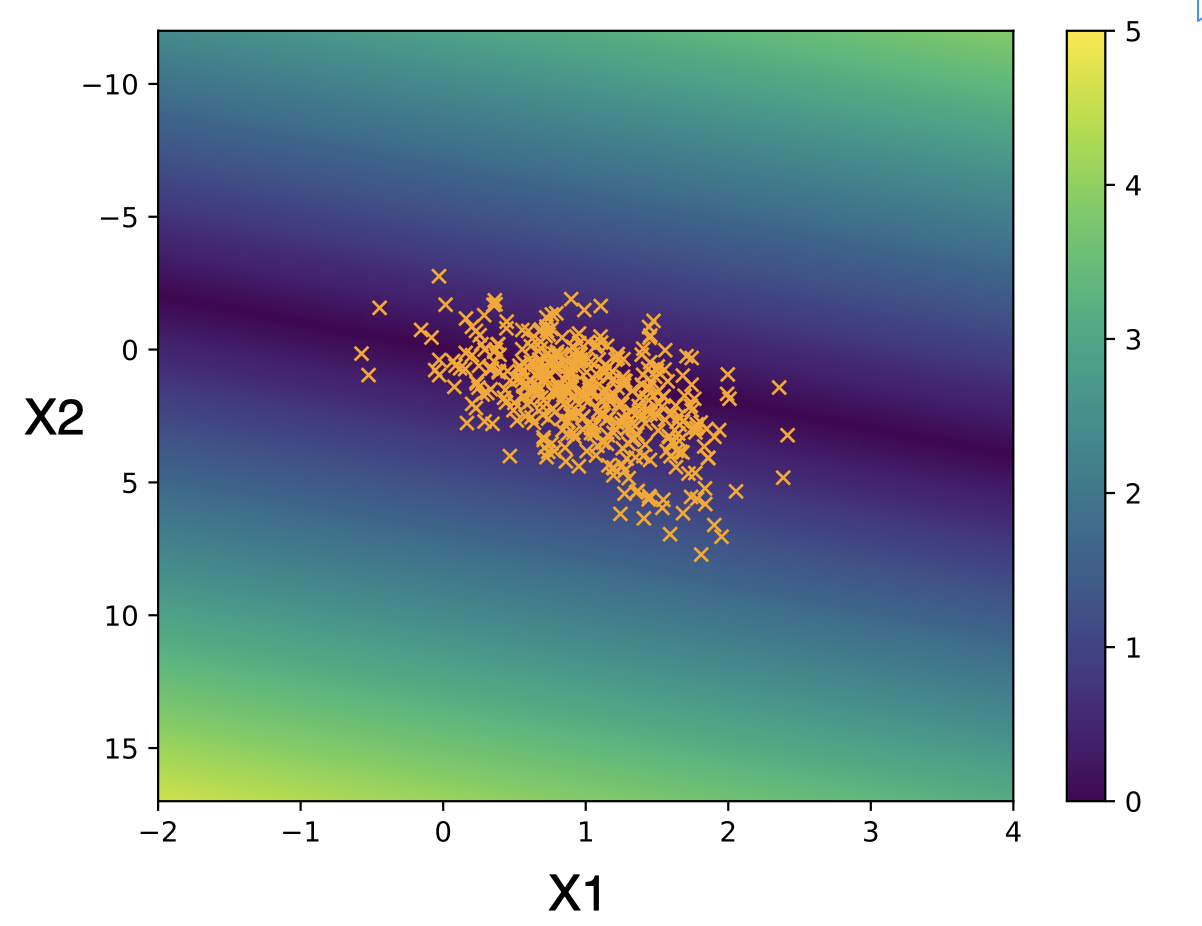} }}  
    \subfloat[CoCo, linear]
    {{\includegraphics[width=0.25\textwidth]{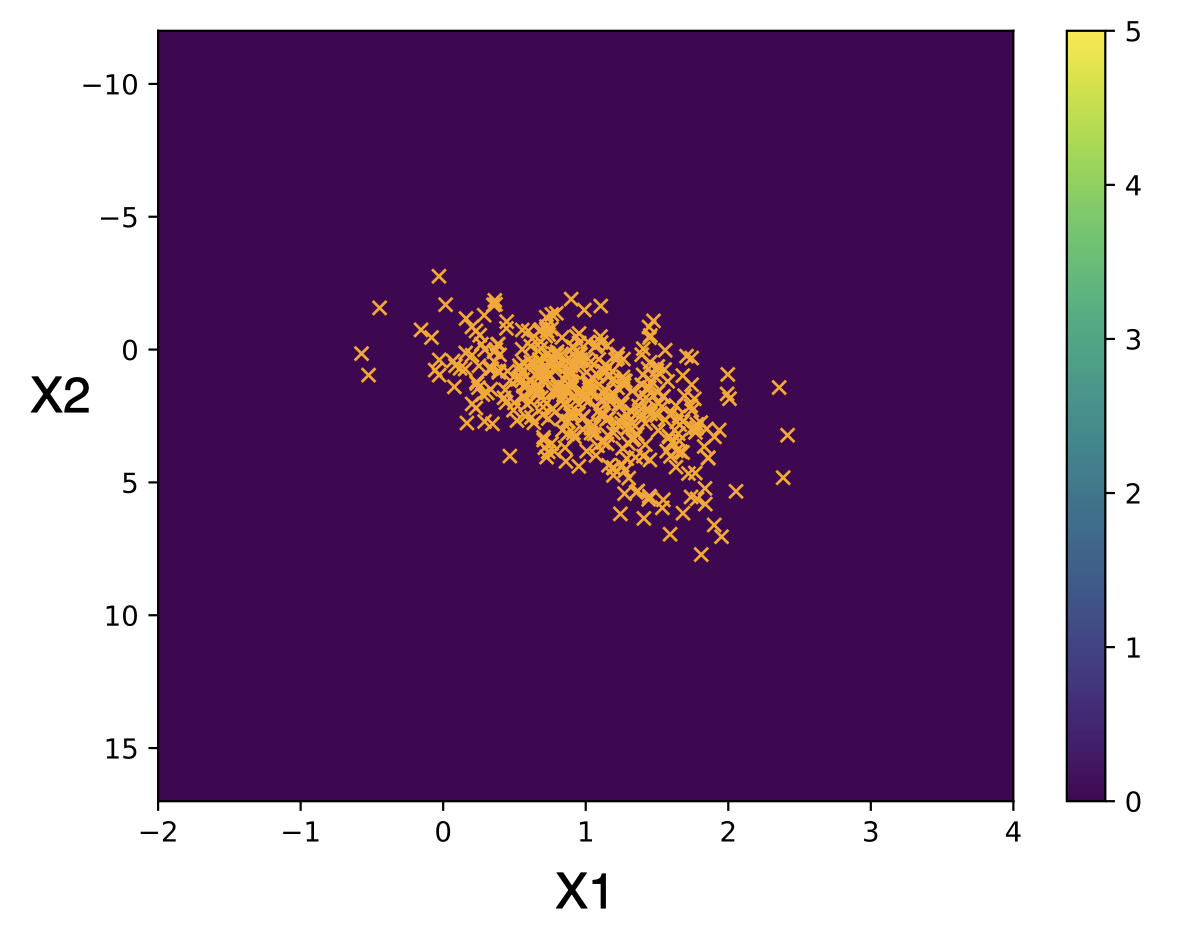} }} 
    \subfloat[ERM, nonlinear]
   {{\includegraphics[width=0.25\textwidth]{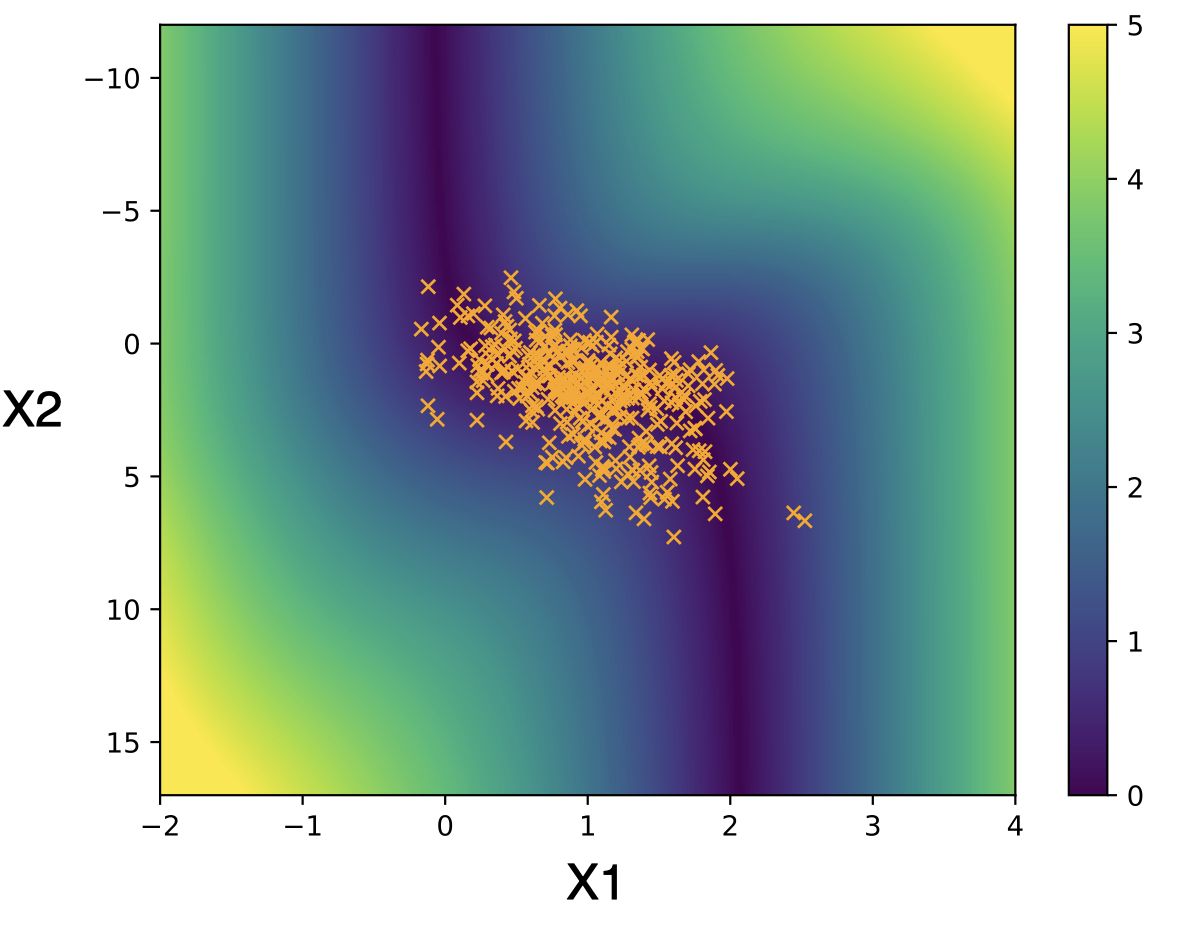} }} 
    \subfloat[CoCo, nonlinear]
    {{\includegraphics[width=0.25\textwidth]{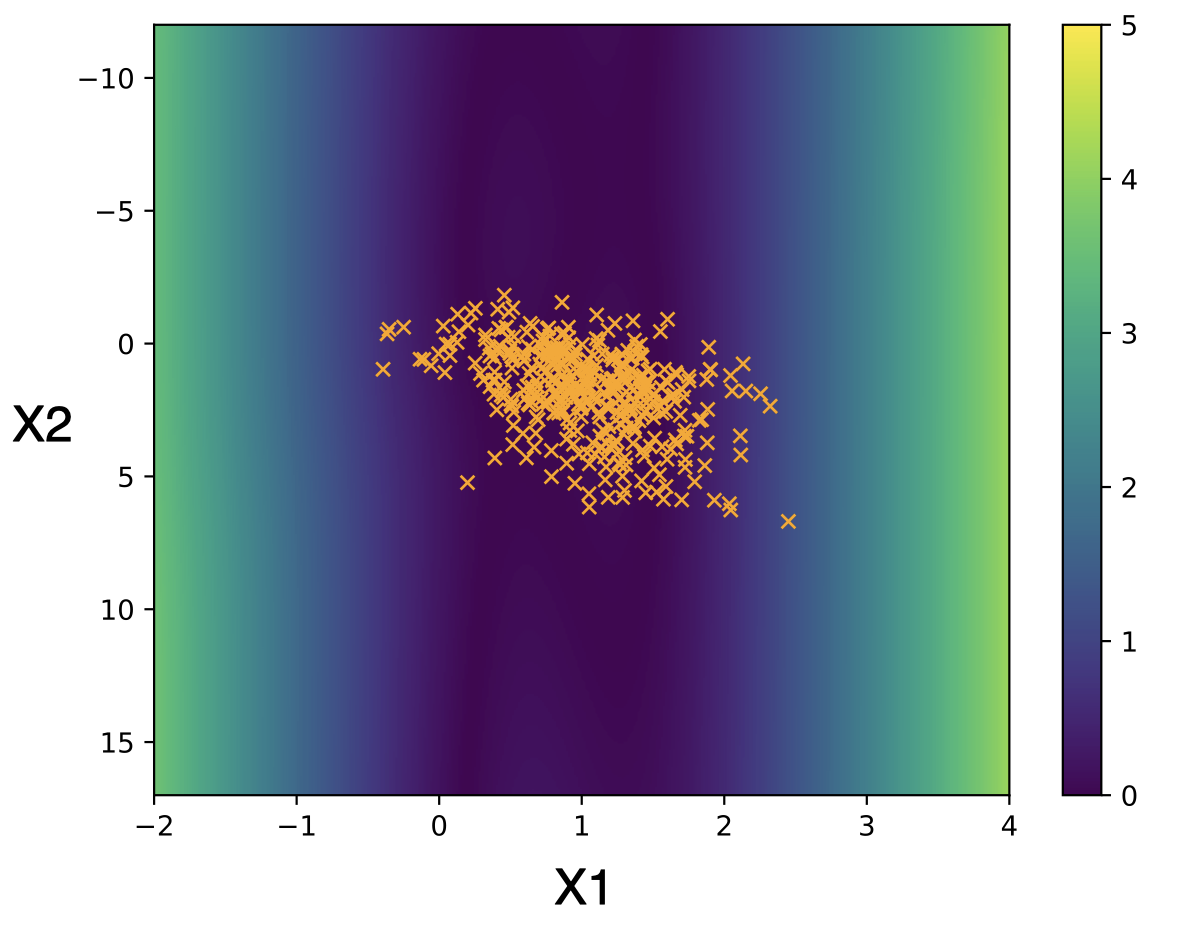} }}
}
    \caption{Prediction accuracy for \gls{CoCO} and  \gls{ERM} with
    linear and nonlinear  predictors.  The heatmap is the prediction error
    $(\hat{y} -
    \E[y|x])^2$, the x-axis, y-axis are the values of input $x_1$ and $x_2$. The orange points are training data %
    from two environments. \gls{CoCO}  has better out-of-sample
    generalization with a wider region of low error (blue region) than ERM for both linear and nonlinear predictors.}
    \label{fig:mis-specify}%
\end{figure}

As shown in \Cref{fig:mis-specify}, when the model is specified correctly as linear, the predictor trained by \gls{ERM} fail to generalize to new values of  $X_2$, but the model trained by \gls{CoCO} can generalize to any input  $(X_1, X_2)$. When the predictor is nonlinear, differing from the DGP,  \gls{ERM} prediction is only accurate for the data interpolating the training points. In contrast,  \gls{CoCO} can make accurate predictions for the inputs with new values of spurious covariate $X_2$.  From the robust prediction view,  \gls{CoCO} prediction can generalize to new environments by avoiding the spurious association for both linear and nonlinear predictors.

\subsection{Gaussian mixture example}
\label{sec:gmm}
We study a multi-class classification problem with a categorical outcome when the inputs contain non-causal covariates. We modify a Gaussian mixture model (GMM) to simulate the data set. The observed covariates are $(\xv_1^e,\xv_2^e)$ and the outcome is $y^e$, where $e$ is the environment index. For each environment $e$, the data are generated with \gls{SEM}
\bac{
& \textstyle \xv_1^e \leftarrow  \sum_{k=1}^K \frac{1}{K} \cN( \muv_k, \Imat) \\ 
&y^e  \leftarrow \text{Categorical}(p_1,\cdots, p_K) \\
& \xv_2^e\leftarrow (1-p^e) \delta_{\uv^e_{y^e}} + p^e \delta_{\uv^e_{k_1}}, 
\label{eq:gmm}
} 
where $p_k = \cN(\xv_1^e; \muv_k,  \Imat) /  \sum_{k'=1}^K  \cN(\xv_1^e; \muv_{k'}, \Imat)$, $ k_1 \sim \text{Multinomial}(1/K,\cdots, 1/K)$. 
Among the covariates, the mapping from the causes $\xv_1^e$ to label $y^e$ is invariant across all $e$, whereas $\xv_2^e$ is predictive to $y^e$ due to spurious associations. %
\begin{figure}[htbp]
    \centering
    \subfloat[Noise scale]
    {{\includegraphics[width=0.33\textwidth]{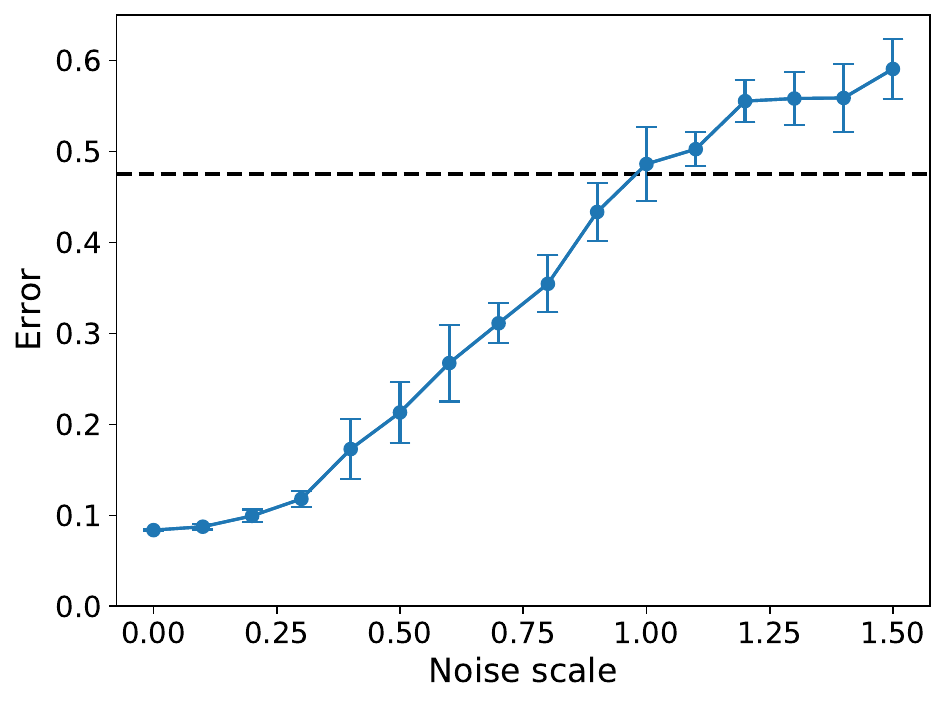} }}  \hspace{-3mm}
    \subfloat[$\#$ of environments]
    {{\includegraphics[width=0.33\textwidth]{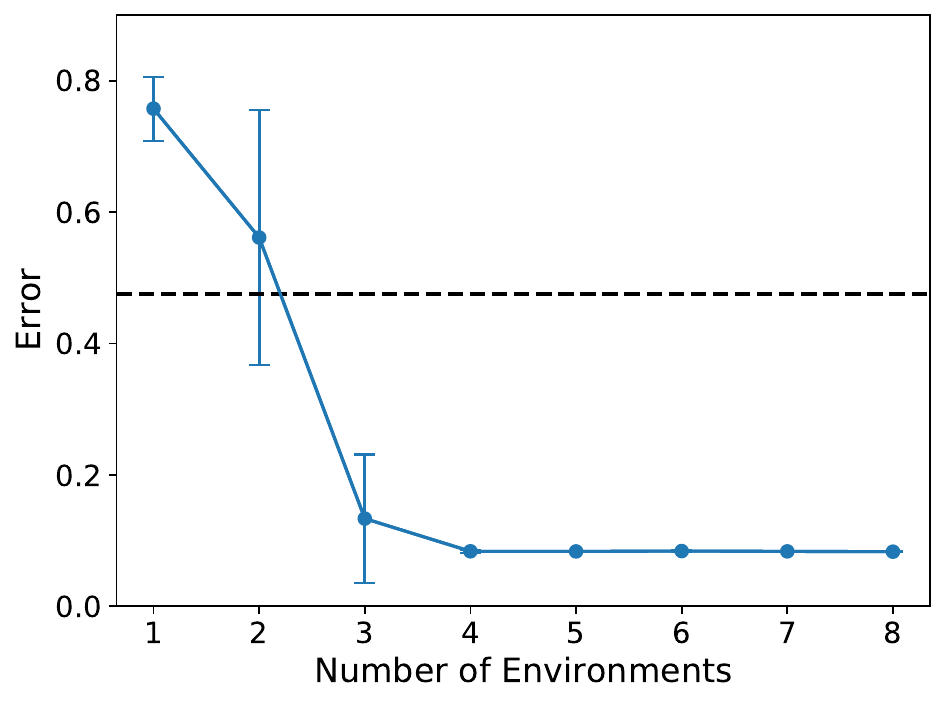} }}  \hspace{-3mm}
        \subfloat[Regularization strength]
    {{\includegraphics[width=0.33\textwidth]{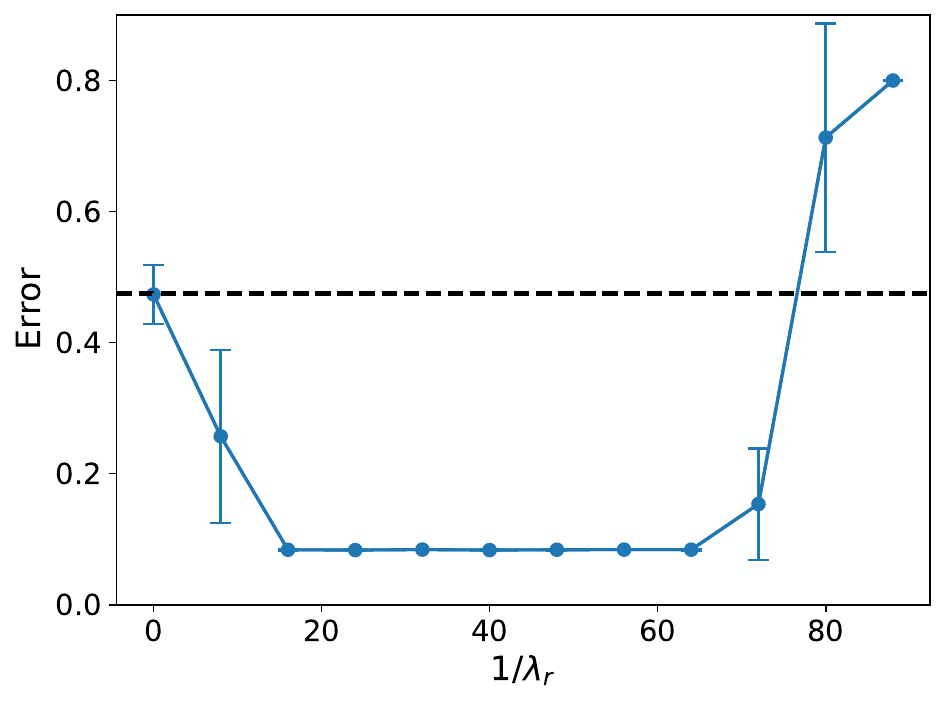} }}
    \caption{ The change of test predictive error of \gls{CoCO} with different
    levels of invariance, number of environments, and the
    hyperparameter. The dashed line is the \gls{ERM} error rate for reference. The error bar is the standard deviation over 5
     trials.}
    \label{fig:knobs}%
\end{figure}

\begin{figure}[htbp]
    \centering
     \subfloat[GMM]
    {{\includegraphics[width=0.38\textwidth]{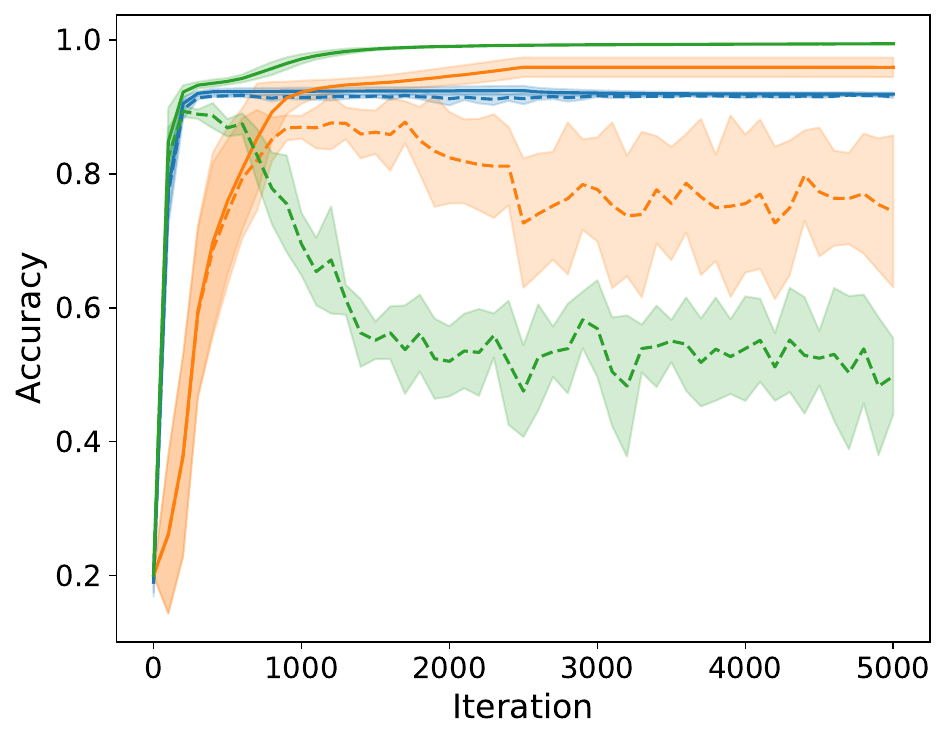} }} \hspace{3mm}
    \subfloat[Colored MNIST ]
    {{\includegraphics[width=0.56\textwidth]{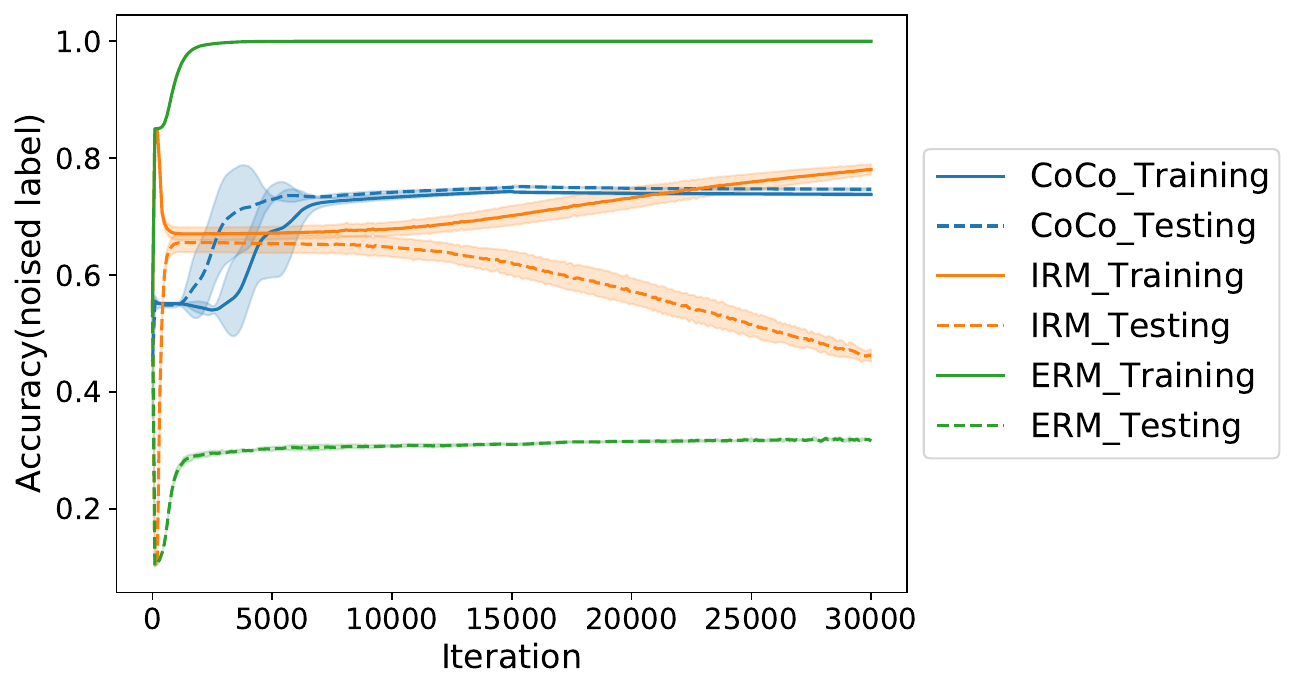} }}  
    \caption{ Trace plot of training and testing accuracy for
    \gls{CoCO}, \gls{IRM} and \gls{ERM} on GMM and Colored MNIST data. In panel (b), the accuracy is measured on predicting the \emph{noised
    label} $y^e$.  \gls{CoCO} has the highest prediction accuracy in a new environment. }
    \label{fig:gmmtrace}%
\end{figure}

The observed data are generated from \Cref{eq:gmm}. The GMM component centers $\muv_k = \sqrt{1.5K} \mathbf{e}_k \in \bR^K$.  To generate the non-causal covariates $\xv_2^e$, we first  generate $K$ random vectors $\{\uv_k^e\}_{k=1}^K$ with $\uv_k^e \sim \prod_{i=1}^{[k/2]} U(0,1)$ for environment $e$. Then for a data point in the component $y^e$, $\xv_2^e $ equals $\uv_{y^e}^e$  with probability $1-p^e$ and equals a random vector from $\{\uv_k^e\}_{k=1}^K$ otherwise. By doing so, $\xv_2^e$ is associated with $y^e$, but the association varies across environments when $\uv^e_{1:K}$ changes with environment $e$.  The \gls{DGP}s that generate the environments are characterized by the values of $\uv_{1:K}^e$ and $p^e$. We set the training environments with $K=5$ and $p^e \in \{0.01,
0.02, \allowbreak \cdots, 0.05\}$ in \Cref{eq:gmm}.  For a validation/test environment $f$ we generate a new set of $\{\uv_k^f\}_{k=1}^K$ and set $p^f=0$. 

We evaluate the test performance by averaging the accuracy over ten testing environments. If the predictor learns to predict based on the causes $\xv_1^e$ instead of the spurious covariates $\xv_2^e$, it can accurately predict  $y^e$ in training and testing environments.

We use a fully connected neural network
with two hidden layers as the predictor and train the model with \Cref{alg:coco2}. %
For both \gls{CoCO} and \gls{IRM}, the penalty weight
 is chosen from ten values equally spaced from 1 to 100  on a log-scale using the validation environments. The weight on the empirical risk term is reduced to 0 after 5k iterations. ~\looseness=-1

\begin{figure}[htbp]
    \centering
    \subfloat[\gls{CoCO}]
    {{\includegraphics[width=0.33\textwidth]{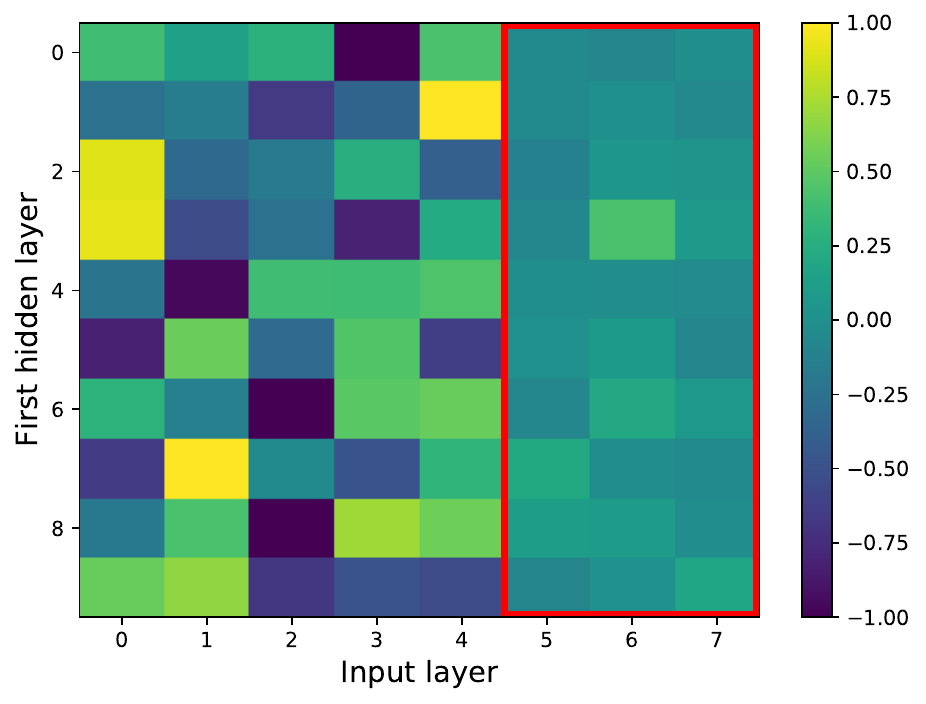} }}  \hspace{-3mm}
    \subfloat[\gls{IRM}]
    {{\includegraphics[width=0.33\textwidth]{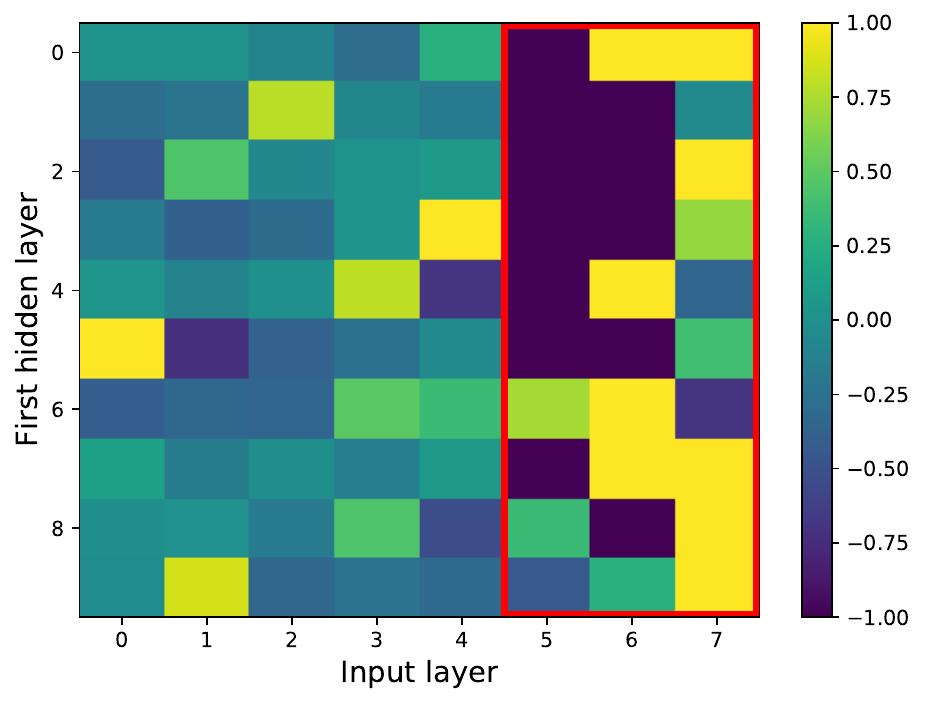} }}  \hspace{-3mm}
        \subfloat[\gls{ERM}]
    {{\includegraphics[width=0.33\textwidth]{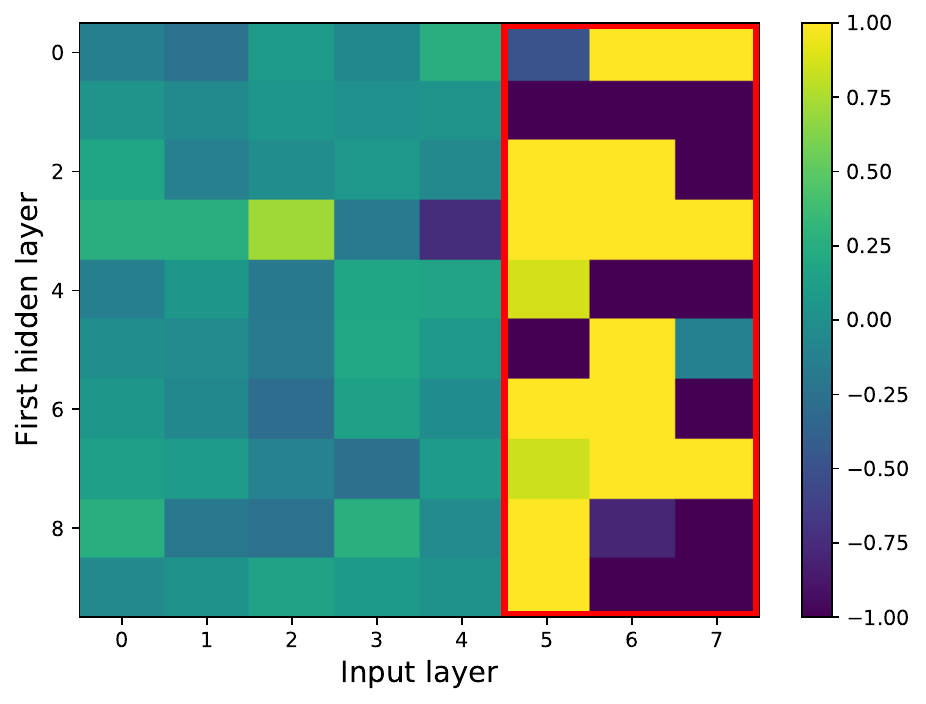} }}
    \caption{ The heatmap for the first layer weight matrix of the
    neural networks trained by \gls{CoCO}, \gls{IRM} and \gls{ERM} on the GMM data.
    The matrix dimension is $10 \times 8$ where the input dimension is
    8, and the first hidden layer dimension is 10. In the input, the
    first five elements are $\xv_1$, and the last three elements are
    $\xv_2$.  Comparing to \gls{IRM} and \gls{ERM}, \gls{CoCO} solution has the weights related to non-causal input $\xv_2$
    (the right block) close to 0. } \label{fig:weights}
\end{figure}

The results are shown in \Cref{fig:weights,fig:knobs,fig:gmmtrace} and  \Cref{tab:accu}. \Cref{fig:gmmtrace} is the trace plot for the
predictive accuracy. The testing
accuracy increases for all the methods in the early stage of training but drops in the later stage for \gls{ERM} and \gls{IRM}. We hypothesize that \gls{ERM} and \gls{IRM} first improve the prediction by utilizing all the 
covariates, including the causal ones. But in the later stage of training, it relies more heavily on the
spurious associations to boost the training accuracy, which harms the accuracy at
the test time.  \Cref{fig:weights} shows the weight matrix that connects the input and the first hidden layer. The model trained by \gls{CoCO}
zeros out the weights associated with the spurious covariates $\xv_2$ (the
right block), aligned with the analysis in
\Cref{prop:nonlinear}. In comparison, the weights obtained by
\gls{IRM} and \gls{ERM} are mostly nonzero, passing the spurious association from the input to the subsequent hidden layers and outputs.

\Cref{tab:accu} summarizes the numerical results. CoCo and V-REx have the highest prediction accuracy in the test environments. In this example, the strong invariance  \Cref{eq:strong-invariance} is satisfied due to the DGP in \Cref{eq:gmm}, which meets V-REx assumption. Hence, V-REx has a high accuracy close to CoCo.  When the data satisfies the strong invariance, it is potentially beneficial to add the V-REx regularization of equal noise variance to the CoCo objective.

\paragraph{Sensitivity to the assumption and hyper-parameter.} In \Cref{fig:knobs}, we study how \gls{CoCO} performs if the invariance assumption is violated and the hyperparameters change. In panel (a), we construct the training environments
by changing the cluster centers $\{\muv_k\}_{k=1}^K$ in \Cref{eq:gmm} to $\{\muv_k +
\epsilonv_k^e\}_{k =1}^K$, $\epsilonv^e_k \sim \cN(\mathbf{0},
\sigma^2 \Imat_K)$. The noise $\epsilonv^e_k$ changes the mapping from
the covariates $\xv_1^e$ to the label $y^e$ across the environments, deviating from the invariance assumption. The
noise scale $\sigma^2$ reflects the magnitude of deviation. Panel (a)
shows that the testing predictive accuracy increases as the invariance tends to hold. Under a moderate deviation from the invariance assumption, the test prediction by \gls{CoCO} remains more accurate than that by ERM. 
In panel (b), we study how the test accuracy varies with the number of environments $M$ in training.  Training environments are constructed by setting the vectors $\uv^e_{k} \sim \prod_{i=1}^{[k/2]} U(0,1)$ for all $k$  and $p^e \in \{0.01,
0.02, \cdots, 0.01M\}$ in \Cref{eq:gmm}.  We find a growing number of environments reduces the testing error monotonically due to the increased environments heterogeneity. In panel (c), we study
how the testing error changes with the penalty weight $\lambda_r$  in
\gls{CoCO} objective~\Cref{eq:coco-erm}. When $\lambda_r$ is large,
the objective is close to the empirical risk, and the test error is
high; when $\lambda_r$ is small, the parameters might collapse to the point $\mathbf{0}$. Between the two extremes, \gls{CoCO}  can learn a model that makes robust predictions in new environments.  %

\subsection{Colored MNIST (CMNIST)}
\label{sec:mnist}
CMNIST is a semi-synthetic data set for
binary classification introduced in
\citet{arjovsky2019invariant}. Based on the MNIST data set,  the image of hand-written digits 0-4 and 5-9 are labeled
as $\tilde{y}=0$ and $\tilde{y}=1$, respectively.  For each
environment, the outcome $y^e$ is generated with 
probability 0.75 as $\tilde{y}$ and with probability 0.25 as $1-\tilde{y}$. We call $\tilde{y}$  the \emph{clean labels} and $y^e$  the \emph{noised labels}.
The digit is colored green with probability $p^e$ if $y^e =1$ and with probability
$1-p^e$ if $y^e =0$; if not colored green, it is colored red. The \gls{DGP}s across the environments differ in the value of $p^e$.  Environments are constructed for the training with $p^e \in \{0.1,0.2\}$, for the validation with $p^e = 0.5$ and for the testing with $p^e = 0.9$.

The predictor takes the colored digit image as the input and the noised label $y^e$ as the target.  The input $\xv \in \bR^{14\times14\times 2}$ where the image has $14 \times 14$ pixels and two color channels.  The relationship between the digit shape and  $y^e$ is genuinely
causal, while the relationship between the color and $y^e$ is spurious. The goal is to learn a predictor that makes the predictions based on the digit shape rather than the color. A predictor using color information cannot accurately predict the noised label $y^e$ at the test time and the clean label $\tilde{y}$ during training and testing.

\begin{table}[t]
 \caption{ Predictive accuracy in training and testing environments
 on the GMM, CMNIST, and Wildlife data. For GMM, the Oracle results are the predictions with causes $\xv_1^e$ instead of $(\xv_1^e, \xv_2^e)$. For CMNIST, the prediction accuracy is reported for both clean label $\tilde{y}$ and noised labels $y^e$; the Oracle is the same function fitted on grey-scale images by \gls{ERM}. \label{tab:accu} }
 \centering
 \setlength{\tabcolsep}{8pt}
\begin{tabular*}{\textwidth}{lrrrrrrr}
\toprule
    &   \multicolumn{2}{c}{GMM} & \multicolumn{3}{c}{ CMNIST} & \multicolumn{2}{c}{Wildlife} \\
   & Train & Test & Train ($\tilde{y}$)& Test ($\tilde{y}$) & Test ($y^e$) &  Train& Test \\
\midrule
\gls{ERM} & 99.4 & 51.0  & 75.8 & 44.4 &31.1& 99.6 & 58.4  \\ 
\gls{IRM} & 95.9 & 75.9 &  81.4 & 70.3 & 46.5&83.4 & {84.9}    \\ %
V-REx & 92.6 & {91.4} &  75.2 & 49.5 & 31.8&96.2 & 67.3  \\
\gls{CoCO} (this paper)   & 91.9 &  \cellcolor[HTML]{d1d0d0}{91.6}  & 93.0 &  \cellcolor[HTML]{d1d0d0}{92.9} &  \cellcolor[HTML]{d1d0d0}{74.7} & 86.1 & 
 \cellcolor[HTML]{d1d0d0}{85.2}  \\   \midrule %
Random Guess   & 20 & 20  & 50 & 50 & 50 &50 & 50  \\
Oracle  & 92.3 & 91.8 & 99.3 & 97.9 & 74.8 & - & -  \\
\bottomrule
\end{tabular*}
\end{table}

\noindent\textbf{Empirical results.} We optimize the predictor, a fully connected neural network with two hidden layers, by \Cref{alg:coco2}. For the algorithms in \Cref{sec:linear-sync}, we compare with ERM, IRM \citep{arjovsky2019invariant}, and  V-REx \citep{krueger2020out} that can generalize to nonlinear models. More baseline results are reported in \Cref{tab:cmnist} in Appendix. 

The weight of the risk term for CoCo and V-REx objectives is chosen on the validation environment from $2\times\{10^{-1}, \cdots, 10^{-5}\}$, and is reduced by a factor of 10  after 15k iterations (half of the total iterations).  We use Adam optimizer  \citep{kingma2014adam} with learning rate $10^{-4}$. %
For \gls{IRM}, we reduce the  learning rate of  the public code to $10^{-4}$ to
ensure stability over long iterations, and we use all the other
hyperparameters and annealing strategies provided in the author's code  \footnote{\url{https://github.com/facebookresearch/InvariantRiskMinimization}}.

\Cref{fig:weights2} in \Cref{sec:addtional} visualizes the weight matrix that connects the input $\xv$ and the first hidden layer. The weight learned by ERM connects the hidden layer to all the inputs, hence encoding non-causal information in the latent representation. In contrast, the weight learned by CoCo sets multiple columns close to zero and is symmetric over the two color channels, potentially removing the dependence of the hidden layer on some non-causal pixels.

The numerical results are shown in \Cref{tab:accu}. The trace plot for the noised label is in \Cref{fig:gmmtrace} (b) and the trace plot for the clean label is in \Cref{fig:wild}  (a) of \Cref{sec:addtional}. The results show that \gls{ERM} predicts labels with accuracy close to 1 in the training but has the lowest accuracy at the testing. The reason might be that its prediction largely depends on the color information rather than the digit shape, whereas the association between color and label $y^e$ changes from training to testing environments.  The testing accuracy of \gls{IRM} increases in the early stage of training but drops in the later stage. We hypothesize
that the model at first improves the prediction by utilizing all the information including that of digit shape, but later it relies more on the color information, which reduces the accuracy at the test time. Similar patterns appear in the prediction of clean label $\tilde{y}$, as shown in \Cref{fig:wild}.

\subsection{Natural image classification}
\label{sec:animal}

In this example, following
\citet{causalsurvey2020}, we adapt the iWildCam 2019 dataset \citep{beery2019iwildcam} that contains wildlife images taken in the wild. The images are collected from
different cameras, each at one of 143 locations. The goal is to classify coyotes and raccoons in images. The data collected in different locations and time usually follow different distributions due to varying physical factors such as landscape, season, vegetation, illumination conditions,  etc. Therefore, the images taken from different cameras can be considered data from heterogeneous environments. 
The physical factors reflected in the image background might be predictive of the species but due to spurious association. Our goal is to learn a predictor that can make accurate predictions in a new environment by training on the data from a limited number of environments. This goal can be achieved if the predictor manages to recognize coyotes and raccoons based on animal pixels instead of using spurious associations from the backgrounds. 

Based on the setting of 
\citet{causalsurvey2020}, 
we use images from two locations as the training data and images from
another location as the test data. We use images from an additional location as the validation data. The inputs are 512-dimensional features
extracted from ResNet18 \citep{he2016deep}, a pre-trained model on the
ImageNet dataset \citep{deng2009imagenet}. The predictor is a fully connected neural network with one hidden layer of size 10. CoCo is trained by \Cref{alg:coco2}. In this example, we find that adding the weak condition \eqref{eq:v1} in  \Cref{alg:coco2} with weight
$\lambda_w$ improves convergence stability with random initialization\footnote{Adding a general condition in \Cref{eq:general} to the strong condition of \Cref{eq:v2}
does not change the optima set, but it may improve the smoothness of the optimization landscape.}. %
The hyperparameters are selected on the validation environment. We set the 
weak condition weight $\lambda_w =10^{-4}$ and the risk regularizer weight $\lambda_r=1$. $\lambda_r$ is reduced to $10^{-5}$ after 100 epochs. The risk regularizer is an inductive bias to encourage nonzero solutions. After the optimizer is sufficiently away from the zero point, annealing the risk regularizer prevents the algorithm from minimizing the objective by reducing the risk function, hence preventing it from using the spurious association.  CoCo is compared with ERM, IRM \citep{arjovsky2019invariant}, and  V-REx \citep{krueger2020out}. All methods are trained by ADAM with a learning rate $10^{-3}$. ~\looseness=-1

The result is summarized in \Cref{tab:accu} and \Cref{fig:wild} (b) in
\Cref{sec:addtional}.  It demonstrates that \gls{ERM} has high accuracy in the training but low
accuracy during testing.  \gls{CoCO} accuracy is slightly higher than \gls{IRM} and is much higher than V-REx and ERM.  Compared to \gls{ERM}, prediction by \gls{CoCO} has a slight drop in training accuracy but has significantly higher testing accuracy. 
\gls{CoCO} has the smallest performance gap between training and testing,
indicating that it largely avoids predicting animal labels via information from
image backgrounds, i.e., information that varies across environments.

%% file: appendix.tex
\section{Proofs} 
\label{sec:proof}
In this section, we present proofs for the results in the main paper. First, we prove the causal optimality results  of the proposed optimization problems.

\begin{proof}[Lemma~\ref{lem:optimal}]
Let the random vector $\xv = (x_1, \cdots, x_p)^\top$ denote the
covariates. The expected mean square error is
\bas{
&\E[(y - \hat{y})^2] \\
=& \E[(\alphav^\top \xv - \betav^\top \xv - \epsilon)^2] \\
=&  (\alphav - \betav)^\top\E[\xv \xv^\top] (\alphav - \betav) - 2  \E[(\alphav - \betav)^\top\xv \epsilon] + \E[\epsilon^2].
}
Since $supp(\alphav) \subset supp(\betav)$, the $(\alphav - \betav)^\top\xv$ is
a linear combination of the true causes as $\sum_{j \in supp(\betav)}
(\alpha_j - \beta_j) x_j$ which is independent of $\epsilon$ by the
\gls{SEM}, thus $\E[(\alphav - \betav)^\top\xv \epsilon] = 0$. Since $\E[\xv
\xv^\top] $ is assumed to be positive definite, the unique optima of the
square error is $\alphav = \betav$.
\end{proof}

\begin{proof}[Lemma~\ref{lem: optima1}]
Recall that $X_{j^*}$ is a known exogenous variable with $X_{j^*} \indep \epsilon$.  We first prove that the optima set of the modified \Cref{eq:thm1} is a subset of that for the non-modified \Cref{eq:coco}, but it still keeps the causal coefficient. This is because $\nabla\,R^e(\alphav)_{j^*}=0$ implies
$\nabla\,R^e(\alphav)_{j^*} \alpha_{j^*}=0$, hence a minimizer of \Cref{eq:thm1} must be a minimizer of \Cref{eq:coco}. 

Next we prove that the causal
coefficient $\betav$ minimizes \Cref{eq:thm1} to zero. It is sufficient to show that $\nabla\,R^e(\betav)_{j^*}=0$ since we already
know $\norm{\nabla R^e(\betav) \circ \betav}_2=0$. It is true because for the DGP in \Cref{eq:lSEM-e}, $\nabla\,R^e(\betav)_{j^*} = -\E[x_{j^*}\epsilon] =
0$. 

Finally, we show that the vector $\zero$ is not an optimum of \Cref{eq:thm1} almost surely 
when $\beta_{j^*} \neq 0$. The zero vector minimizes \Cref{eq:thm1} if and only
if $\nabla\,R^e(\zero)_{j^*}=0$. For the linear model, 
it requires $\sum_{j \in S} \E[x^e_jx^e_{j^*}] \beta_j =0$ for all $e
\in \cE$. By the independent causal mechanisms principle \citep{scholkopf2019causality,scholkopf2021toward}, the causal coefficients $\betav$ (the mechanism) are independent of the distribution of causes $\xv_S$. It means for the zero vector to be an optimum, $\betav_S$ has to fall in the intersection of $|\cE|$
hyperplanes in $\bR^{|S|}$ which has measure zero, hence the probability is zero. 
\end{proof}

\begin{proof}[Lemma~\ref{lem:opt1}]
\Cref{sec:estimation} shows the feasible directions $\mathcal{U} = \text{span}
\{\mathbf{e}_j: j \in S\}$. Therefore, the causal parameter $\betav$ itself is a feasible direction with $\betav
\in \mathcal{U}$. 

The first order condition implies the optima of the \Cref{eq:constrained} sets the directional derivative to zero in the direction of $\betav$, i.e.,
$\ip{\nabla  R(\alphav)} {\betav} = 0$.  By \Cref{lem:optimal},
plugging $\betav$ into $\alphav$  produces $\ip{\nabla
R(\betav)}{\betav} = 0$,  which means
\ba{
\betav \in \argmin_{\alphav}  ~(\ip{\nabla  R(\alphav)}{\alphav})^2.
}

Similarly, any partition
$\mathcal{P}$ of the set $\idx{p}$ gives a necessary condition  that admits 
\ba{
\betav \in \argmin_{\alphav}  ~ \sum_{A \in \mathcal{P}}(\ip{\nabla
R(\alphav)_A}{\alphav_A})^2.
}
\end{proof}

\begin{proof}[Proposition~\ref{prop:irm}] The statement follows directly from the Cauchy–Schwarz inequality$$\sum_{e \in \mathcal{E}}p \norm{\nabla R^e(\alphav) \circ \alphav}^2_2 \geq  \sum_{e \in \mathcal{E}} (\ip{\nabla  R^e(\alphav)}{\alphav})^2  \geq 0$$.  
\end{proof}

\begin{proof}[Lemma~\ref{prop:subset}]
By \Cref{eq:grad_thm1}, the sub-vector of the gradient is 
\ba{
\nabla R(\alphav)_H = W_H(\alphav - \betav) - \mathbf{s}_H
\label{eq-prop-1}
}
By \Cref{eq:part-gradient}, 
\ba{
 \nabla_{\alphav_H} R_H(\alphav_H) = W_{HH}(\alphav_H - \betav_H) - W_{HH^c} \betav_{H^c} - \mathbf{s}_H.
 \label{eq-prop-2}
}
By \Cref{eq-prop-1}, suppose $\alphav$ is  a CoCo objective optima with $\nabla R(\alphav)_H = 0$ and $\alphav_{\backslash H} = \zero$, $\alphav$  also satisfies $ \nabla_{\alphav_H} R_H(\alphav_H)$, so its nonzero elements are the ERM solution when regressing with input $\xv_H$. On the other hand, suppose $\xv_H$ is ERM solution with input $\xv_H$, $\alphav = (\alphav_H, \zero)$  will be a CoCo objective optima  by \Cref{eq-prop-2}. 
\end{proof}

The following proofs are for the identification results in main paper \Cref{sec:identify}. 

\begin{proof}[Theorem~\ref{thm:lm}]
Let $s_j^e = \E[X_j^e\epsilon] = cov(X_j^e, \epsilon)$, $\mathbf{s}^e = (s_1^e, \cdots, s_p^e)^T$. By the data generating process, $s_j^e = 0$ for $j \in \{1,\cdots, K\}$. Let 
\ba{
g^e(\alphav) = \norm{\nabla R^e(\alphav) \circ \tilde{\alphav}]}_2,~~ f(\alphav) = \frac{1}{|\cE|} \sum_{e \in \mathcal{E}} g^e(\alphav). 
\label{eq:coco-obj}
}
where $f(\alphav)$ is CoCo objective. Direct computation shows
\ba{
\nabla R^e(\alphav) = W^e(\alphav - \betav) - \mathbf{s}^e
\label{eq:grad_thm1}
}
Notice $f(\alphav) \geq 0$ and by the structual equation model, due to independence of the exogenous noise $\epsilon$ and causes $\text{Pa}(Y)$, we have $\mathbf{s}^e \circ \betav =\bzero$. Hence for $\alphav^* = \betav$, $f(\alphav^*) = 0$. This guarantees the existence of a solution as causal coefficient $\betav$. To prove the identification, it is sufficient to prove that for all $\alphav \neq \alphav^*$, $f(\alphav) > 0$. We use proof by contradiction.

Let $H = \text{supp}(\tilde{\alphav})$ and $H^c$ as its compoment set in $\idx{p}$. We assume $f(\alphav) = 0$ and $\alphav \neq \betav$ and deduce a contradiction. Since $f(\alphav) = 0$, for all $e$, $\norm{g^e(\alphav)} = 0$. Since $g^e(\alphav) = \nabla R^e(\alphav) \circ \tilde{\alphav}$, it means $\nabla R^e(\alphav)_H = \bzero$, for all $e$. However, by the characterization of the feasible set in Section~\ref{sec:estimation}, Assumption A2) implies that there does not exist $\alphav \in \bR^p$,  $\alphav \neq \betav$, such that $\nabla R^e(\alphav)_H = \mathbf{0}$, $\forall$ $e \in \cE$. Otherwise, the set $H$ is an invariant set and both $\alphav$ and $\betav$ are invariant estimations, which violates Assumption A2). Hence for $\alphav \neq \betav$, there exists an environment  $e' \in \cE$ with $\nabla R^{e'}(\alphav)_H \neq \bzero$. This yields a contradiction.
\end{proof}

\begin{proof}[Corollary~\ref{cor:expand}]
The claim i) is trivial. To see why the claim ii) holds, we prove its equivalent contrapositive statement.  If assumption (A2) does not hold  for $\cE_2$, it means there exists $\alphav$, $\nabla\bR^e(\alphav)_{H} = \mathbf{0}$ for all $e \in \cE_2$, which  also applies to all $e \in \cE_1$ since $\cE_1 \subset \cE_2$. Hence the assumption (A2) does not hold for $\cE_1$. 
\end{proof}

\begin{proof}[Theorem~\ref{thm:weak}]
We prove the statement by contradiction. Using notations in \Cref{eq:coco-obj}, suppose for $\alphav^*$, $f(\alphav^*)=0$ and $\alphav^*_{\cC} \neq \betav_{\cC} $. Then let set $H = supp(\alphav^*) \cup \cC$. Since $f(\alphav^*)=0$, we have $\nabla R^e(\alphav^*)_H = \mathbf{0}$, for all $e$. By \Cref{eq:part-gradient}, this means
\ba{
W^e_{HH}(\alphav^*_H - \betav_H) = W^e_{HH^c} \betav_{H^c}  + \mathbf{s}_H^e,~~\forall~e
\label{eq:thm4}
}
Denoting $\deltav = \alphav^*_H - \betav_H$, we have $\deltav \neq \mathbf{0}$. Then \Cref{eq:thm4} contradicts with the assumption (A2'). 
\end{proof}

\begin{proof}[Corollary~\ref{thm:suff}]
Since the do intervention satisfies validity assumption (A1), to prove the identification of the treatment effect for the variable of interest, say $x_{j^*}$, by \Cref{thm:weak} it is sufficient to show that it satisfies the weak effectiveness assumption (A2'). We suppose the environments $\cE$ violate assumption (A2'), that is $\exists~H \subset \idx{p},  \deltav \in \bR^{|H|}, \deltav \neq \bzero$ such that $\Wmat_H^{\cE} \deltav = \thetav_H^{\cE}$, and yield a contradiction.  

By For notation convenience, denote $\delta_{\sigma(j)}$ as the element of $\deltav$ associated with the column of $\Wmat_H^{\cE}$ that consists of the elements $\{W_{ij}^e; e \in \cE, i \in H\}$; here $\sigma:H \mapsto \idx{|H|}$ is a bijection by definition. Suppose $\E[X_{j^*}] \neq 0$.

Consider the set of variables $\mathcal{S}_{\deltav} = \{j : j \in H, x_{\sigma(j)} \neq 0\} \cup \{j: j \in H^c, \beta_j \neq 0\}$. Since $\deltav \neq \mathbf{0}$, the set $\mathcal{S}_{\deltav}$ is non-empty.  We consider the youngest node $X_j$, $j \in \mathcal{S}_{\deltav}$ that there is no direct path  from $X_j$ to any other node with index in $\mathcal{S}_{\deltav}$.  There are two possible cases. When $j \in H \cap \mathcal{S}_{\deltav}$,  for $e, e' \in \cE$ with $\cI^e = \cI^{e'} = \{j\}$, we have the linear equation $W_{jH}^e \deltav = \theta_j^e$ in the linear system $\Wmat_H^{\cE} \deltav = \thetav_H^{\cE}$ as
\ba{
\sum_{k \in H} \delta_{\sigma(k)} \E[X_j^e X_k^e] = \sum_{t \in H^c} \beta_t \E[X_j^e X_t^e] +  \E[X_j^e \epsilon^e]. 
\label{eq:pf-thm5}
}
For the do intervention $X_j^e \leftarrow a_j^e$,  we have $\E[X_j^e X_k^e] = a_j^e \E[X_k^e] =a_j^e  \mu_k^e$, and since  $X_j$ is the youngest node, \Cref{eq:pf-thm5} becomes
\ba{
\delta_{\sigma(j)} a_j^e  + \sum_{k \in H, k \neq j} \delta_{\sigma(k)} \mu_k = \sum_{t \in H^c} \beta_t \mu_t
\label{eq:suff_pf1}
}
for environment $e'$, we  have
\ba{
\delta_{\sigma(j)} a_j^{e'}  + \sum_{k \in H, k \neq j} \delta_{\sigma(k)} \mu_k = \sum_{t \in H^c} \beta_t \mu_t
\label{eq:suff_pf2}
}
Since  $a_j^e \neq  a_j^{e'}$, Eqs.~\eqref{eq:suff_pf1}, \eqref{eq:suff_pf2} are inconsistent and yield a contradiction. When $j \in H^c \cap \mathcal{S}_{\deltav}$, analogously we have the linear equation $W_{j^*H}^e \deltav = \theta_{j^*}^e$ as
\ba{
\sum_{k \in H} \delta_{\sigma(k)} W_{j^*k}  = \sum_{t \in H^c, t\neq j} \beta_t W_{j^*t} + \beta_j \mu_{j^*} a_j^e
\label{eq:suff_pf3}
}
and equation $W_{j^*H}^{e'} \deltav = \theta_{j^*}^{e'}$ as
\ba{
\sum_{k \in H} \delta_{\sigma(k)} W_{j^*k}  = \sum_{t \in H^c, t\neq j} \beta_t W_{j^*t} + \beta_j \mu_j^* a_j^{e'}
\label{eq:suff_pf4}
}
Since  $a_j^e \neq  a_j^{e'}$, Eqs.~\eqref{eq:suff_pf3}, \eqref{eq:suff_pf4}  yield a contradiction.
\end{proof}

\begin{proof}[\Cref{prop:checking}]
Suppose \Cref{assp:eff-w} does not hold, then by \Cref{eq:inconsistent} there exists $H \subset \idx{p}$, $ \cC \subset H$ and $\deltav \in \bR^{|H|}$, $\deltav \neq \mathbf{0}$, such that $\Wmat^e_{HH} \deltav + \Wmat^e_{H H^c}(- \betav_{H^c}) = \mathbf{s}_H^e$ for all $e \in \cE$. Since $x_i \indep \epsilon$, for $i \in \cC$, we know $s_i^e = 0$.  Letting $\vv_{H} = \deltav$, $\vv_{H^c} = - \betav_{H^c}$, we have 
\ba{
\Wmat^{\cE}_{\mathcal{C} \cP} \vv= \Wmat^{\cE}_{\mathcal{C} H} \vv_H +  \Wmat^{\cE}_{\mathcal{C} H^c} \vv_{H^c} = \mathbf{s}_{\cC}^{\cE} = \mathbf{0},
}
 which cannot pass the checking step since $\vv \neq \mathbf{0}$.  
\end{proof}

\begin{proof} [Proposition~\ref{prop:nonlinear}]
By the construction of $ \Lambda$, $\Bmat^* =  \Amat
\Lambda$ is a matrix where the $j$-th column $B^*_j = \mathbf{0}$ if
$j \notin S$. Similar to the proof of \Cref{lem:optimal},  we can
compute the $L_2$ risk as
\bas{
& \E[(y - \hat{y})^2] \\ =& \E[(f_{\gammav}(\Bmat\xv) -
f_{\gammav^*}(\Bmat^*\xv) - \epsilon)^2] \\ 
= &
\E[(f_{\gammav}(\Bmat\xv) - f_{\gammav^*}(\Bmat^*\xv) )^2] -
2\E[((f_{\gammav}(\Bmat\xv) - f_{\gammav^*}(\Bmat^*\xv) )\epsilon] +
\E[\epsilon^2]. } 
Due to the constraints, $B_j = B_j^* = \mathbf{0}$, $\Bmat\xv \indep
\epsilon$, $\Bmat^*\xv \indep \epsilon$, therefore the second term is
zero.  Then the $L_2$ risk reaches its minimum as $\E[\epsilon^2]$
when $\Bmat = \Bmat^*$,  $\gammav = \gammav^*$.
\end{proof}

\begin{proof}[\Cref{prop:pred}]
Since $f_{\mathcal E}(\alphav')=0$, $\forall e \in \mathcal{E}$
\begin{align*}
   ||\nabla R^e(\alphav') \circ \alphav'||_2 = 0.
\end{align*}
Let $\pi = \text{supp}(\alphav')$, we have $(\nabla R^e(\alphav'))_\pi = 0$. This means for $j \in \pi$,
\begin{align*}
    0 = \frac{\partial}{\partial \alpha_{j}} R^{e}(\alphav') = \int p^e(\xv,y)\frac{\partial}{\partial \alpha_{j}} L(y, \hat{y}(\xv;\alphav))|_{\alphav=\alphav'} d\xv dy  %
\end{align*}

For $j \in \pi$, 
\begin{align*}
    \frac{\partial}{\partial \alpha_{j}} R^{\gamma}(\alphav) =& \frac{\partial}{\partial \alpha_{j}} E_{\xv,y \sim p^\gamma(\xv,y)} L(y, \hat{y}(\xv;\alphav)) \\
    =& \sum_{e \in \mathcal{E}} w_e \int p^e(\xv,y)\frac{\partial}{\partial \alpha_{j}} L(y, \hat{y}(\xv;\alphav)) d\xv dy.
\end{align*}
Plug in $\alphav = \alphav'$ we have $\frac{\partial}{\partial \alpha_{j}} R^{\gamma}(\alphav)|_{\alphav=\alphav'} = 0$. 
\end{proof}

\section{A Summary of algorithms on multiple environments}

We summarize the properties of CoCo and several representative causal algorithms that leverage data from multiple environments in \Cref{tab:assumptions}.
\label{sec:summary}

\begin{table}[h]
 \caption{  Comparing causal algorithms with multiple environments. \emph{Gnr. interv.}: allow a general type of intervention as long as the invariance in \Cref{eq:invariance} or \Cref{eq:strong-invariance} is satisfied. \emph{nl. model}: has been applied to the nonlinear predictive function. \emph{scalability}: computational efficiency in scaling up to high dimensional problems. \emph{uneq. variance}: allow the variance of exogenous noise of the outcome to vary across environments. \emph{unm.  cf.}: allow unmeasured confounding between the true causes and the outcome. 
 \label{tab:assumptions} }
 \centering
 \resizebox{\columnwidth}{!}{%
\begin{tabular}{lccccc}
\toprule
  & \makecell{gnr. interv.} &  \makecell{nl. model} & scalability &  \makecell{uneq. variance} &   \makecell{unm.  cf.}  \\
  CoCo (this paper) & \ding{51} & \ding{51} & \ding{218}  & \ding{51}&  \ding{55} \\
  IRM \citep{arjovsky2019invariant} & \ding{51} & \ding{51} & \ding{218}  & \ding{51}&  \ding{55} \\
  V-REx \citep{krueger2020out} & \ding{51} & \ding{51} & \ding{218}  & \ding{55}&  \ding{55} \\
  RVP \citep{xie2020risk} & \ding{51} & \ding{51} & \ding{218}  & \ding{55}&  \ding{55}  \\
  group-DRO \citep{sagawa2019distributionally}  & \ding{51} & \ding{51} & \ding{218}  & \ding{55}&  \ding{55}  \\
  ICP \citep{peters2016causal} &\ding{51}&\ding{55}& \ding{216}&\ding{55}   & \ding{55}  \\
  Causal Dantzig \citep{rothenhausler2019causal} &\ding{55}  &\ding{55}  & \ding{218}& \ding{51} &  \ding{51} \\
  LRE \citep{ghassami2017learning} & \ding{51} &\ding{55}& \ding{216}& \ding{51} &  \ding{55} \\
  MC \citep{ghassami2018multi} &\ding{51} &\ding{55}& \ding{216}& \ding{51} &  \ding{55}  \\
\bottomrule
\end{tabular}
}
\end{table}

\section{Analytic case studies}
This section include two concrete cases. One analytically demonstrate how optimization-based methods can estimate causal coefficients, and compare CoCo, IRM and ERM. The other case is an instance for ineffective interventions. 

\subsection{An example of optimization-based estimation}
\label{sec:example}

To illustrative the discussion in \Cref{sec:irm}, we study \gls{ERM}, \gls{IRM} and \gls{CoCO} on a specific example. The DGP follows the Case 1 of \Cref{tab:sems},
\bas{
    x_2^e \leftarrow &\cN(m_2^e, (\gamma^e)^2) \\
    x_1^e \leftarrow & \cN(m_1^e, (\gamma^e)^2) \\
    y^e \leftarrow &3x_1^e + 2x_2^e  + \cN(0,1) \\
    x_3^e  \leftarrow &\gamma^ey^e + \cN(0, (\gamma^e)^2).    
}
We consider the DGP of two environments corresponding to parameters  $(m_1^{(1)}, m_2^{(1)}, \gamma^{(1)}) = (2,0.5,2)$, $(m_1^{(2)}, m_2^{(2)}, \gamma^{(2)}) = (3,-1,0.5)$. 

Consider the predictor $\hat{y} = \alphav^\top \xv^e$ and the risk function $R^e( \alphav)=\frac12 \bE[(y^e-\alphav^\top \xv^e)^2]$. The gradient of the risk function is 
\begin{align*}
\nabla_{\alphav} R^e( \alphav) = \Big( &(\alpha_1-3) \bE[x_1^2] + (\alpha_2-2) \bE[x_1x_2] + \alpha_3 \bE[x_1x_3], \\
&(\alpha_2-2) \bE[x_2^2] + (\alpha_1-3) \bE[x_1x_2] + \alpha_3 \bE[x_2x_3], \\
&\alpha_3 \bE[x_3^2] + (\alpha_1-3) \bE[x_1x_3] + (\alpha_2-2) \bE[x_2x_3] - \gamma \Big). 
\end{align*}
And the moments are 
\begin{align*}
\bE[x_1 x_3] &= 3\gamma(m_1^2 + \gamma^2) + 2\gamma(m_1 m_2), \\
\bE[x_2 x_3] &= 3\gamma(m_1 m_2) + 2\gamma(m_2^2 + \gamma^2), \\
\bE[x_1^2] &= m_1^2 + \gamma^2, \\
\bE[x_2^2] &= m_2^2 + \gamma^2, \\
\bE[x_1 x_2] &= m_1 m_2, \\
\bE[y^2] &= 9\bE[x_1^2] + 12\bE[x_1 x_2] + 4\bE[x_2^2] + 1, \\
\bE[x_3^2] &= \gamma^2 (\bE[y^2] + 1).
\end{align*}
The CoCo optima for each environment is given by solving $\alphav$ with the system of equations
\bas{
&\alpha_1(\alpha_1-3) \bE[x_1^2] + \alpha_1(\alpha_2-2) \bE[x_1x_2] + \alpha_1\alpha_3 \bE[x_1x_3] = 0, \\ 
&\alpha_2(\alpha_2-2) \bE[x_2^2] + \alpha_2(\alpha_1-3) \bE[x_1x_2] + \alpha_2\alpha_3 \bE[x_2x_3] = 0, \\
&\alpha_3^2 \bE[x_3^2] + \alpha_3(\alpha_1-3) \bE[x_1x_3] + \alpha_3(\alpha_2-2) \bE[x_2x_3] - \gamma\alpha_3 = 0.
}
The optima of the IRM regularization for each environment forms the quadric surface
\bas{
&\alpha_1(\alpha_1-3) \bE[x_1^2] + \alpha_1(\alpha_2-2) \bE[x_1x_2] + \alpha_1\alpha_3 \bE[x_1x_3] +  
\alpha_2(\alpha_2-2) \bE[x_2^2] + \alpha_2(\alpha_1-3) \bE[x_1x_2] \\
& \qquad + \alpha_2\alpha_3 \bE[x_2x_3] + 
\alpha_3^2 \bE[x_3^2] + \alpha_3(\alpha_1-3) \bE[x_1x_3] + \alpha_3(\alpha_2-2) \bE[x_2x_3] - \gamma\alpha_3 = 0.
}
 
Empirically, we sample $10^5$ data points of each environment. The ERM solution for the two environments is $\alphav_{\text{ERM}} = (2.815,  1.778,  0.043)$.  The empirical solutions of IRMv1 with various $\lambda$ values are in \Cref{tab:irmcase}, which is obtained with a fixed random seed. The stepsize is 0.01. We consider random initialization or an idealized initialization at the causal coefficient $\betav$.  For $\lambda \in \{0.1,1,10\}$,  both initializations converges within $3 \times 10^4$ iterations. 
For a large $\lambda=1000$, the convergence is slow with $8.8 \times 10^5$ iterations for the random initialization and $1.7 \times 10^5$ iterations for the initialization at $\betav$, likely due to the weak regularization of the empirical risk and the identification issue as discussed in \Cref{sec:irm}. The observation that the idealized initialization does not stay at $\betav$ indicates $\betav$ does not have the lowest empirical risk among the optima set of the IRM regularization. For $\lambda = + \infty$, the optimization will stay at $\betav$ for the idealized initialization since $\betav$ belongs to the optima set of the IRM regularization, but such initialization is not feasible in practice. 
The empirical solution of CoCo is $(3.001, 2.004, 0.000)$ which is close to the true causal coefficient $(3,2,0)$.

\begin{table}[]
 \centering
\begin{tabular}{lcc}
\toprule
$\lambda$ & Random initialization         & Idealized initialization at $\betav$  \\
\midrule 
0.1       & (-0.372,   -0.104, 0.547) & (2.923,   1.842, 0.019) \\
1         & (-0.842, -0.390,  0.621)  & (2.932, 1.847, 0.017)   \\
10        & (-0.877,   -0.411, 0.626) & (2.933, 1.848, 0.017)   \\
1000 & (-0.830, -0.300, 0.613) & (2.954,  1.896,   0.012) \\
\bottomrule       
\end{tabular}
\caption{The convergence points of IRMv1 for the case study. \label{tab:irmcase}}
\end{table}

\subsection{An example of the ineffective intervention}
\label{sec:nonidentify}
Consider environments indexed by $\gamma^e \in \{1,2,3\}$, and SEM as:
\bac{
&x_2^e \leftarrow \cN(0,(\gamma^e/2)^2) \\
&x_1^e \leftarrow x_2^e + U(-\gamma^e,\gamma^e) + 1  \\
& y^e \leftarrow 2 x_1 +  1.5 x_2 +  \cN(0,1) \\
& x_3^e \leftarrow 0.5 \cdot y^e + \cN(0,1).
}
The predictor is linear as:
\ba{
\hat{y}^e(\alphav) = \alpha_1 x_1^e + \alpha_2 x_2^e + \alpha_3 x_3^e.
}
Ideally, we want to identify the causal coefficients $\betav = (2, 1.5,0)$. However, in this example, a straightforward calculation shows the point  $\hat{\alphav} = (1.6, 1.2, 0.4)$ minimizes the risk function $\E[(1/2)(y^e - \hat{y}^e)^2]$ for each environment. This means $\nabla R^e(\alphav)|_{\alphav=\hat{\alphav}} = 0$, and  hence $\hat{\alphav}$ minimizes the objective \Cref{eq:thm1}. Both $\betav$ and $\hat{\alphav}$ belong to the set of optima of objective~\eqref{eq:thm1}, which cannot be distinguished under given interventions. Loosely speaking, $\betav$ and $\hat{\alphav}$ are equally invariant over these three environments.

\begin{table}[htbp]
\centering
\caption{Test accuracy (percent) for noised label $y^e$ on the ColoredMNIST. Results are reported from the original papers. The baseline methods compared with CoCo (this paper) are V-IRMG,  F-IRMG \citep{ahuja2020invariant}, ReBias \citep{bahng2020learning}, DecAug \citep{bai2021decaug},  MM-REx \citep{krueger2020out}, PAIR \citep{chen2022pareto}, Fishr \citep{rame2022fishr}. ~\looseness=-1
}
\resizebox{\textwidth}{!}{%
\begin{tabular}{@{\extracolsep{\fill}\quad}lcccccccc@{}}
\toprule
 &  V-IRMG &    F-IRMG&    ReBias&    DecAug&   MM-REx  & PAIR&Fishr& CoCo\\ \midrule
Test accu. & 49.06(3.4) &  59.91(2.7) &  29.40(0.3) &  69.60(2.0) &  66.1(1.5) &68.4(1.1)&73.8(1.0)& 74.7(0.3)\\
\bottomrule
\end{tabular}%
}
\label{tab:cmnist}
\end{table}

\section{Additional simulation results} 
\label{sec:addtional}
\begin{figure}[htbp]
    \centering
     \subfloat[Colored MNIST] 
    {{\includegraphics[width=0.4\textwidth]{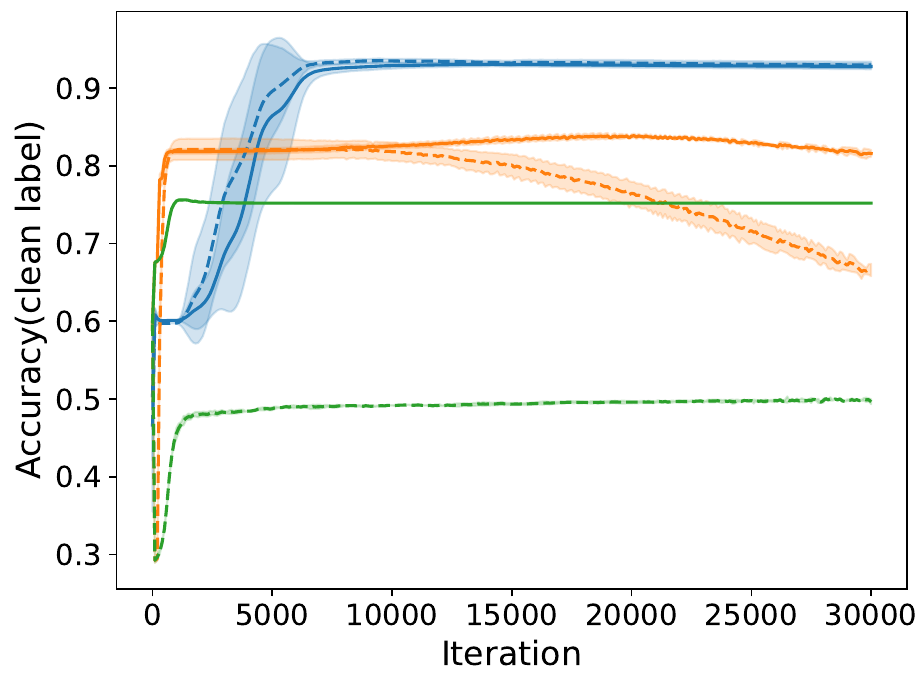} }}  \hspace{5mm}
    \subfloat[Wildlife]
    {{\includegraphics[width=0.42\textwidth]{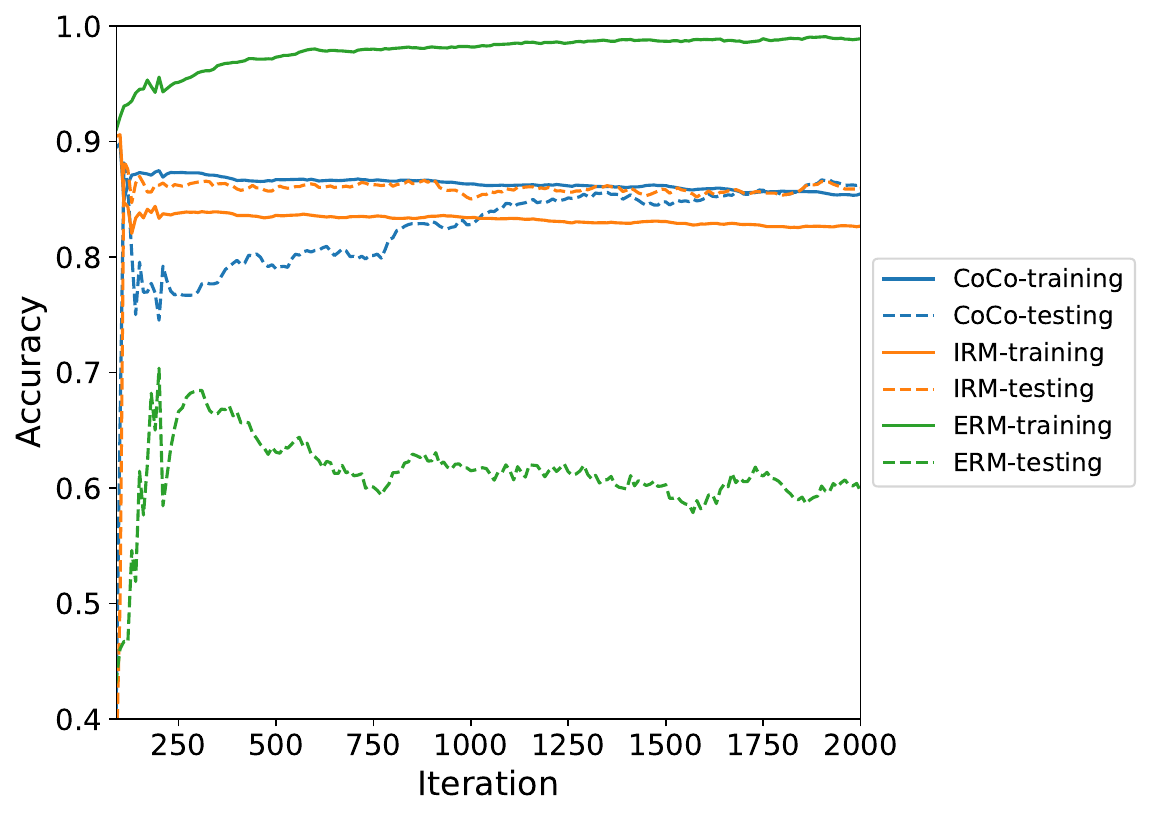} }}
    \caption{ Trace plot of training and testing accuracy for
    \gls{CoCO}, \gls{IRM} and \gls{ERM} on Color-MNIST and Wildlife data.
    In panel (a), the accuracy is measured on predicting the \emph{clean
    label} $\tilde{y}$.  \gls{CoCO} has high accuracy in both training and testing environments. }
    \label{fig:wild}%
\end{figure}

\begin{figure}[htbp]
    \centering
    \subfloat[Two environments]
   {{\includegraphics[width=0.48\textwidth]{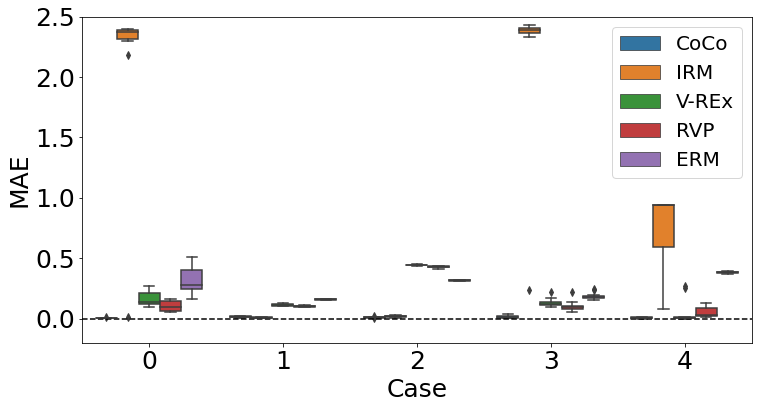} }}  
    \subfloat[Four environments]
    {{\includegraphics[width=0.48\textwidth]{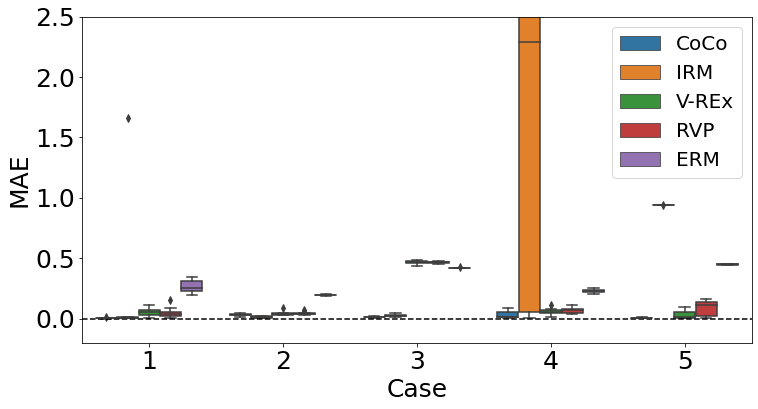} }} 
    \caption{ The boxplot for the mean absolute error (MAE) of the estimations for
causal parameters $\betav$ (lower the better) over two (left) and four environments (right).  \gls{CoCO} estimation is close to the true causal coefficients across
all settings.  \gls{CoCO} has a more accurate estimation comparing to RVP \citep{xie2020risk}, V-REx \citep{krueger2020out},  IRM \citep{arjovsky2019invariant} and ERM. Each case is run with ten independent trials. }
    \label{fig:4env}%
\end{figure}

\begin{figure}[htbp]
    \centering
    \subfloat[Weights by CoCo]
    {{\includegraphics[width=0.35\textwidth]{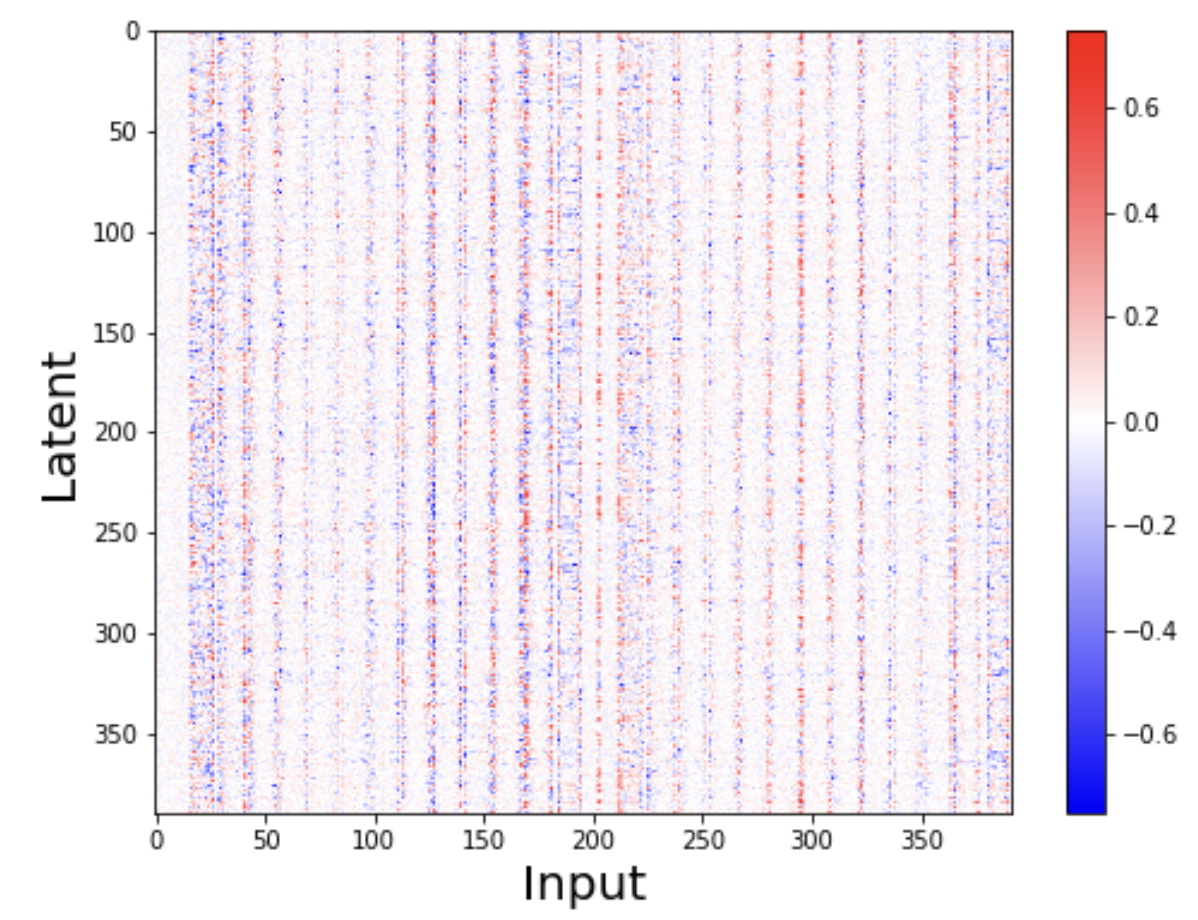} }}    ~~~
     \subfloat[Weights by CoCo (stacked color channels)]
    {{\includegraphics[width=0.25\textwidth]{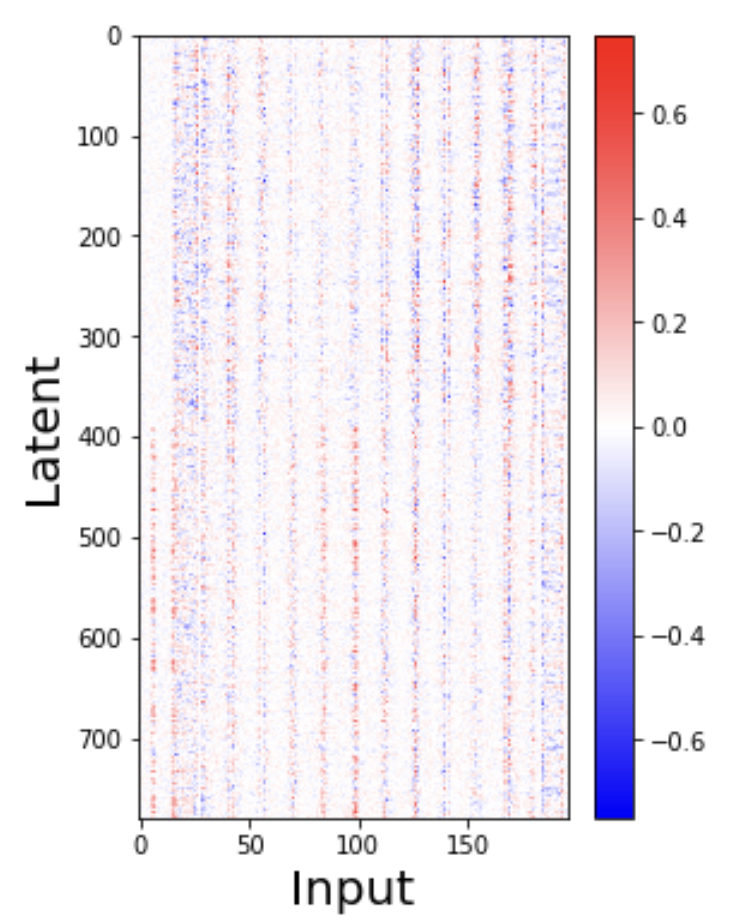} }} ~~~
        \subfloat[Weights by ERM]
    {{\includegraphics[width=0.35\textwidth]{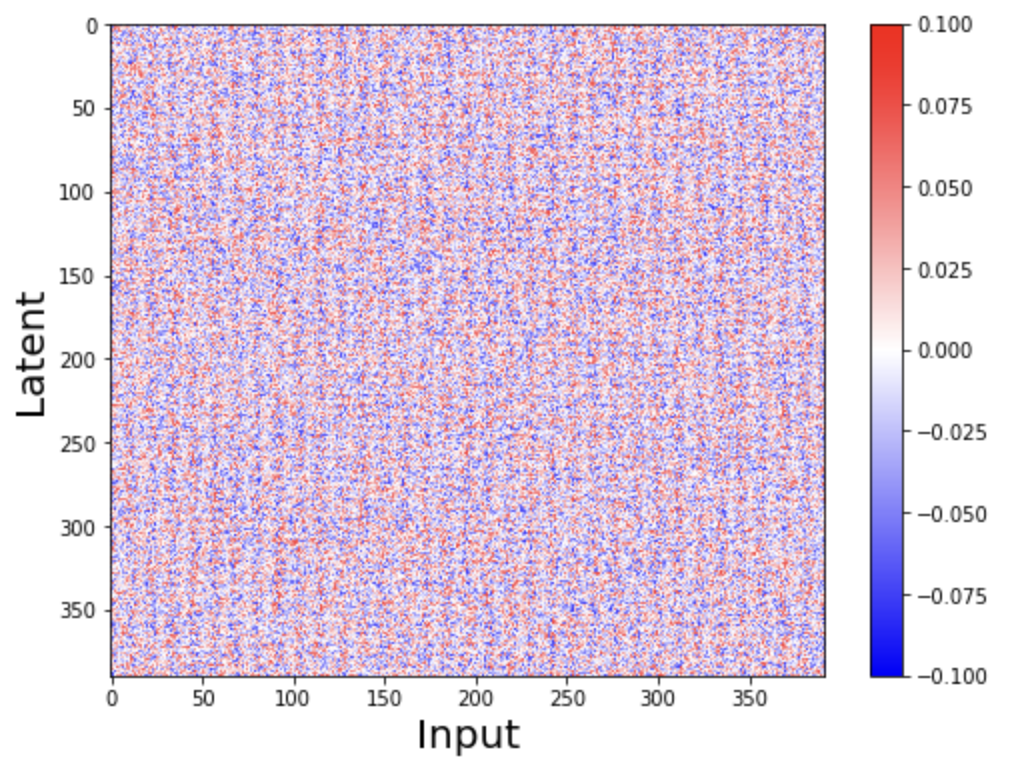} }}
    \caption{The weight matrix that connects the input and the first hidden layer for the CMNIST data. (a) The weights learned by CoCo set multiple columns close to zero, removing the dependence of the hidden layer on a set of pixels. (b) Stacking the weights corresponding to the two image channels (the left and right half of (a)). CoCo selects the same pixels over two color channels hence removing the color information from the hidden layer. (c) The weights  by ERM connect the hidden layer to all the inputs, so the prediction depends on the spurious association. }
    \label{fig:weights2}%
\end{figure}

This section contains experimental results additional to  \Cref{sec:experiment} in the main paper. \Cref{fig:4env} contain the MAEs for the linear synthetic data (\Cref{sec:linear-sync}). It shows the mean and minimal MAEs over ten random trials for two and four environments. \Cref{fig:weights2} shows the weight parameters learned by CoCo and ERM on CMNIST (\Cref{sec:mnist}), which provides explanations of why CoCo could learn a causal representation.  \Cref{tab:cmnist} contains more baseline results for comparison on the ColorMNIST data. \Cref{fig:wild} shows the trace plot of the predictive accuracy for the clean label $\tilde{y}$ prediction on the CMNIST data  (\Cref{sec:mnist}) and for the prediction on the iWildCam data (\Cref{sec:animal}). ~\looseness=-1

%% file: causal_optimization.bbl
\begin{thebibliography}{65}
\providecommand{\natexlab}[1]{#1}
\providecommand{\url}[1]{\texttt{#1}}
\expandafter\ifx\csname urlstyle\endcsname\relax
  \providecommand{\doi}[1]{doi: #1}\else
  \providecommand{\doi}{doi: \begingroup \urlstyle{rm}\Url}\fi

\bibitem[Ahuja et~al.(2020)Ahuja, Shanmugam, Varshney, and
  Dhurandhar]{ahuja2020invariant}
Kartik Ahuja, Karthikeyan Shanmugam, Kush Varshney, and Amit Dhurandhar.
\newblock Invariant risk minimization games.
\newblock In \emph{International Conference on Machine Learning}, pages
  145--155. PMLR, 2020.

\bibitem[Angrist et~al.(1996)Angrist, Imbens, and
  Rubin]{angrist1996identification}
Joshua~D Angrist, Guido~W Imbens, and Donald~B Rubin.
\newblock Identification of causal effects using instrumental variables.
\newblock \emph{Journal of the American statistical Association}, 91\penalty0
  (434):\penalty0 444--455, 1996.

\bibitem[Arjovsky et~al.(2019)Arjovsky, Bottou, Gulrajani, and
  Lopez-Paz]{arjovsky2019invariant}
Martin Arjovsky, L{\'e}on Bottou, Ishaan Gulrajani, and David Lopez-Paz.
\newblock Invariant risk minimization.
\newblock \emph{arXiv preprint arXiv:1907.02893}, 2019.

\bibitem[Bae et~al.(2021)Bae, Choi, and Lee]{bae2021meta}
Jun-Hyun Bae, Inchul Choi, and Minho Lee.
\newblock Meta-learned invariant risk minimization.
\newblock \emph{arXiv preprint arXiv:2103.12947}, 2021.

\bibitem[Bahng et~al.(2020)Bahng, Chun, Yun, Choo, and Oh]{bahng2020learning}
Hyojin Bahng, Sanghyuk Chun, Sangdoo Yun, Jaegul Choo, and Seong~Joon Oh.
\newblock Learning de-biased representations with biased representations.
\newblock In \emph{International Conference on Machine Learning}, pages
  528--539. PMLR, 2020.

\bibitem[Bai et~al.(2021)Bai, Sun, Hong, Zhou, Ye, Ye, Chan, and
  Li]{bai2021decaug}
Haoyue Bai, Rui Sun, Lanqing Hong, Fengwei Zhou, Nanyang Ye, Han-Jia Ye,
  S-H~Gary Chan, and Zhenguo Li.
\newblock Decaug: Out-of-distribution generalization via decomposed feature
  representation and semantic augmentation.
\newblock In \emph{Proceedings of the AAAI Conference on Artificial
  Intelligence}, volume~35, pages 6705--6713, 2021.

\bibitem[Beery et~al.(2019)Beery, Morris, and Perona]{beery2019iwildcam}
Sara Beery, Dan Morris, and Pietro Perona.
\newblock The {iWildCam} 2019 challenge dataset.
\newblock \emph{arXiv preprint arXiv:1907.07617}, 2019.

\bibitem[Bertsimas et~al.(2016)Bertsimas, King, and
  Mazumder]{bertsimas2016best}
Dimitris Bertsimas, Angela King, and Rahul Mazumder.
\newblock Best subset selection via a modern optimization lens.
\newblock \emph{The Annals of Statistics}, 44\penalty0 (2):\penalty0 813, 2016.

\bibitem[Brouillard et~al.(2020)Brouillard, Lachapelle, Lacoste,
  Lacoste-Julien, and Drouin]{brouillard2020differentiable}
Philippe Brouillard, S{\'e}bastien Lachapelle, Alexandre Lacoste, Simon
  Lacoste-Julien, and Alexandre Drouin.
\newblock Differentiable causal discovery from interventional data.
\newblock \emph{Advances in Neural Information Processing Systems}, 33, 2020.

\bibitem[B{\"u}hlmann et~al.(2020)]{buhlmann2020invariance}
Peter B{\"u}hlmann et~al.
\newblock Invariance, causality and robustness.
\newblock \emph{Statistical Science}, 35\penalty0 (3):\penalty0 404--426, 2020.

\bibitem[Chen et~al.(2022)Chen, Zhou, Bian, Xie, Ma, Zhang, Yang, Han, and
  Cheng]{chen2022pareto}
Yongqiang Chen, Kaiwen Zhou, Yatao Bian, Binghui Xie, Kaili Ma, Yonggang Zhang,
  Han Yang, Bo~Han, and James Cheng.
\newblock Pareto invariant risk minimization.
\newblock \emph{arXiv preprint arXiv:2206.07766}, 2022.

\bibitem[Christiansen et~al.(2020)Christiansen, Pfister, Jakobsen, Gnecco, and
  Peters]{christiansen2020causal}
Rune Christiansen, Niklas Pfister, Martin~Emil Jakobsen, Nicola Gnecco, and
  Jonas Peters.
\newblock A causal framework for distribution generalization.
\newblock \emph{arXiv e-prints}, pages arXiv--2006, 2020.

\bibitem[Cloudera(2020)]{causalsurvey2020}
Cloudera.
\newblock Causality for machine learning, 2020.
\newblock URL \url{https://ff13.fastforwardlabs.com}.

\bibitem[Dawid et~al.(2010)Dawid, Didelez, et~al.]{dawid2010identifying}
A~Philip Dawid, Vanessa Didelez, et~al.
\newblock Identifying the consequences of dynamic treatment strategies: A
  decision-theoretic overview.
\newblock \emph{Statistics Surveys}, 4:\penalty0 184--231, 2010.

\bibitem[Deng et~al.(2009)Deng, Dong, Socher, Li, Li, and
  Fei-Fei]{deng2009imagenet}
Jia Deng, Wei Dong, Richard Socher, Li-Jia Li, Kai Li, and Li~Fei-Fei.
\newblock Imagenet: A large-scale hierarchical image database.
\newblock In \emph{2009 IEEE Conference on Computer Vision and Pattern
  Recognition}, pages 248--255. Ieee, 2009.

\bibitem[Eberhardt and Scheines(2007)]{eberhardt2007interventions}
Frederick Eberhardt and Richard Scheines.
\newblock Interventions and causal inference.
\newblock \emph{Philosophy of Science}, 74\penalty0 (5):\penalty0 981--995,
  2007.

\bibitem[Efron(2020)]{efron2020prediction}
Bradley Efron.
\newblock Prediction, estimation, and attribution.
\newblock \emph{Journal of the American Statistical Association}, 115\penalty0
  (530):\penalty0 636--655, 2020.

\bibitem[Elwert and Winship(2014)]{elwert2014endogenous}
Felix Elwert and Christopher Winship.
\newblock Endogenous selection bias: The problem of conditioning on a collider
  variable.
\newblock \emph{Annual review of sociology}, 40:\penalty0 31--53, 2014.

\bibitem[Ghassami et~al.(2017)Ghassami, Salehkaleybar, Kiyavash, and
  Zhang]{ghassami2017learning}
Amir~Emad Ghassami, Saber Salehkaleybar, Negar Kiyavash, and Kun Zhang.
\newblock Learning causal structures using regression invariance.
\newblock In \emph{Neural Information Processing Systems}, pages 3015--3025,
  2017.

\bibitem[Ghassami(2020)]{ghassami2020causal}
Amiremad Ghassami.
\newblock \emph{Causal discovery beyond Markov equivalence}.
\newblock PhD thesis, University of Illinois at Urbana-Champaign, 2020.

\bibitem[Ghassami et~al.(2018)Ghassami, Kiyavash, Huang, and
  Zhang]{ghassami2018multi}
AmirEmad Ghassami, Negar Kiyavash, Biwei Huang, and Kun Zhang.
\newblock Multi-domain causal structure learning in linear systems.
\newblock In \emph{Neural Information Processing Systems}, pages 6269--6279,
  2018.

\bibitem[Golub et~al.(1999)Golub, Hansen, and O'Leary]{golub1999tikhonov}
Gene~H Golub, Per~Christian Hansen, and Dianne~P O'Leary.
\newblock Tikhonov regularization and total least squares.
\newblock \emph{SIAM journal on matrix analysis and applications}, 21\penalty0
  (1):\penalty0 185--194, 1999.

\bibitem[Guo et~al.(2021)Guo, Zhang, Liu, and Kiciman]{guo2021out}
Ruocheng Guo, Pengchuan Zhang, Hao Liu, and Emre Kiciman.
\newblock Out-of-distribution prediction with invariant risk minimization: The
  limitation and an effective fix.
\newblock \emph{arXiv preprint arXiv:2101.07732}, 2021.

\bibitem[Haavelmo(1944)]{haavelmo1944probability}
Trygve Haavelmo.
\newblock The probability approach in econometrics.
\newblock \emph{Econometrica: Journal of the Econometric Society}, pages
  iii--115, 1944.

\bibitem[Hastie et~al.(2009)Hastie, Tibshirani, Friedman, and
  Friedman]{hastie2009elements}
Trevor Hastie, Robert Tibshirani, Jerome~H Friedman, and Jerome~H Friedman.
\newblock \emph{The elements of statistical learning: data mining, inference,
  and prediction}, volume~2.
\newblock Springer, 2009.

\bibitem[He et~al.(2016)He, Zhang, Ren, and Sun]{he2016deep}
Kaiming He, Xiangyu Zhang, Shaoqing Ren, and Jian Sun.
\newblock Deep residual learning for image recognition.
\newblock In \emph{IEEE Conference on Computer Vision and Pattern Recognition},
  pages 770--778, 2016.

\bibitem[Heinze-Deml and Meinshausen(2021)]{heinze2021conditional}
Christina Heinze-Deml and Nicolai Meinshausen.
\newblock Conditional variance penalties and domain shift robustness.
\newblock \emph{Machine Learning}, 110\penalty0 (2):\penalty0 303--348, 2021.

\bibitem[Heinze-Deml et~al.(2018)Heinze-Deml, Peters, and
  Meinshausen]{heinze2018invariant}
Christina Heinze-Deml, Jonas Peters, and Nicolai Meinshausen.
\newblock Invariant causal prediction for nonlinear models.
\newblock \emph{Journal of Causal Inference}, 6\penalty0 (2), 2018.

\bibitem[Huang et~al.(2019)Huang, Zhang, Gong, and Glymour]{huang2019causal}
Biwei Huang, Kun Zhang, Mingming Gong, and Clark Glymour.
\newblock Causal discovery and forecasting in nonstationary environments with
  state-space models.
\newblock In \emph{International Conference on Machine Learning}, pages
  2901--2910. PMLR, 2019.

\bibitem[Huang et~al.(2020)Huang, Zhang, Gong, and Glymour]{huang2020causal}
Biwei Huang, Kun Zhang, Mingming Gong, and Clark Glymour.
\newblock Causal discovery from multiple data sets with non-identical variable
  sets.
\newblock In \emph{Proceedings of the AAAI Conference on Artificial
  Intelligence}, volume~34, pages 10153--10161, 2020.

\bibitem[Imbens and Rubin(2015)]{imbens2015causal}
Guido~W Imbens and Donald~B Rubin.
\newblock \emph{Causal inference in statistics, social, and biomedical
  sciences}.
\newblock Cambridge University Press, 2015.

\bibitem[Kamath et~al.(2021)Kamath, Tangella, Sutherland, and
  Srebro]{kamath2021does}
Pritish Kamath, Akilesh Tangella, Danica~J Sutherland, and Nathan Srebro.
\newblock Does invariant risk minimization capture invariance?
\newblock \emph{arXiv preprint arXiv:2101.01134}, 2021.

\bibitem[Kingma and Ba(2014)]{kingma2014adam}
Diederik~P Kingma and Jimmy Ba.
\newblock Adam: A method for stochastic optimization.
\newblock \emph{arXiv preprint arXiv:1412.6980}, 2014.

\bibitem[Krueger et~al.(2020)Krueger, Caballero, Jacobsen, Zhang, Binas, Priol,
  and Courville]{krueger2020out}
David Krueger, Ethan Caballero, Joern-Henrik Jacobsen, Amy Zhang, Jonathan
  Binas, Remi~Le Priol, and Aaron Courville.
\newblock Out-of-distribution generalization via risk extrapolation (rex).
\newblock \emph{arXiv preprint arXiv:2003.00688}, 2020.

\bibitem[Lu et~al.(2021)Lu, Wu, Hern{\'a}ndez-Lobato, and
  Sch{\"o}lkopf]{lu2021nonlinear}
Chaochao Lu, Yuhuai Wu, Jo{\'s}e~Miguel Hern{\'a}ndez-Lobato, and Bernhard
  Sch{\"o}lkopf.
\newblock Nonlinear invariant risk minimization: A causal approach.
\newblock \emph{arXiv preprint arXiv:2102.12353}, 2021.

\bibitem[Marban(1969)]{marban1969directional}
Jorge~A Marban.
\newblock \emph{Directional derivatives in classical optimization}.
\newblock PhD thesis, University of Florida, 1969.

\bibitem[Mooij et~al.(2020)Mooij, Magliacane, and Claassen]{mooij2020joint}
Joris~M Mooij, Sara Magliacane, and Tom Claassen.
\newblock Joint causal inference from multiple contexts.
\newblock \emph{Journal of Machine Learning Research}, 21\penalty0
  (99):\penalty0 1--108, 2020.

\bibitem[M{\"u}ller et~al.(2020)M{\"u}ller, Schmier, Ardizzone, Rother, and
  K{\"o}the]{muller2020learning}
Jens M{\"u}ller, Robert Schmier, Lynton Ardizzone, Carsten Rother, and Ullrich
  K{\"o}the.
\newblock Learning robust models using the principle of independent causal
  mechanisms.
\newblock \emph{arXiv preprint arXiv:2010.07167}, 2020.

\bibitem[Pearl(2009)]{pearl2009causality}
J.~Pearl.
\newblock \emph{Causality: Models, Reasoning and Inference}.
\newblock Cambridge University Press, 2009.

\bibitem[Peters and B{\"u}hlmann(2014)]{peters2014identifiability}
Jonas Peters and Peter B{\"u}hlmann.
\newblock Identifiability of gaussian structural equation models with equal
  error variances.
\newblock \emph{Biometrika}, 101\penalty0 (1):\penalty0 219--228, 2014.

\bibitem[Peters et~al.(2016)Peters, B{\"u}hlmann, and
  Meinshausen]{peters2016causal}
Jonas Peters, Peter B{\"u}hlmann, and Nicolai Meinshausen.
\newblock Causal inference by using invariant prediction: {I}dentification and
  confidence intervals.
\newblock \emph{Journal of the Royal Statistical Society. Series B (Statistical
  Methodology)}, pages 947--1012, 2016.

\bibitem[Pfister et~al.(2019)Pfister, B{\"u}hlmann, and
  Peters]{pfister2019invariant}
Niklas Pfister, Peter B{\"u}hlmann, and Jonas Peters.
\newblock Invariant causal prediction for sequential data.
\newblock \emph{Journal of the American Statistical Association}, 114\penalty0
  (527):\penalty0 1264--1276, 2019.

\bibitem[Rame et~al.(2022)Rame, Dancette, and Cord]{rame2022fishr}
Alexandre Rame, Corentin Dancette, and Matthieu Cord.
\newblock Fishr: Invariant gradient variances for out-of-distribution
  generalization.
\newblock In \emph{International Conference on Machine Learning}, pages
  18347--18377. PMLR, 2022.

\bibitem[Rojas-Carulla et~al.(2018)Rojas-Carulla, Sch{\"o}lkopf, Turner, and
  Peters]{rojas2018invariant}
Mateo Rojas-Carulla, Bernhard Sch{\"o}lkopf, Richard Turner, and Jonas Peters.
\newblock Invariant models for causal transfer learning.
\newblock \emph{Journal of Machine Learning Research}, 19\penalty0
  (1):\penalty0 1309--1342, 2018.

\bibitem[Rosenfeld et~al.(2020)Rosenfeld, Ravikumar, and
  Risteski]{rosenfeld2020risks}
Elan Rosenfeld, Pradeep Ravikumar, and Andrej Risteski.
\newblock The risks of invariant risk minimization.
\newblock \emph{arXiv preprint arXiv:2010.05761}, 2020.

\bibitem[Rothenh{\"a}usler et~al.(2019)Rothenh{\"a}usler, B{\"u}hlmann, and
  Meinshausen]{rothenhausler2019causal}
Dominik Rothenh{\"a}usler, Peter B{\"u}hlmann, and Nicolai Meinshausen.
\newblock Causal dantzig: fast inference in linear structural equation models
  with hidden variables under additive interventions.
\newblock \emph{The Annals of Statistics}, 47\penalty0 (3):\penalty0
  1688--1722, 2019.

\bibitem[Rothenh{\"a}usler et~al.(2021)Rothenh{\"a}usler, Meinshausen,
  B{\"u}hlmann, and Peters]{rothenhausler2018anchor}
Dominik Rothenh{\"a}usler, Nicolai Meinshausen, Peter B{\"u}hlmann, and Jonas
  Peters.
\newblock Anchor regression: Heterogeneous data meet causality.
\newblock \emph{Journal of the Royal Statistical Society: Series B (Statistical
  Methodology)}, 83\penalty0 (2):\penalty0 215--246, 2021.

\bibitem[Rudin et~al.(1964)]{rudin1964principles}
Walter Rudin et~al.
\newblock \emph{Principles of mathematical analysis}, volume~3.
\newblock McGraw-hill New York, 1964.

\bibitem[Sagawa et~al.(2019)Sagawa, Koh, Hashimoto, and
  Liang]{sagawa2019distributionally}
Shiori Sagawa, Pang~Wei Koh, Tatsunori~B Hashimoto, and Percy Liang.
\newblock Distributionally robust neural networks for group shifts: On the
  importance of regularization for worst-case generalization.
\newblock \emph{arXiv preprint arXiv:1911.08731}, 2019.

\bibitem[Sch{\"o}lkopf(2019)]{scholkopf2019causality}
Bernhard Sch{\"o}lkopf.
\newblock Causality for machine learning.
\newblock \emph{arXiv preprint arXiv:1911.10500}, 2019.

\bibitem[Sch{\"o}lkopf et~al.(2012)Sch{\"o}lkopf, Janzing, Peters, Sgouritsa,
  Zhang, and Mooij]{scholkopf2012causal}
Bernhard Sch{\"o}lkopf, Dominik Janzing, Jonas Peters, Eleni Sgouritsa, Kun
  Zhang, and Joris Mooij.
\newblock On causal and anticausal learning.
\newblock \emph{arXiv preprint arXiv:1206.6471}, 2012.

\bibitem[Sch{\"o}lkopf et~al.(2021)Sch{\"o}lkopf, Locatello, Bauer, Ke,
  Kalchbrenner, Goyal, and Bengio]{scholkopf2021toward}
Bernhard Sch{\"o}lkopf, Francesco Locatello, Stefan Bauer, Nan~Rosemary Ke, Nal
  Kalchbrenner, Anirudh Goyal, and Yoshua Bengio.
\newblock Toward causal representation learning.
\newblock \emph{Proceedings of the IEEE}, 109\penalty0 (5):\penalty0 612--634,
  2021.

\bibitem[Shi et~al.(2020)Shi, Veitch, and Blei]{shi2020invariant}
Claudia Shi, Victor Veitch, and David Blei.
\newblock Invariant representation learning for treatment effect estimation.
\newblock \emph{arXiv preprint arXiv:2011.12379}, 2020.

\bibitem[Shmueli et~al.(2010)]{shmueli2010explain}
G.~Shmueli et~al.
\newblock To explain or to predict?
\newblock \emph{Statistical Science}, 25\penalty0 (3):\penalty0 289--310, 2010.

\bibitem[Spirtes et~al.(2000)Spirtes, Glymour, Scheines, and
  Heckerman]{spirtes2000causation}
Peter Spirtes, Clark~N Glymour, Richard Scheines, and David Heckerman.
\newblock \emph{Causation, prediction, and search}.
\newblock Springer, 2000.

\bibitem[Tian and Pearl(2001)]{tian2001causal}
Jin Tian and Judea Pearl.
\newblock Causal discovery from changes.
\newblock In \emph{Proceedings of the Seventeenth conference on Uncertainty in
  artificial intelligence}, pages 512--521, 2001.

\bibitem[Tibshirani(1996)]{tibshirani1996regression}
Robert Tibshirani.
\newblock Regression shrinkage and selection via the lasso.
\newblock \emph{Journal of the Royal Statistical Society: Series B
  (Methodological)}, 58\penalty0 (1):\penalty0 267--288, 1996.

\bibitem[Winkler et~al.(2019)Winkler, Fink, Toberer, Enk, Deinlein,
  Hofmann-Wellenhof, Thomas, Lallas, Blum, Stolz,
  et~al.]{winkler2019association}
Julia~K Winkler, Christine Fink, Ferdinand Toberer, Alexander Enk, Teresa
  Deinlein, Rainer Hofmann-Wellenhof, Luc Thomas, Aimilios Lallas, Andreas
  Blum, Wilhelm Stolz, et~al.
\newblock Association between surgical skin markings in dermoscopic images and
  diagnostic performance of a deep learning convolutional neural network for
  melanoma recognition.
\newblock \emph{JAMA dermatology}, 155\penalty0 (10):\penalty0 1135--1141,
  2019.

\bibitem[Wright(1921)]{wright1921correlation}
Sewall Wright.
\newblock Correlation and causation.
\newblock \emph{Journal of agricultural research}, 20:\penalty0 557--580, 1921.

\bibitem[Xie et~al.(2020)Xie, Chen, Liu, and Li]{xie2020risk}
Chuanlong Xie, Fei Chen, Yue Liu, and Zhenguo Li.
\newblock Risk variance penalization: From distributional robustness to
  causality.
\newblock \emph{arXiv preprint arXiv:2006.07544}, 2020.

\bibitem[{Yin} et~al.(2020){Yin}, {Ho}, {Yan}, {Qian}, and
  {Zhou}]{yin2020probabilistic}
Mingzhang {Yin}, Nhat {Ho}, Bowei {Yan}, Xiaoning {Qian}, and Mingyuan {Zhou}.
\newblock {Probabilistic Best Subset Selection by Gradient-Based Optimization}.
\newblock \emph{arXiv e-prints}, 2020.

\bibitem[Yu et~al.(2019{\natexlab{a}})Yu, Liu, and Li]{yu2019learning}
Kui Yu, Lin Liu, and Jiuyong Li.
\newblock Learning markov blankets from multiple interventional data sets.
\newblock \emph{IEEE transactions on neural networks and learning systems},
  31\penalty0 (6):\penalty0 2005--2019, 2019{\natexlab{a}}.

\bibitem[Yu et~al.(2019{\natexlab{b}})Yu, Liu, Li, Ding, and Le]{yu2019multi}
Kui Yu, Lin Liu, Jiuyong Li, Wei Ding, and Thuc~Duy Le.
\newblock Multi-source causal feature selection.
\newblock \emph{IEEE transactions on pattern analysis and machine
  intelligence}, 42\penalty0 (9):\penalty0 2240--2256, 2019{\natexlab{b}}.

\bibitem[Zhang et~al.(2020)Zhang, Lyle, Sodhani, Filos, Kwiatkowska, Pineau,
  Gal, and Precup]{zhang2020invariant}
Amy Zhang, Clare Lyle, Shagun Sodhani, Angelos Filos, Marta Kwiatkowska, Joelle
  Pineau, Yarin Gal, and Doina Precup.
\newblock Invariant causal prediction for block {MDP}s.
\newblock In \emph{International Conference on Machine Learning}, pages
  11214--11224. PMLR, 2020.

\bibitem[Zhao and Yu(2006)]{zhao2006model}
Peng Zhao and Bin Yu.
\newblock On model selection consistency of lasso.
\newblock \emph{The Journal of Machine Learning Research}, 7:\penalty0
  2541--2563, 2006.

\end{thebibliography}
